\documentclass[aps,prb,reprint,letterpaper,superscriptaddress,floatfix]{revtex4-2}

\usepackage{amsmath,amssymb,amsthm,bm}
\usepackage{graphicx}
\usepackage{xcolor}
\usepackage{tabularx}
\usepackage{array}
\usepackage[colorlinks=true,linkcolor=blue,citecolor=blue,urlcolor=blue]{hyperref}

\newcommand{\bk}{\mathbf{k}}
\newcommand{\bQ}{\mathbf{Q}}
\newcommand{\bD}{\mathbf{D}}
\newcommand{\bM}{\mathbf{M}}

\newcommand{\bR}{\mathbf{R}}
\newcommand{\tr}{\operatorname{Tr}}

\begin{document}

\title{Layer-Parity Control of the Superconducting Diode Effect in Monolayer NbSe$_2$}

\author{Yusuf Wicaksono}
\email{Contact author: WICAKSONO.Yusuf@nims.go.jp}
\affiliation{International Center for Materials Nanoarchitectonics (WPI-MANA), National Institute for Materials Science, 1-1 Namiki, Tsukuba, Ibaraki 305-0044, Japan}

\author{Toshikaze Kariyado}
\email{Contact author: KARIYADO.Toshikaze@nims.go.jp}
\affiliation{International Center for Materials Nanoarchitectonics (WPI-MANA), National Institute for Materials Science, 1-1 Namiki, Tsukuba, Ibaraki 305-0044, Japan}

\date{\today}

\begin{abstract}
The superconducting diode effect is often associated with finite-momentum
pairing, although model studies indicate that rectification reflects the
asymmetry of the free energy about its minimum rather than the equilibrium
pair momentum itself. Using a Wannier-based, self-consistent finite-$\mathbf Q$
Bogoliubov--de Gennes approach with a Fermi-surface-adaptive quadrature that
converges the critical currents, we show that the layer parity and
orientation of magnetic proximity select the diode response of monolayer
NbSe$_2$. For in-plane exchange, the equilibrium pair momentum
$\mathbf Q_{\rm eq}$ and the diode efficiency are both linear in the
layer-odd exchange contrast: parallel proximity on the two Se sublayers is
reciprocal, and antiparallel proximity doubles the response. An
out-of-plane component separates the two observables. It leaves
$\mathbf Q_{\rm eq}$ nearly unchanged but, through pair breaking, suppresses
the efficiency by more than an order of magnitude and, at low temperature,
reverses it. Purely out-of-plane
exchange, for which $C_3$ symmetry forbids a helical term, keeps
$\mathbf Q_{\rm eq}=0$ yet produces a trigonal diode effect, controlled by the layer-even exchange and
larger than the in-plane one above $0.65$~meV at $0.58$~K. Magnetic layer parity
thus selects between a helical and a trigonal diode channel. 
This trigonal channel provides a candidate intrinsic contribution to the
field-free rectification reported in NbSe$_2$/CrBr$_3$.
The model predicts that two-sided out-of-plane proximity 
cancels it when antiparallel and, at the same
exchange per layer, enhances it about fourfold when parallel, in the
computed range.
\end{abstract}

\maketitle


\emph{Introduction.}
The most direct manifestation of nonreciprocal superconductivity is the
superconducting diode effect (SDE), for which the positive and negative
critical currents are unequal,
$K_c^+\neq |K_c^-|$
\cite{AndoNature20,NadeemNatRevPhys23,DavydovaSciAdv24,ShafferLevchenkoArXiv25}.
The SDE has been studied in superconductors with broken inversion and
time-reversal symmetries, where spin--orbit coupling and Zeeman or exchange
fields can favor finite-momentum pairing
\cite{WakatsukiSciAdv17,HoshinoPRB18,YuanFuPNAS22,DaidoPRL22,
HeNJP22,IlicPRL22}, which connects the SDE to a helical state with an
equilibrium pair momentum $Q_{\rm eq}\neq0$.
Ginzburg--Landau and model calculations have shown, however, that the
diode efficiency is governed by the asymmetry of the free energy about its
minimum rather than by the finite momentum itself
\cite{HeNJP22,DaidoYanasePRB22,ShafferLevchenkoArXiv25}; with threefold
rotation symmetry, a cubic term can rectify at zero equilibrium pair momentum
~\cite{ShafferLevchenkoArXiv25}.
The two quantities probe different parts of the same free-energy landscape:
$Q_{\rm eq}$ is fixed by its minimum,
$\partial_Q\Omega_{\rm SC}(Q_{\rm eq})=0$, whereas the critical currents are
extrema of the supercurrent $K(Q)\propto\partial_Q\Omega_{\rm SC}(Q)$ near
directional depairing. For Ising superconductors, the in-plane diode effect
has been analyzed in effective models of the $K$ and $K'$ pockets with a
substrate that breaks the horizontal
mirror~\cite{BankierPRB25,LevchenkoKhodasArXiv26}.
How this distinction plays out on the full multiband Fermi surface, and how
layer-resolved exchange selects between these mechanisms, remains open.

In monolayer $1H$-NbSe$_2$, broken inversion symmetry and strong
spin--orbit coupling generate Ising spin--valley locking, while
superconductivity survives in the atomically thin limit
\cite{UgedaNatPhys15,XiNatPhys15,delaBarreraNatComm18}.
In few-layer NbSe$_2$, the diode effect is driven by the out-of-plane field
component and has been attributed to valley-Zeeman coupling; an in-plane
field alone does not rectify, and its influence remains
unexplained \cite{BauriedlNatCommun22}.
Magnetic proximity replaces the field by an interfacial exchange.
On CrPS$_4$, an A-type antiferromagnet whose layers are magnetized out of
plane \cite{PengAdvMater20}, NbSe$_2$ shows a proximity-induced diode
effect in weak out-of-plane fields, and in CrPS$_4$/NbSe$_2$/CrPS$_4$ spin
valves the response depends on the relative alignment of the two magnets;
this was discussed in terms of finite-momentum pairing \cite{YunPRR23}.
In NbSe$_2$/CrBr$_3$~\cite{KezilebiekeNature20}, also magnetized out of
plane, a large field-free diode effect, with polarity reversals traced to
switching of the interfacial magnetic layer, has instead been attributed to
a cubic term that leaves the equilibrium condensate at zero pair momentum
\cite{SunArXiv26}.
Which exchange orientations and layer configurations rectify, and whether a
finite pair momentum is involved, thus remains unresolved.
Because van der Waals interfaces produce strongly layer-resolved exchange
\cite{ZhongNatNano20,MatsuokaNatCommun22,WicaksonoPRB26}, the exchange on
the two Se sublayers can differ, making its layer parity a control
parameter.

Here we combine a fully relativistic Wannier description of monolayer
NbSe$_2$ with a self-consistent finite-$\mathbf Q$ Bogoliubov--de Gennes
treatment that resolves the gapless depairing regime.
The horizontal mirror assigns each diode channel to one magnetic layer
parity.
Layer-odd in-plane exchange produces a helical response, in which
$\mathbf Q_{\rm eq}$ and the diode efficiency $\eta$ are both linear in
the exchange contrast over the range studied; an out-of-plane component
suppresses and, at low
temperature, reverses $\eta$ but leaves $\mathbf Q_{\rm eq}$ nearly
unchanged.
Layer-even out-of-plane exchange produces a trigonal diode effect with
$\mathbf Q_{\rm eq}=0$ that exceeds the helical one at strong exchange.

\emph{Model and finite-$Q$ superconductivity.}
We describe monolayer $1H$-NbSe$_2$ using a fully relativistic Wannier
Hamiltonian~\cite[Sec.~S2.1]{SMnote} constructed from Nb $d$ and Se $p$
orbitals \cite{qespresso,qe2,Mostofi2008,Pizzi2020}: five Nb $d$ orbitals
and three Se $p$ orbitals on each chalcogen sublayer, or 22 spinor Wannier
states.
Magnetic proximity is modeled by a layer-selective exchange field acting on
the Se orbitals,
\begin{equation}
    H_{\rm ex}
    =
    \sum_{\ell=t,b}
    P_\ell\,\mathbf{D}_\ell\cdot\boldsymbol{\sigma},
    \label{eq:Hex}
\end{equation}
where $P_\ell$ projects onto the top ($t$) or bottom ($b$) Se sublayer,
$\boldsymbol{\sigma}$ are the Pauli matrices, and
$\mathbf{D}_\ell=D_\ell(\sin\theta_\ell\cos\phi_\ell,
\sin\theta_\ell\sin\phi_\ell,\cos\theta_\ell)$
with $\mathbf a_1\parallel\hat{\mathbf x}$.
The layer-even and layer-odd components are
$\mathbf D_\pm=(\mathbf D_b\pm\mathbf D_t)/2$, and the in-plane exchange
contrast is $\Delta\mathbf D_\parallel=2\mathbf D_{-,\parallel}$, with $x$
component $\Delta D_x=D_{b,x}-D_{t,x}$.
We use $|\mathbf D_\ell|\le1.84$~meV, with $1$~meV as the reference.
The bottom Se sublayer carries on average $0.10$--$0.16$ of the
Fermi-surface weight, so that $D_{b,z}=1$~meV shifts the states by
$0.10$--$0.20$~meV, the spin Zeeman energy of a $1.7$--$3.5$~T field, and
in-plane exchange by less than $0.04$~meV~\cite[Sec.~S2.2]{SMnote}, below
the few-meV proximity splittings of transition-metal dichalcogenides on
magnetic insulators~\cite{ZhaoNatNano17,ZollnerPRB19}.

To describe current-carrying states, Cooper pairs acquire a center-of-mass
wave vector $\mathbf Q$. With $\mathbf{k}_{\pm}=\pm\mathbf{k}+\mathbf{Q}/2$,
the Bogoliubov--de Gennes Hamiltonian
\cite{deGennesBook66,FuldeFerrellPR64,DaidoPRL22,YuanFuPNAS22}
in the Nambu basis $(c_{\mathbf{k}_{+}},c^\dagger_{\mathbf{k}_{-}})^T$ is
\begin{equation}
H_{\rm BdG}(\mathbf{k},\mathbf{Q})
=
\begin{pmatrix}
H_e(\mathbf{k}_{+})-\mu & \hat{\Delta}\\
\hat{\Delta}^{\dagger} & -H_e^{*}(\mathbf{k}_{-})+\mu
\end{pmatrix},
\label{eq:BdG}
\end{equation}
where $H_e(\mathbf{k})=H_0(\mathbf{k})+H_{\rm ex}$.
We assume a common local spin-singlet amplitude on the Nb $d_{z^2}$,
$d_{x^2-y^2}$, and $d_{xy}$ orbitals,
$\hat{\Delta}=\Delta P_{\rm pair}\otimes i\sigma_y$, and determine
$\Delta^\star(\mathbf Q)$ self-consistently for each $\mathbf Q$.
This ansatz, used in models of Ising superconductivity in
NbSe$_2$~\cite{HeCommPhys18} and consistent with phonon-mediated
pairing~\cite{LianNanoLett18}, covers a fraction $0.73$ of the Fermi-level
weight and preserves the crystal symmetries, so it does not bias the
selection rules~\cite[Sec.~S6.3]{SMnote}.
It retains induced triplet correlations but no attractive triplet channel,
which can reshape the in-plane diode response~\cite{LevchenkoKhodasArXiv26}.
The interaction is calibrated in pristine NbSe$_2$ to
$\Delta^\star=0.4$~meV at $T=0.12$~K, motivated by the gap fitted to
tunneling spectra of monolayer NbSe$_2$ at $0.34$~K~\cite{DreherACSNano21},
and then kept fixed; because pairing acts only on these Nb orbitals, the
model spectral gap is smaller, $0.24$~meV~\cite[Sec.~S2.3]{SMnote}.
Unless stated otherwise, $\mu_0=-2.4083$~eV and $T=0.58$~K
($k_BT=5\times10^{-5}$~eV), about $0.3\,T_c$ for the model
$T_c\approx1.9$~K estimated from $\Delta^\star(T)$~\cite[Sec.~S6.2]{SMnote}.
We use the unreconstructed monolayer without charge-density-wave order.
The selection rules assume mirror-equivalent interfaces; the computed
magnitudes are predictions of this effective model rather than
parameter-free values for a specific
heterostructure~\cite[Secs.~S2.2 and S6.4]{SMnote}.
The critical states are gapless~\cite[Sec.~S6.1]{SMnote}
and contain Bogoliubov Fermi surfaces~\cite{AkbariPRR22,ZhuScience21},
whose quasiparticle back-flow is poorly resolved on uniform meshes: with up to $1803\times1803$ points, $\eta$ scatters by more than
one percentage point~\cite[Sec.~S3.1]{SMnote}. A Fermi-surface-adaptive
quadrature (effective resolution $5427\times5427$ at two levels; three
levels for out-of-plane states) gives an estimated accuracy of about $0.1$
percentage points in $\eta$ for the tested cases;
higher-precision differences below compare states on a common
quadrature~\cite[Secs.~S3 and S4.6]{SMnote}.

On the selected self-consistent branch, we require
$\Delta\Omega(\mathbf Q)=
\Omega_{\rm SC}(\mathbf Q)-\Omega_{\rm N}(\mathbf Q)<0$,
where $\Omega$ is the grand potential per primitive cell of area
$A_{\rm uc}=10.1443~\text{\AA}^2$; for representative states, the selected
gap minimizes $\Omega_{\rm SC}(\Delta)$ over the scanned amplitudes of the
scalar-gap ansatz at the equilibrium and critical
momenta~\cite[Sec.~S6.5]{SMnote}.
For a scan $\mathbf Q=Q\hat{\mathbf n}$, the sheet supercurrent and diode
efficiency are
\begin{equation}
  \begin{aligned}
    K(Q)&=\frac{2e}{\hbar A_{\rm uc}}
    \frac{d\Omega_{\rm SC}(Q)}{dQ},
    \qquad K(Q_{\rm eq})=0,\\
    \eta&=\frac{K_c^+-|K_c^-|}{K_c^++|K_c^-|},
  \end{aligned}
  \label{eq:observables}
\end{equation}
where $K_c^+>0$ and $K_c^-<0$ are the extrema of $K(Q)$ along the branch
($1$~A/m $=1~\mu$A/$\mu$m) and $Q_{\rm eq}$ minimizes $\Omega_{\rm SC}$
along the scan.
These extrema are intrinsic depairing currents of the uniform state;
device switching can also involve vortices, edge barriers, Meissner
currents, and orbital effects~\cite{HouPRL23,KochanArXiv23}.
The expansion about $Q=0$~\cite{HeNJP22,YuanFuPNAS22,DaidoYanasePRB22} is
\begin{equation}
    \Omega_{\rm SC}(Q)
    =
    \Omega_0+\gamma_1Q+\gamma_2Q^2+\gamma_3Q^3+\gamma_4Q^4+\cdots,
    \label{eq:OmegaExpansion}
\end{equation}
which, for a small shift and $\gamma_2>0$, gives
$Q_{\rm eq}\simeq-\gamma_1/(2\gamma_2)$.
By contrast, $\eta$ depends on the part of the free energy
that is odd about this minimum, near the depairing regions.
A rigid shift of an even free energy produces no such asymmetry.
In general, writing $q=Q-Q_{\rm eq}$, the leading allowed odd
term is cubic, with coefficient $\gamma_3+4\gamma_4Q_{\rm eq}$
within the quartic truncation.
A finite equilibrium momentum therefore does not by itself imply a diode
effect.

\begin{figure*}[tb]
\includegraphics[width=\textwidth]{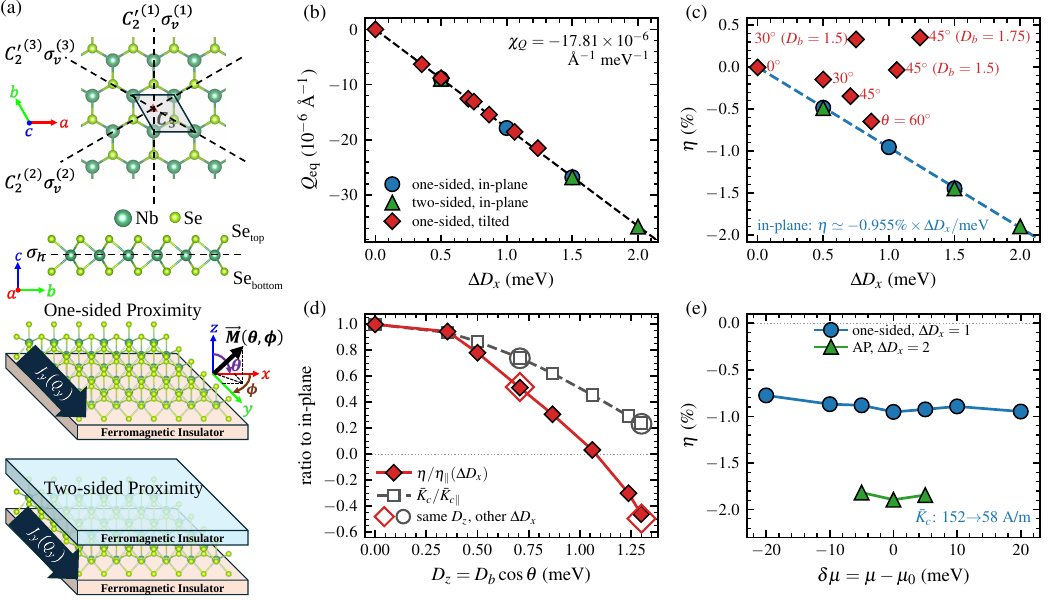}
\caption{
Helical response and its separation from rectification.
(a) Crystal structure of monolayer $1H$-NbSe$_2$ and one- and two-sided
magnetic-proximity geometries; $(\theta,\phi)$ parameterize the proximity
magnetization. Dashed lines mark the vertical mirrors $\sigma_v$ and the
$C_2'$ axes; with $\mathbf a_1\parallel\hat{\mathbf x}$ (zigzag), they run
along $\hat{\mathbf y}$ and its $C_3$ images.
(b) Equilibrium pair momentum versus the in-plane exchange contrast
$\Delta D_x=D_{b,x}-D_{t,x}$ for one-sided and two-sided in-plane textures
and for tilted one-sided textures (filled diamonds; Table~S6), all
scanned along $\hat{\mathbf y}$. The dashed line is
$Q_{\rm eq}=\chi_Q\Delta D_x$.
(c) Diode efficiency for the same states (except $\theta=45^\circ$,
$D_b=0.5$~meV, which lies on the in-plane line). In-plane textures (circles and
triangles) follow a single line; tilted textures (diamonds, labeled by
$\theta$) have a smaller magnitude and, for $D_z\gtrsim1.1$~meV, the
opposite sign.
(d) Efficiency of the tilted textures relative to the in-plane line at the
same $\Delta D_x$, and their mean critical current
$\bar K_c=(K_c^++|K_c^-|)/2$ relative to the in-plane value, versus the
out-of-plane component $D_z$. Large open symbols: states that repeat a
$D_z$ with a different $\Delta D_x$.
(e) Diode efficiency versus carrier filling $\delta\mu=\mu-\mu_0$ for the
one-sided in-plane state ($D_b=1$~meV) and the antiparallel (AP) texture.
All results at $k_BT=5\times10^{-5}$~eV on the converged quadrature
(three-level for the tilted textures with $D_z>1$~meV).}
\label{fig:fig1}
\end{figure*}

\emph{Helical response and layer parity.}
Figure~\ref{fig:fig1}(b) shows $Q_{\rm eq}$ for one-sided, two-sided, and
tilted configurations scanned along $\hat{\mathbf y}$.
All follow $Q_{\rm eq}=\chi_Q\Delta D_x$ with
$\chi_Q=-17.8\times10^{-6}~\text{\AA}^{-1}\,{\rm meV}^{-1}$, to within
$0.53\times10^{-6}~\text{\AA}^{-1}$: within this accuracy and the
parameter range studied, the helical
displacement is insensitive to the layer-even exchange and the
out-of-plane component.
Symmetry fixes this form.
The horizontal mirror $\sigma_h$ exchanges the two Se sublayers and
reverses in-plane axial vectors, so parallel in-plane exchange
($\Delta D_x=0$) is invariant under $\sigma_h\mathcal T$, which maps
$\mathbf Q\rightarrow-\mathbf Q$ and makes the state
reciprocal~\cite[Sec.~S1]{SMnote}.
The vertical mirror $M_x$ ($x\rightarrow-x$) combined with time reversal
similarly makes configurations with $D_{t,x}=D_{b,x}=0$ reciprocal along
$\hat{\mathbf y}$.
The only term linear in both $\mathbf Q$ and the exchange is then
$\mathbf Q\cdot(\hat{\mathbf z}\times\Delta\mathbf D_\parallel)$, giving
$\mathbf Q_{\rm eq}\simeq
\chi_Q\hat{\mathbf z}\times\Delta\mathbf D_\parallel$
\cite[Sec.~S4.4]{SMnote}; scans along $\mathbf b_1$ reproduce its
projections to within $0.02\times10^{-6}~\text{\AA}^{-1}$.

For in-plane exchange, the diode efficiency follows the same rule
[Fig.~\ref{fig:fig1}(c)],
\begin{equation}
  \eta\simeq-0.955\%\times\Delta D_x/{\rm meV},
  \label{eq:eta_linear}
\end{equation}
over $0.5\leq\Delta D_x\leq2$~meV; configurations with the same
$\Delta D_x$ but different layer-even exchange agree to within $0.005$
percentage points~\cite[Sec.~S4.2]{SMnote}.
The mean critical current, $\bar K_c=(K_c^++|K_c^-|)/2\simeq96$~A/m, stays
close to its pristine value because Ising spin--valley locking protects the
condensate against in-plane exchange.
Along $\mathbf b_1$, $\eta$ follows the projection to within a relative
$0.04$~\cite[Sec.~S4.4]{SMnote}.
Antiparallel exchange (AP, $D_{b,x}=-D_{t,x}=1$~meV) gives
$Q_{\rm eq}=-35.7\times10^{-6}~\text{\AA}^{-1}$ and $\eta=-1.90\%$, twice
the bottom-only values, and reversing all exchange fields reverses
both~\cite[Sec.~S4.3]{SMnote}.

\begin{figure*}[t]
\includegraphics[width=\textwidth]{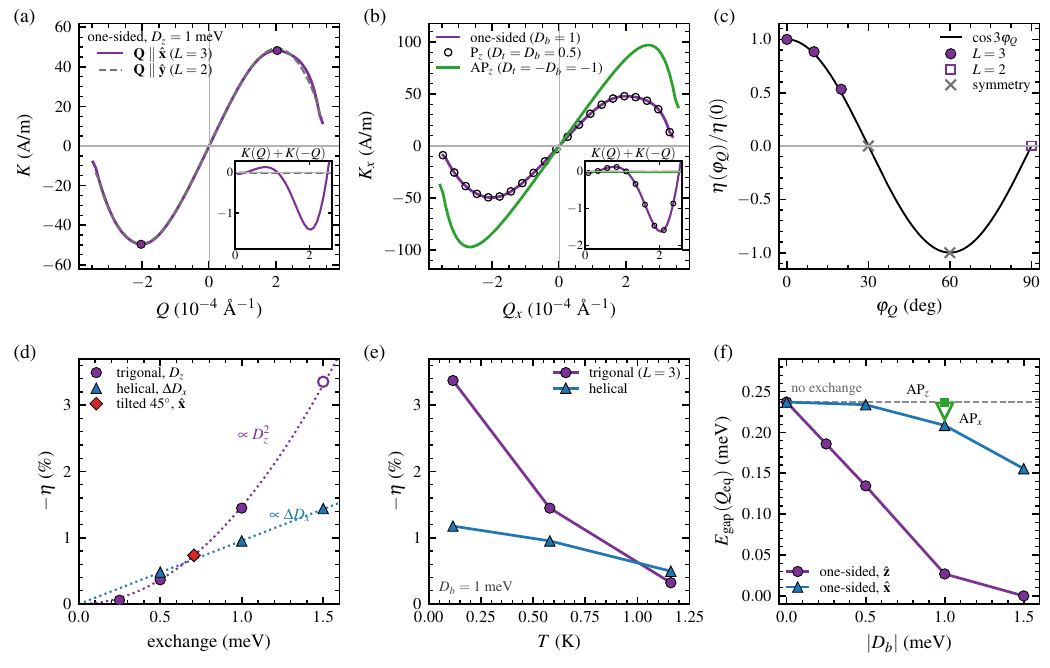}
\caption{
Trigonal diode channel of out-of-plane exchange.
(a) Sheet current for the one-sided out-of-plane texture ($D_z=1$~meV)
scanned along the mirror normal $\hat{\mathbf x}$ (filled circles: $K_c^\pm$)
and along the mirror line $\hat{\mathbf y}$ (two-level quadrature). Inset: $K(Q)+K(-Q)$, twice the
even-in-$Q$ part, which vanishes along $\hat{\mathbf y}$.
(b) Layer parity, all on the two-level quadrature: the parallel texture
P$_z$ reproduces the one-sided state with the same layer-even exchange
$D_{+,z}=0.5$~meV, while the antiparallel texture AP$_z$ ($D_{+,z}=0$) is
reciprocal and not depaired. Inset as in (a).
(c) Efficiency versus scan direction, normalized to $\varphi_Q=0$; crosses
are fixed by symmetry.
(d) $-\eta$ versus exchange for the trigonal channel (one-sided $D_z$, along
$\hat{\mathbf x}$; open circle: two-level quadrature) and the helical channel
(one-sided $\Delta D_x$, along $\hat{\mathbf y}$). The diamond is the tilted
texture $\theta=45^\circ$, $D_b=1$~meV, scanned along $\hat{\mathbf x}$ and
plotted at its $D_z$. Dotted lines: $1.47\%\times(D_z/{\rm meV})^2$ and
$0.955\%\times\Delta D_x/{\rm meV}$.
(e) $-\eta$ versus temperature for $D_b=1$~meV.
(f) Spectral gap at the equilibrium momentum for one-sided out-of-plane
($\hat{\mathbf z}$) and in-plane ($\hat{\mathbf x}$) textures and for the antiparallel textures $D_t=-D_b$
($\mathbf Q_{\rm eq}=0$ for the out-of-plane textures); dashed line: no
exchange.
$k_BT=5\times10^{-5}$~eV unless noted; three-level quadrature for the
trigonal channel unless noted.}
\label{fig:fig2}
\end{figure*}

An out-of-plane component breaks this correspondence.
The tilted configurations in Fig.~\ref{fig:fig1}(b) and (c) keep the
helical momentum of their in-plane component but have a much smaller
$|\eta|$.
Figure~\ref{fig:fig1}(d) shows that the suppression is governed by
$D_z=D_b\cos\theta$: configurations with equal $D_z$ but different in-plane
components give nearly identical ratios~\cite[Sec.~S4.1]{SMnote}.
For example, $\theta=45^\circ$ with $D_b=1.5$~meV
($\Delta D_x=D_z\simeq1.06$~meV) has
$Q_{\rm eq}=-18.6\times10^{-6}~\text{\AA}^{-1}$, close to the in-plane
state with $\Delta D_x=1$~meV ($-17.9\times10^{-6}~\text{\AA}^{-1}$), but
an efficiency smaller by a factor of $27$, $-0.034\%$ against $-0.93\%$
($16$ with two-level refinement), although its slightly larger
$\Delta D_x$ alone would increase $|\eta|$.
Spin--valley locking does not protect the condensate against $D_z$, which
is pair breaking: in this state it reduces the spectral gap at $Q_{\rm eq}$ from $0.21$~meV
(in-plane) to $7~\mu$eV and $\bar K_c$ by a factor of $2.2$
[Fig.~\ref{fig:fig1}(d)]~\cite[Sec.~S4.1]{SMnote}.

The small net asymmetry results from compensation between pocket
contributions.
The pocket-resolved current asymmetry, dominated by quasiparticle
back-flow and evaluated at momenta mirrored about
$Q_{\rm eq}$~\cite[Sec.~S4.1]{SMnote}, comes predominantly from the
$\Gamma$ pocket for in-plane exchange ($-1.90$~A/m), opposed by the $K$
and $K'$ pockets ($+0.10$~A/m), on which the effective models of
Refs.~\cite{BankierPRB25,LevchenkoKhodasArXiv26} focus.
In the tilted configuration above, $D_z$ depairs the $K$ and $K'$ pockets
and raises their contribution to $+1.39$~A/m, nearly canceling that of the
$\Gamma$ pocket ($-1.40$~A/m).
A larger $D_z$ reverses the balance and the diode polarity near
$D_z\approx1.1$~meV [Fig.~\ref{fig:fig1}(c) and (d)], where the spectrum at
$Q_{\rm eq}$ becomes gapless: $\theta=30^\circ$ with $D_b=1.5$~meV
($D_z\simeq1.30$~meV) gives $\eta=+0.33\%$, now dominated by the $K$ and
$K'$ pockets~\cite[Secs.~S4.1 and S6.5]{SMnote}, as for competing bands in
Rashba superconductors~\cite{DaidoPRL22}.
The reversal persists on cooling to $0.12$~K but not at $1.16$~K, where
thermal smearing suppresses the net asymmetry~\cite[Sec.~S6.2]{SMnote}.
$Q_{\rm eq}$ nevertheless stays near the in-plane line: it is set at the
free-energy minimum, whereas the critical currents lie beyond the gap
closing~\cite[Secs.~S3.1 and S6.1]{SMnote}.

Carrier filling and temperature also separate the two observables.
Over $-20\leq\delta\mu\leq20$~meV [Fig.~\ref{fig:fig1}(e)], $Q_{\rm eq}$
changes by $0.27\times10^{-6}~\text{\AA}^{-1}$ and $\bar K_c$ by a factor
of $2.6$, from $152$ to $58$~A/m, while $\eta$ stays between $-0.95\%$ and
$-0.78\%$.
Cooling from $k_BT=10^{-4}$ to $10^{-5}$~eV instead raises $|\eta|$ from
$0.50\%$ to $1.18\%$, while $Q_{\rm eq}$ changes by
$0.06\times10^{-6}~\text{\AA}^{-1}$~\cite[Sec.~S6.2]{SMnote}.

\emph{Rectification without helical momentum.}
For purely out-of-plane exchange, the magnetic point group of a one-sided
configuration is $3m'$: $C_3$ and the vertical mirrors combined with time
reversal.
$M_x\mathcal T$ maps $(Q_x,Q_y)\rightarrow(Q_x,-Q_y)$ and allows a linear
term only along $\hat{\mathbf x}$, which $C_3$ forbids, so $\mathbf Q=0$ is
a stationary point of the free energy.
The leading odd term is cubic and, to first order in the exchange, reads
\begin{equation}
\Omega^{\rm odd}_{\rm SC}
\propto D_z(Q_x^3-3Q_xQ_y^2)
=D_z|\mathbf Q|^3\cos3\varphi_Q,
\label{eq:trig}
\end{equation}
with $\varphi_Q$ measured from $\hat{\mathbf x}$, so a diode effect is
allowed with $\mathbf Q_{\rm eq}=0$, provided this state remains the
minimum~\cite{ShafferLevchenkoArXiv25,SunArXiv26}.
It is maximal along $\hat{\mathbf x}$ and its $C_3$ images and vanishes
along the mirror lines, including $\hat{\mathbf y}$.

The calculation confirms these predictions [Fig.~\ref{fig:fig2}(a) and
(c)].
For bottom-only exchange with $D_b=1$~meV and $\theta=0^\circ$, the
$\hat{\mathbf y}$ scan is reciprocal within numerical accuracy, whereas the
$\hat{\mathbf x}$ scan gives
$|Q_{\rm eq}|<0.1\times10^{-6}~\text{\AA}^{-1}$ and $\eta=-1.45\%$.
Along every scanned direction ($0^\circ$, $10^\circ$, $20^\circ$, and
$90^\circ$ from $\hat{\mathbf x}$, which with $\pm\mathbf Q$, $C_3$, and
$M_x\mathcal T$ cover all inequivalent directions in $10^\circ$ steps), the
free energy is lowest at $\mathbf Q=0$ over the whole branch.
Since $C_3$ makes the quadratic term isotropic, $\mathbf Q=0$ is a local
minimum in the full $\mathbf Q$ plane, which we identify with the
equilibrium within the finite-$\mathbf Q$ ansatz~\cite[Sec.~S4.6]{SMnote}.
At $\varphi_Q=10^\circ$ and $20^\circ$, $\eta(\varphi_Q)/\eta(0)=0.884$ and
$0.532$, close to $\cos3\varphi_Q=0.866$ and $0.500$; the small excesses
fit a symmetry-allowed $\cos9\varphi_Q$ harmonic~\cite[Sec.~S4.6]{SMnote}.

Why can the efficiency be larger without any helical momentum, $-1.45\%$
against $-0.93\%$ for in-plane exchange of the same strength?
First, $Q_{\rm eq}$ is set by $\gamma_1$ in
Eq.~(\ref{eq:OmegaExpansion}), which $C_3$ forbids here, whereas
rectification requires only an odd part of $\Omega_{\rm SC}$ about its
minimum, entering through $\gamma_3$; a finite $\mathbf Q_{\rm eq}$ is thus
neither necessary nor sufficient for a diode effect.
Second, $\eta=\Delta K/(2\bar K_c)$, with $\Delta K=K_c^+-|K_c^-|$,
measures the current asymmetry relative to the critical current.
The out-of-plane state has a smaller asymmetry, $\Delta K=-1.42$ against
$-1.80$~A/m, but a halved mean critical current, $49.0$ against
$96.9$~A/m: acting as a pair-breaking Zeeman field along the Ising axis,
which spin--valley locking does not suppress, the exchange reduces the
spectral gap at $\mathbf Q=0$ to $27~\mu$eV and lowers the critical
current much more than its asymmetry.
Microscopically, the exchange adds to the Ising splitting in one valley and
subtracts from it in the other, so the odd contributions of the trigonally
warped $K$ and $K'$ pockets no longer cancel and leave a cubic term of the
form of Eq.~(\ref{eq:trig}).
The same mechanism underlies magnetochiral anisotropy in gated MoS$_2$ and
has been invoked for the field-driven diode effect in few-layer NbSe$_2$
\cite{WakatsukiSciAdv17,HoshinoPRB18,BauriedlNatCommun22,BankierPRB25};
here the proximity exchange alone provides it, as proposed for
NbSe$_2$/CrBr$_3$~\cite{SunArXiv26}.

The horizontal mirror leaves out-of-plane axial components invariant and
exchanges the two Se sublayers, so $\sigma_h\mathcal T$ reverses
$\mathbf Q$, $\mathbf D_{-,\parallel}$, and $D_{+,z}$ but keeps
$\mathbf D_{+,\parallel}$ and $D_{-,z}$.
Only the layer-odd in-plane and the layer-even out-of-plane exchange can
therefore rectify.
A uniform Zeeman field is layer-even: its out-of-plane component rectifies,
whereas an in-plane field alone does not, consistent with few-layer
NbSe$_2$~\cite{BauriedlNatCommun22}; in-plane rectification requires
breaking $\sigma_h\mathcal T$, by one-sided proximity or a
substrate~\cite{BankierPRB25,LevchenkoKhodasArXiv26}.
Exchange on the Nb orbitals, which lie in the mirror plane, is layer-even:
in-plane, it cannot rectify on its own; out-of-plane, it adds to the
trigonal channel~\cite[Sec.~S6.4]{SMnote}.

Two-sided out-of-plane configurations confirm this selection
[Fig.~\ref{fig:fig2}(b)]~\cite[Sec.~S4.6]{SMnote}.
Parallel configurations with $D_{t,z}=D_{b,z}=0.25$ and $0.5$~meV
reproduce the bottom-only states with the same $D_{+,z}$,
$D_{b,z}=0.5$ and $1$~meV, to within $2\times10^{-5}$ of the maximum
$\Delta^\star$ and $3.2\times10^{-5}$ of the maximum $|K|$.
The antiparallel configuration $D_{b,z}=-D_{t,z}=1$~meV is reciprocal
($|\eta|<10^{-3}\%$), and its order parameter and spectral gap match the
no-exchange values to within $1~\mu$eV.
Layer-odd out-of-plane exchange couples only states of opposite
$\sigma_h$ eigenvalue and shifts nondegenerate band energies
only at second order~\cite[Sec.~S4.6]{SMnote}.
Layer parity thus selects the two channels oppositely: the helical channel
requires layer-odd in-plane exchange, and the trigonal channel layer-even
out-of-plane exchange.

Figure~\ref{fig:fig2}(d)--(f) quantifies the trigonal channel.
For one-sided exchange,
$-\eta\simeq1.47\%\times D_z|D_z|/{\rm meV}^2$ over
$0.5\leq D_z\leq1.5$~meV; at smaller exchange, $|\eta|$ falls below this
law ($0.058\%$ at $D_z=0.25$~meV); there the current asymmetry changes
sign along the branch as the spectrum becomes gapless, and no linear term is
resolved~\cite[Sec.~S4.6]{SMnote}.
At equal one-sided exchange, 
the trigonal efficiency exceeds the helical one in magnitude
above $0.65$~meV at $0.58$~K [Fig.~\ref{fig:fig2}(d)].
A tilted configuration with $D_x=D_z=0.71$~meV, scanned along
$\hat{\mathbf x}$ through $\mathbf Q_{\rm eq}$, gives $\eta=-0.735\%$,
against $-0.729\%$ interpolated for purely out-of-plane exchange: the
in-plane component leaves the trigonal channel intact, whereas $D_z$
strongly suppresses the helical one [Fig.~\ref{fig:fig1}(d)].

The spectral gap explains this asymmetry [Fig.~\ref{fig:fig2}(f)].
Out-of-plane exchange lowers $E_{\rm gap}(0)$ approximately linearly over
$0\leq D_{+,z}\leq0.5$~meV, by $0.42$~meV per meV of $D_{+,z}$ ($D_{+,z}=D_b/2$ for one-sided configurations), from
$0.24$~meV to $27~\mu$eV at $D_b=1$~meV and to zero at $1.5$~meV; in-plane
exchange reduces the gap at $\mathbf Q_{\rm eq}$ only to $0.21$ and
$0.16$~meV at $D_b=1$ and $1.5$~meV, and AP$_z$ leaves it unchanged.
At $D_z=1.5$~meV, the superconducting solution still minimizes
$\Omega_{\rm SC}(\Delta)$ over the scanned scalar-gap amplitudes at the
tested momenta~\cite[Sec.~S6.5]{SMnote}.

At fixed $D_b=1$~meV, the two channels also respond differently to
temperature [Fig.~\ref{fig:fig2}(e)].
Cooling from $k_BT=10^{-4}$ to $10^{-5}$~eV ($1.16$ to $0.12$~K) raises
the trigonal $|\eta|$ tenfold, from $0.32\%$ to $3.37\%$, against a factor
of $2.4$ for the helical channel, which is larger at $1.16$~K.
The trigonal $|\Delta K|$ grows $35$-fold and $\bar K_c$ $3.4$-fold,
consistent with thermal smearing of exchange-split quasiparticle structure
on the scale of tens of $\mu$eV; at $1.16$~K the spectrum is gapless
already at $\mathbf Q=0$.
The ranking at $0.58$~K survives a larger pairing strength: recalibrating
the coupling to $\Delta^\star=0.6$~meV, which brings the spectral gap
closer to the measured one, leaves $\mathbf Q_{\rm eq}$ nearly unchanged,
lowers the helical $|\eta|$ to $0.79\%$, and keeps the trigonal efficiency
at $-1.51\%$~\cite[Sec.~S6.3]{SMnote}.

\emph{Conclusion.}
We have shown that the layer parity and orientation of magnetic exchange
select the superconducting diode channel in monolayer NbSe$_2$, and that
the equilibrium helical momentum and the diode efficiency probe different
features of the same finite-$\mathbf Q$ free-energy landscape.
For in-plane exchange, both follow
$\hat{\mathbf z}\times\Delta\mathbf D_\parallel$: parallel proximity is
reciprocal, and antiparallel proximity approximately doubles the one-sided
response.
An out-of-plane component leaves the helical momentum nearly unchanged but,
by depairing the $K$ and $K'$ pockets, suppresses the rectification by more
than an order of magnitude and, at low temperature, reverses it;
temperature also changes the efficiency at nearly constant
$\mathbf Q_{\rm eq}$.
A finite helical momentum is therefore not a quantitative proxy for the
diode efficiency.
Purely out-of-plane exchange realizes the opposite limit, a trigonal diode
effect with $\mathbf Q_{\rm eq}=0$, larger than the helical one above
$0.65$~meV at $0.58$~K because the same exchange also lowers the critical
current.
Layer-resolved proximity thus controls both whether and by which mechanism
a monolayer superconductor rectifies, within an effective model of
mirror-equivalent interfaces.

Existing heterostructures can test these rules.
The trigonal channel is a candidate intrinsic contribution to the
field-free rectification with a zero-momentum condensate in
NbSe$_2$/CrBr$_3$~\cite{SunArXiv26}, whose reported efficiencies, up to
$36\%$, exceed our depairing values tenfold, which motivates quantitative
modeling of the specific heterostructure and of possible device
contributions.
Its fingerprints are opposite polarities for zigzag directions $60^\circ$
apart, reciprocity along the armchair directions, a polarity set by the
interfacial magnetic configuration rather than the net moment, consistent
with the history-dependent reversals, and a superlinear growth with exchange
and strong enhancement on cooling.
Two-sided proximity yields a diode spin valve whose on and off states
interchange with the magnetization orientation.
For out-of-plane magnetization, the parallel state rectifies,
with approximately four times the one-sided diode efficiency
in magnitude at $0.5$--$0.75$~meV per layer, and the
antiparallel state is reciprocal, as can be tested in
CrPS$_4$/NbSe$_2$/CrPS$_4$~\cite{YunPRR23}; for in-plane magnetization the
roles are reversed.
In inversion-symmetric even-layer $2H$ stacks with equivalent interfaces,
parallel configurations are reciprocal for either orientation, which
reverses the out-of-plane rule~\cite[Sec.~S6.4]{SMnote}.

\emph{Data availability.} The data that support the findings of this
article are available from the authors upon reasonable request.

\emph{Acknowledgments.}
The calculations in this study were performed on the Numerical Materials
Simulator at NIMS. This work was supported by JSPS KAKENHI Grant
No.~JP24K06968 from MEXT, Japan.

\nocite{PBEsol,paw,pslibrary,mv_smearing,fr,Mustofi2012,Mostofi2014,ZhaoNatNano17,ZollnerPRB19,DreherACSNano21,BogoliubovJETP58,deGennesBook66}
\bibliography{refs_arxiv}

\clearpage
\onecolumngrid
\begin{center}
{\large\bfseries Supplemental Material for ``Layer-Parity Control of the
Superconducting Diode Effect in Monolayer NbSe$_2$''}\\[0.8em]
Yusuf Wicaksono and Toshikaze Kariyado\\[0.3em]
{\small\itshape International Center for Materials Nanoarchitectonics (WPI-MANA),
National Institute for Materials Science, 1-1 Namiki, Tsukuba, Ibaraki 305-0044, Japan}
\end{center}
\vspace{0.5em}
\begin{quote}
\small
This Supplemental Material provides the symmetry analysis, numerical
implementation, convergence tests, and additional results
supporting the main-text conclusions on the superconducting diode effect
(SDE) in magnetically proximitized monolayer NbSe$_2$.
The material is organized as follows.
Section~S1 derives the symmetry constraints associated with the horizontal
mirror $\sigma_h$ and the combined antiunitary operation
$\sigma_h\mathcal T$, with particular emphasis on the reciprocal parallel
two-sided configuration and the layer-parity selection rules.
Section~S2 summarizes the fully relativistic Wannier Hamiltonian,
layer-selective proximity exchange, finite-$\bQ$ Bogoliubov--de Gennes
(BdG) formulation, self-consistent pairing procedure, thermodynamic
potential, supercurrent, and critical-current extraction.
Section~S3 establishes numerical convergence and the reciprocity floor,
including the Fermi-surface-adaptive Brillouin-zone quadrature and its
refinement-level convergence, $Q$-grid resolution, self-consistency
criteria, and the no-exchange and parallel-state checks.
Section~S4 presents the complete results for the helical response
(angular and exchange-strength scans, a pocket-resolved decomposition of
the current asymmetry, layer-even exchange, parallel/antiparallel and
single-layer textures, and the vector form of the response) and for the trigonal channel generated by layer-even
out-of-plane exchange.
Section~S5 gives the chemical-potential dependence of the critical currents
and the diode efficiency.
Finally, Section~S6 collects superconducting-stability and robustness
checks, including a representative finite-$Q$ branch, the temperature
dependence, the sensitivity to the magnitude of the superconducting gap,
the pairing subspace, the scope of the proximity model, and the stability
of the selected solutions within the scalar-gap ansatz.
Reference numbers refer to the list at the end of the main text.
\end{quote}
\vspace{0.5em}
\twocolumngrid

\renewcommand{\theHfigure}{S\arabic{figure}}
\renewcommand{\theHtable}{S\arabic{table}}
\renewcommand{\theHequation}{S\arabic{equation}}
\setcounter{page}{1}
\renewcommand{\thepage}{S\arabic{page}}
\setcounter{figure}{0}
\renewcommand{\thefigure}{S\arabic{figure}}

\setcounter{table}{0}
\renewcommand{\thetable}{S\arabic{table}}

\setcounter{equation}{0}
\renewcommand{\theequation}{S\arabic{equation}}


\section*{S1. Symmetry constraints, magnetic layer parity, and reciprocity}

This section establishes and numerically verifies the symmetry
constraint responsible for reciprocity in the equal parallel
in-plane (P) configuration used in the main text.
We first derive the constraint at the normal-state and
Bogoliubov--de Gennes (BdG) Hamiltonian levels, then obtain the corresponding
selection rules for the finite-$Q$ free energy. The numerical calculations
are finally used to distinguish symmetry-enforced reciprocity from an
accidental cancellation of the positive- and negative-current branches.

Throughout this Supplemental Material, $\bk$ and $\bQ$ denote physical
wave vectors in $\text{\AA}^{-1}$. Each scan varies the pair momentum
along a fixed in-plane unit vector $\hat{\mathbf n}$, $\bQ=Q\hat{\mathbf n}$,
with $\hat{\mathbf n}=\hat{\mathbf y}$ (armchair) for the in-plane exchange
and $\hat{\mathbf n}=\hat{\mathbf x}$ (zigzag) for the out-of-plane exchange
unless stated otherwise; $Q\equiv Q_y$ or $Q_x$ accordingly.
Grand potentials are energies per primitive cell, and $K$ denotes the
two-dimensional sheet current along $\hat{\mathbf n}$ in A/m,
$K=(2e/\hbar A_{\rm uc})\,d\Omega_{\rm SC}/dQ$, with $e>0$. When the
in-plane case is discussed explicitly we write $K_y$ and $Q_y$.
The diode efficiency $\eta$ is quoted in percent and its
differences in percentage points. Relative changes of other
quantities are expressed as dimensionless ratios or fractions,
while absolute changes retain the corresponding units.
The dimensionless fractional coordinates $\boldsymbol\kappa$ and $\mathbf q$
are defined in Sec.~S2.3.

\subsection*{S1.1. Horizontal mirror symmetry of monolayer
\texorpdfstring{$1H$-NbSe$_2$}{1H-NbSe2}}

Monolayer $1H$-NbSe$_2$ possesses a horizontal mirror $\sigma_h$ whose
mirror plane passes through the Nb layer. It exchanges the top and bottom Se
sublayers,
\begin{equation}
  \sigma_h:\quad (x,y,z)\rightarrow(x,y,-z).
  \label{eq:S1_mirror_coordinates}
\end{equation}
Because the proximity magnetization is an axial vector, its transformation
under an improper spatial operation differs from that of a polar vector.
For $\sigma_h$,
\begin{equation}
  (M_x,M_y,M_z)
  \xrightarrow{\sigma_h}
  (-M_x,-M_y,M_z).
  \label{eq:S1_axial_mirror}
\end{equation}
Thus an in-plane magnetic moment changes sign under $\sigma_h$, while the
top and bottom Se sublayers are interchanged. Time reversal acts as
\begin{equation}
  \mathcal T:\quad
  \bM\rightarrow-\bM,\qquad
  \bk\rightarrow-\bk,\qquad
  \bQ\rightarrow-\bQ .
  \label{eq:S1_time_reversal}
\end{equation}
For the in-plane configurations considered here, the additional sign
reversal of the moment under $\mathcal T$ compensates the sign change
generated by $\sigma_h$.

For collinear in-plane proximity exchange along $\hat{\mathbf x}$,
let $D_t$ and $D_b$ denote the signed exchange components on
the top and bottom Se sublayers. We introduce the layer-even
and layer-odd combinations
\begin{equation}
  D_{+}=\frac{D_b+D_t}{2},
  \qquad
  D_{-}=\frac{D_b-D_t}{2}
  =\frac{\Delta D}{2},
  \label{eq:S1_exchange_parity}
\end{equation}
where $\Delta D=D_b-D_t=\Delta D_x$ is the exchange contrast
used in the main text.
Since $\sigma_h$ exchanges the two sublayers while $\sigma_h\mathcal T$
restores the direction of an in-plane axial exchange field,
\begin{equation}
  \sigma_h\mathcal T:~
  D_{+}\rightarrow D_{+},
  ~
  D_{-}\rightarrow-D_{-},
  ~
  Q_y\rightarrow-Q_y .
  \label{eq:S1_exchange_transform}
\end{equation}

\subsection*{S1.2. Hamiltonian-level origin of reciprocity in the
parallel configuration}

It is useful to make the reciprocity constraint explicit already at the
Hamiltonian level. Let
\begin{equation}
  \mathcal A\equiv\sigma_h\mathcal T=U_{\mathcal A}\mathcal K
  \label{eq:S1_A_def}
\end{equation}
denote the combined antiunitary operation, where $\mathcal K$ is complex
conjugation and $U_{\mathcal A}$ contains the spin-orbital action of
$\sigma_h\mathcal T$. The nonmagnetic relativistic Wannier Hamiltonian is
invariant under both $\sigma_h$ and $\mathcal T$. Including the
layer-selective in-plane exchange therefore gives
\begin{equation}
  \mathcal A
  H_e(\bk;D_{+},D_{-})
  \mathcal A^{-1}
  =
  H_e(-\bk;D_{+},-D_{-}).
  \label{eq:S1_normal_H_symmetry}
\end{equation}

The pairing channel used in this work is also compatible with
$\mathcal A$. The active Nb orbitals
$d_{z^2}$, $d_{x^2-y^2}$, and $d_{xy}$ are even under $\sigma_h$, while
the local spin-singlet matrix $i\sigma_y$ is invariant under time reversal
up to the superconducting gauge phase. 
Choosing the scalar gap $\Delta$ real and a compatible phase
convention for $U_{\mathcal A}$, the pairing matrix satisfies
\begin{equation}
  U_{\mathcal A}\hat{\Delta}^{*}U_{\mathcal A}^{T}
  =\hat{\Delta}.
\end{equation}
The corresponding Nambu-space antiunitary operator can then
be written as
\begin{equation}
  \widetilde{\mathcal A}
  =
  \begin{pmatrix}
    U_{\mathcal A} & 0\\
    0 & U_{\mathcal A}^{*}
  \end{pmatrix}\mathcal K ,
  \label{eq:S1_Nambu_A}
\end{equation}
and the finite-$Q_y$ BdG Hamiltonian obeys
\begin{multline}
  \widetilde{\mathcal A}\,
  H_{\rm BdG}(\bk,Q_y;D_{+},D_{-})\,
  \widetilde{\mathcal A}^{-1} \\
  =
  H_{\rm BdG}(-\bk,-Q_y;D_{+},-D_{-}).
  \label{eq:S1_BdG_symmetry}
\end{multline}
Consequently the quasiparticle spectra are paired according to
\begin{equation}
  E_n(\bk,Q_y;D_{+},D_{-})
  =
  E_n(-\bk,-Q_y;D_{+},-D_{-}),
  \label{eq:S1_spectrum_symmetry}
\end{equation}
up to a relabeling of BdG eigenstates. Because the pairing functional and the Brillouin-zone sum are
invariant under $\bk\rightarrow-\bk$, the thermodynamic
potentials of symmetry-related self-consistent solutions satisfy
\begin{equation}
  \Omega_{\rm SC}(Q_y;D_{+},D_{-})
  =
  \Omega_{\rm SC}(-Q_y;D_{+},-D_{-})
   .
  \label{eq:S1_general_symmetry}
\end{equation}
This relation also holds after minimizing over the scalar
pairing amplitude at each $Q_y$.

The equal parallel configuration has $D_t=D_b$ and hence $D_{-}=0$.
Equation~\eqref{eq:S1_general_symmetry} therefore reduces directly to
\begin{equation}
  \Omega_{\rm SC}^{P}(Q_y)
  =
  \Omega_{\rm SC}^{P}(-Q_y).
  \label{eq:S1_P_even}
\end{equation}
This result is exact within the ideal mirror-equivalent model. In
particular, every odd power of $Q_y$ is forbidden:
\begin{equation}
  \Omega_{\rm SC}^{P}(Q_y)
  =
  \Omega_0+\gamma_2Q_y^2+\gamma_4Q_y^4+\gamma_6Q_y^6+\cdots .
  \label{eq:S1_P_expansion}
\end{equation}
The thermodynamic current is therefore strictly odd,
\begin{align}
  K_y^{P}(Q_y)
  &=\frac{2e}{\hbar A_{\rm uc}}
    \frac{\partial\Omega_{\rm SC}^{P}}{\partial Q_y} \nonumber\\
  &=\frac{2e}{\hbar A_{\rm uc}}
    \left(2\gamma_2Q_y+4\gamma_4Q_y^3+6\gamma_6Q_y^5+\cdots\right),
  \label{eq:S1_P_current}
\end{align}
so that
\begin{equation}
  K_y^{P}(-Q_y)=-K_y^{P}(Q_y).
  \label{eq:S1_P_current_odd}
\end{equation}
The same symmetry maps every stable superconducting solution at
$+Q_y$ to one at $-Q_y$. For the symmetry-connected branch
considered here, the current extrema therefore satisfy
\begin{equation}
  K_c^-=-K_c^+,
  \qquad
  \eta=0 .
  \label{eq:S1_P_eta_zero}
\end{equation}
Equation~\eqref{eq:S1_P_even} alone permits either a unique minimum at
$Q_y=0$ or a pair of degenerate minima at $\pm Q_0$. In the present P
calculation the central minimum is unique; hence
$Q_{\rm eq}=0$ within numerical accuracy.

\subsection*{S1.3. Numerical verification and reciprocity floor}

To quantify the numerical accuracy with which the symmetry is satisfied, we
evaluate the symmetry-forbidden parts of the superconducting free energy and
current,
\begin{align}
  \Omega_{\rm SC}^{\rm odd}(Q_y)
  &=
  \frac{\Omega_{\rm SC}(Q_y)-\Omega_{\rm SC}(-Q_y)}{2},
  \label{eq:S1_Omega_odd}
  \\
  K_y^{\rm even}(Q_y)
  &=
  \frac{K_y(Q_y)+K_y(-Q_y)}{2}.
  \label{eq:S1_J_even}
\end{align}
Both quantities vanish when symmetry enforces
$\Omega_{\rm SC}(Q_y)=\Omega_{\rm SC}(-Q_y)$, as in the
pristine and equal parallel in-plane configurations.

\begin{figure}[t]
  \centering
  \includegraphics[width=\columnwidth]{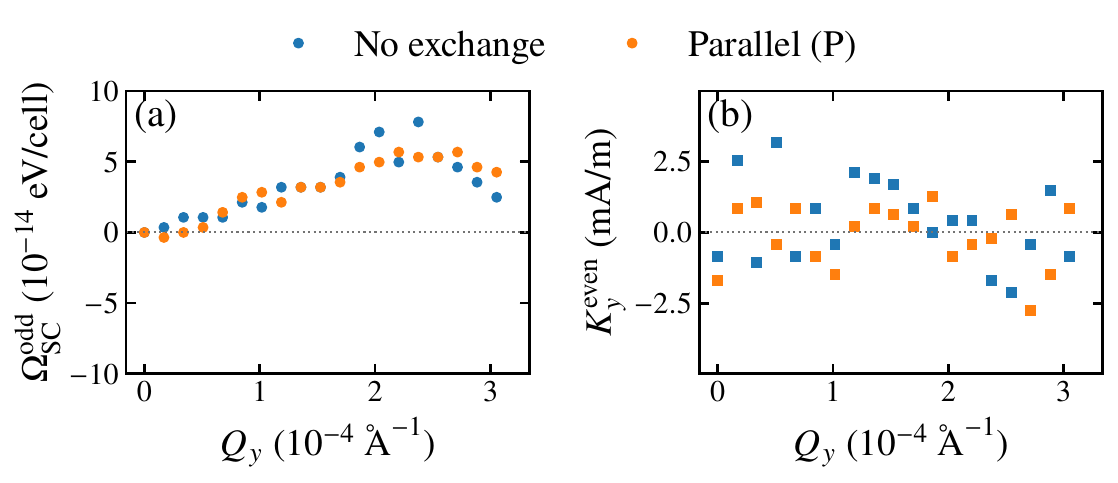}
  \caption{Numerical reciprocity controls for pristine and parallel
  magnetic configurations (uniform $201\times201+{\rm third9}$ mesh,
  $k_BT=10^{-5}$~eV).
  (a) Odd part of the superconducting thermodynamic potential,
  $\Omega_{\rm SC}^{\rm odd}(Q_y)
  =[\Omega_{\rm SC}(Q_y)-\Omega_{\rm SC}(-Q_y)]/2$, for pristine
  (no-exchange) monolayer NbSe$_2$ and for equal parallel (P) in-plane
  exchange on the top and bottom Se sublayers.
  (b) Corresponding even part of the thermodynamic current,
  $K_y^{\rm even}(Q_y)=[K_y(Q_y)+K_y(-Q_y)]/2$.
  Only $Q_y\geq0$ is shown because the plotted combinations already encode
  the comparison between $\pm Q_y$. The momentum axis is in
  $10^{-4}~\text{\AA}^{-1}$; the ordinate units are
  $10^{-14}$~eV/cell in (a) and mA/m in (b). In both cases the residual signals
  remain at the numerical reciprocity floor: the odd free-energy component
  is of order $10^{-14}$~eV/cell and the forbidden even-current component is of
  order $10^{-5}$ of the total current scale.}
  \label{fig:S1_reciprocity}
\end{figure}

Figure~\ref{fig:S1_reciprocity} shows that the pristine and P
configurations exhibit essentially the same residual symmetry breaking.
For the no-exchange calculation,
$\max|\Omega_{\rm SC}^{\rm odd}|=8.17\times10^{-14}$~eV/cell, while the P state
gives $7.82\times10^{-14}$~eV/cell. These values define the numerical
reciprocity floor of the finite-$Q$ thermodynamic calculation. 
The current provides a complementary test:
the maximum forbidden even component is only
$3.27\times10^{-5}$ and $3.15\times10^{-5}$ of the corresponding total
current scale for the no-exchange and P states, respectively. Notably, the
P-state residual is not larger than the pristine control on either measure.

The residual equilibrium shifts and diode efficiencies are comparably
small. As summarized in Table~\ref{tab:S1_reciprocity}, the extracted
equilibrium momenta are
$Q_{\rm eq}=1.84\times10^{-9}$ and $3.68\times10^{-9}~\text{\AA}^{-1}$ for the
no-exchange and P states, respectively, while the corresponding efficiencies
are only $6.9\times10^{-4}\%$ and $5.3\times10^{-4}\%$. Thus the P-state
values are at the same numerical scale as the nominally reciprocal pristine
system. This establishes that the apparent residual $Q_{\rm eq}$ and
$\eta$ of the P state are numerical-resolution effects rather than a
physically resolved nonreciprocal signal.

\begin{table*}[t]
\caption{Numerical reciprocity controls at $\mu=\mu_0$ (uniform
$201\times201+{\rm third9}$ mesh, $k_BT=10^{-5}$~eV).
The quantities
$\Omega_{\rm SC}^{\rm odd}
=[\Omega_{\rm SC}(Q)-\Omega_{\rm SC}(-Q)]/2$
and
$K^{\rm even}=[K(Q)+K(-Q)]/2$
vanish for the symmetry-protected configurations considered here. 
The last column normalizes the forbidden
even-current component by the maximum magnitude of the physical current on
the same finite-$Q$ branch.}
\label{tab:S1_reciprocity}
\begin{ruledtabular}
\begin{tabular}{lcccc}
State &
$Q_{\rm eq}$ ($\text{\AA}^{-1}$) &
$\eta$ (\%) &
$\max|\Omega_{\rm SC}^{\rm odd}|$ (eV/cell) &
$\max|K^{\rm even}|/\max|K|$\\
\hline
No exchange  & $+1.84\times10^{-9}$ & $+0.0007$ &
$8.17\times10^{-14}$ & $3.27\times10^{-5}$ \\
Parallel (P) & $+3.68\times10^{-9}$ & $+0.0005$ &
$7.82\times10^{-14}$ & $3.15\times10^{-5}$ \\
\end{tabular}
\end{ruledtabular}
\end{table*}

\subsection*{S1.4. Layer-odd selection rules and consequences}

Equation~\eqref{eq:S1_general_symmetry} gives a general selection rule for
the finite-$Q_y$ free energy. 
At fixed $D_{+}$, within a region where the selected
self-consistent free energy is analytic, write
\begin{equation}
  \Omega_{\rm SC}(Q_y,D_-)
  =
  \sum_{m,n\geq0}
  c_{mn}\,D_-^{m}Q_y^{n}.
  \label{eq:S1_general_expansion}
\end{equation}
Applying Eq.~\eqref{eq:S1_general_symmetry} to each monomial gives
\begin{equation}
  c_{mn}
  =
  (-1)^{m+n}c_{mn},
  \label{eq:S1_coefficient_constraint}
\end{equation}
and therefore
\begin{equation}
  c_{mn}\neq0
  \quad\Longrightarrow\quad
  m+n~\text{is even}.
  \label{eq:S1_selection_rule}
\end{equation}
This compact rule contains both the P-state reciprocity and the leading
layer-parity response. When $D_-=0$, all odd powers of $Q_y$ vanish,
recovering Eq.~\eqref{eq:S1_P_even}. When $D_-\neq0$, terms such as
$D_-Q_y$, $D_-Q_y^3$, and $D_-^3Q_y$ are symmetry allowed.

Equivalently, if the free energy is written as \cite{HeNJP22}
\begin{equation}
  \Omega_{\rm SC}(Q_y,D_-)
  =
  \sum_{n\geq0}\gamma_n(D_-)Q_y^n,
  \label{eq:S1_gamma_expansion}
\end{equation}
then
\begin{equation}
  \gamma_n(D_-)
  =
  (-1)^n\gamma_n(-D_-).
  \label{eq:S1_gamma_parity}
\end{equation}
With $Q_y$ in $\text{\AA}^{-1}$, $\gamma_n$ has units of
energy times length to the power $n$, per cell. If a fractional-coordinate
fit is written as $\Omega=\sum_n\widetilde\gamma_n q_2^n$, then
$\gamma_n=\widetilde\gamma_n/|\mathbf b_2|^n$.
Thus even-$n$ coefficients are even functions of $D_-$, whereas odd-$n$
coefficients are odd functions of $D_-$. In particular,
\begin{gather}
  \gamma_1(D_-)
  =
  \lambda_1D_-+\mathcal O(D_-^3), \nonumber
  \\
  \gamma_3(D_-)
  =
  \lambda_3D_-+\mathcal O(D_-^3).
  \label{eq:S1_odd_coefficients}
\end{gather}
Keeping the leading terms,
\begin{multline}
  \Omega_{\rm SC}
  =
  \Omega_0
  +a_2Q_y^2+a_4Q_y^4 \\
  +\lambda_1D_-Q_y
  +\lambda_3D_-Q_y^3+\cdots .
  \label{eq:S1_low_order_expansion}
\end{multline}
Here $a_2$ and $a_4$ are the even coefficients at $D_-=0$. For weak layer
asymmetry and a stable central minimum,
\begin{equation}
  Q_{\rm eq}
  \simeq
  -\frac{\lambda_1D_-}{2a_2}
  =
  -\frac{\lambda_1}{4a_2}\Delta D .
  \label{eq:S1_Qeq_linear}
\end{equation}
The coefficients here depend on the fixed layer-even exchange $D_+$.
Equivalently, $Q_{\rm eq}(D_+,D_-)=\chi(D_+)D_-+\mathcal O(D_-^3)$,
with $\chi(D_+)=-\lambda_1(D_+)/[2a_2(D_+)]$; per unit $\Delta D=2D_-$ the
slope is $\chi/2$, i.e., $\chi_Q$ of Sec.~S4. The textures in Fig.~1(b) of
the main text have different $D_+$, so a common linear law
$Q_{\rm eq}=\chi_Q\Delta D_x$ is not required by the fixed-$D_+$ selection
rule alone. Section~S4.2 shows numerically that $\chi(D_+)$ is in fact
independent of $D_+$ to a relative $10^{-3}$ over the in-plane textures
considered.

The selection rule also makes clear why the equilibrium helical momentum
does not determine the diode efficiency. 
The term $\lambda_1D_-Q_y$ controls
the leading displacement of the free-energy
minimum, whereas the critical-current asymmetry depends on the full
nonlinear free-energy landscape and on the termination of the
self-consistent superconducting branch near the positive- and negative-$Q_y$
depairing regions. Higher-order odd terms such as $\lambda_3D_-Q_y^3$ and
the pair breaking discussed in Sec.~S4.1 can therefore change the magnitude,
and even the sign, of the diode response without a comparable change in $Q_{\rm eq}$.

A one-sided configuration has $D_-\neq0$, whereas the antiparallel
configuration $D_t=-D_b$ maximizes $|D_-|$ for fixed
$|D_t|=|D_b|$. At fixed $D_+$, reversal of the layer-odd channel gives,
for the corresponding central-minimum branch,
\begin{equation}
  Q_{\rm eq}(D_-)
  =
  -Q_{\rm eq}(-D_-),
  \label{eq:S1_Qeq_odd}
\end{equation}
where the fixed $D_+$ argument is suppressed. Full magnetic reversal instead
maps $(D_+,D_-)\to(-D_+,-D_-)$ and relates the two free-energy landscapes
by time reversal with $Q_y\to-Q_y$. It therefore reverses the corresponding
$Q_{\rm eq}$ and diode polarity. For the AP texture, $D_+=0$, these two
relations coincide.

\subsection*{S1.5. Symmetry interpretation}

The analytical and numerical results lead to three conclusions. 
First, equal parallel in-plane exchange on mirror-equivalent
top and bottom Se sublayers is reciprocal by the antiunitary
$\sigma_h\mathcal T$ symmetry:
BdG eigenvalues
at $(\bk,Q_y)$ are paired with those at $(-\bk,-Q_y)$, up to band
relabeling. After the Brillouin-zone sum, the self-consistent free energy is
even in $Q_y$ and the corresponding thermodynamic current is odd. Figure~\ref{fig:S1_reciprocity} and
Table~\ref{tab:S1_reciprocity} verify this constraint down to the same
numerical floor as pristine NbSe$_2$. Second, the exchange contrast
$\Delta D=2D_-$ parametrizes the symmetry-breaking layer-odd channel. The
selection rule in Eq.~\eqref{eq:S1_selection_rule} requires the leading
helical displacement to be odd in $\Delta D$ at fixed $D_+$, giving a leading
linear response near a stable central minimum. Third, once
$\sigma_h\mathcal T$ is broken, symmetry allows both finite pair momentum and
nonreciprocal critical currents but does not lock their magnitudes or signs together.
The diode efficiency depends on the nonlinear structure of the critical
states, which is analyzed in Secs.~S4 and S5.


\section*{S2. Wannier--BdG model and finite-\texorpdfstring{$\bQ$}{Q} numerical formulation}

This section documents the microscopic model, superconducting calibration,
finite-$\bQ$ Bogoliubov--de Gennes (BdG) construction, branch-selection criteria,
and current evaluation used throughout the main text.  The purpose is to make
clear which ingredients are fixed by the normal-state first-principles model,
which ingredients are phenomenological proximity or pairing terms, and how the
critical currents are extracted from the self-consistent finite-$Q$ branch.

\subsection*{S2.1. Fully relativistic Wannier Hamiltonian}

The normal-state electronic structure is represented by a fully relativistic
Wannier Hamiltonian for monolayer $1H$-NbSe$_2$.  We use the same
first-principles and Wannier90 protocol as in our previous work on
layer-selective proximity responses in monolayer $1H$-Nb$X_2$
\cite{WicaksonoPRB26}.  Briefly, the underlying density-functional
calculations were performed using \textsc{Quantum ESPRESSO}
\cite{qespresso,qe2}, with the PBEsol functional \cite{PBEsol} and fully
relativistic PAW pseudopotentials \cite{paw,fr,pslibrary}; spin--orbit coupling
was included self-consistently.  The $3\times3$ charge-density-wave
reconstruction observed in low-temperature monolayer NbSe$_2$
\cite{UgedaNatPhys15} was not included.  Self-consistent calculations used a
$42\times42\times1$ $k$ mesh, wave-function and charge-density cutoffs of
90~Ry and 720~Ry, respectively, and Marzari--Vanderbilt cold smearing
\cite{mv_smearing} of width $0.01$~Ry.  The in-plane lattice constant was
$a=3.423$~\AA.  The spinor Wannierization employed 50 Kohn--Sham bands,
22 spinor Wannier functions, an outer disentanglement window
$[-8,3]$~eV, a frozen window $[-8,2]$~eV (both windows
specified on the same energy scale as the chemical potential
$\mu_0$ below), and a $60\times60\times1$ uniform
Wannierization grid.

Maximally localized spinor Wannier functions (MLWFs) were constructed using \textsc{wannier90}
\cite{Mustofi2012,Pizzi2020,Mostofi2008,Mostofi2014}, starting from the fully relativistic
Bloch states. 
The Wannier basis consists of five Nb $d$ orbitals and three Se $p$ orbitals
on each chalcogen sublayer, giving 11 spatial orbitals and 22 spinor Wannier
states.  The real-space Hamiltonian is Fourier transformed as
\begin{equation}
  H_0(\bk)
  =
  \sum_{\bR} H_0(\bR) e^{i\bk\cdot\bR},
  \label{eq:S2_H0_Fourier}
\end{equation}
where $\bk$ is a physical wave vector and
$\bR=\sum_i n_i\mathbf a_i$ is a real-space lattice translation, with
$n_i\in\mathbb Z$. The fractional-coordinate implementation is given in
Sec.~S2.3.  The orbital ordering used in the present calculation is
listed in Table~\ref{tab:S2_wannier_basis}.  Throughout the superconducting
calculation, the chemical potential is measured relative to the same Wannier
Hamiltonian, with the reference value $\mu_0=-2.4083~{\rm eV}$.

\begin{figure}[t]
  \centering
  \includegraphics[width=\columnwidth]{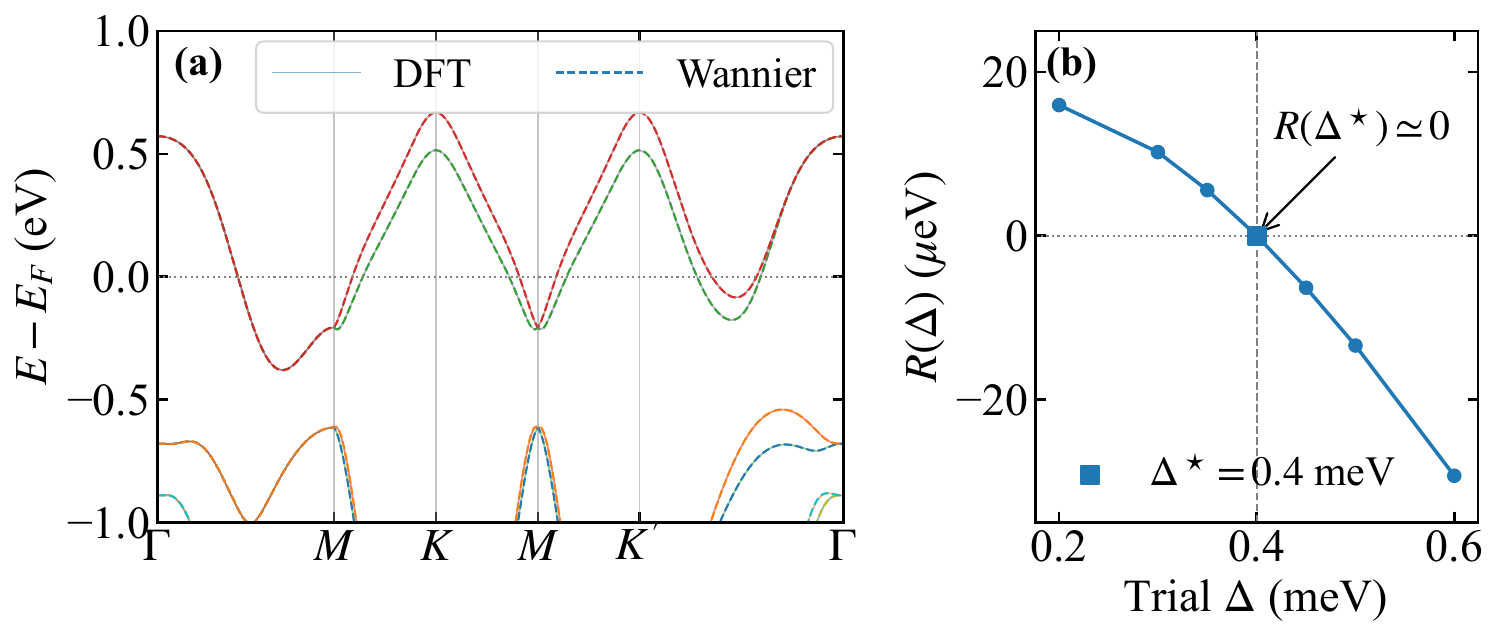}
  \caption{Validation and superconducting calibration of the
  Wannier--BdG model.
  (a) Comparison between fully relativistic DFT bands and the 22-state
  Wannier interpolation along $\Gamma$--$M$--$K$--$M$--$K^\prime$--$\Gamma$,
  with energies shifted by $E_F=\mu_0=-2.4083$~eV.  Within
  $|E-E_F|\leq1$~eV, the nearest-band RMS deviation is 0.85~meV.
  (b) Scalar gap-equation residual
  $R(\Delta)=g_{\rm SC}F_{\rm pair}(\Delta,Q=0)-\Delta$ for pristine
  monolayer NbSe$_2$, evaluated on the uniform
  $201\times201+{\rm third9}$ mesh at $k_BT=10^{-5}$~eV.  The calibrated
  coupling $g_{\rm SC}=0.500$~eV gives the target pairing amplitude
  $\Delta^\star=0.4$~meV with
  $R(\Delta^\star)=-3.42\times10^{-12}$~eV.}
  \label{fig:S2_model_calibration}
\end{figure}

Figure~\ref{fig:S2_model_calibration}(a) compares the 22-state Wannier
Hamiltonian with the low-energy DFT manifold used in the BdG calculation.
After shifting both DFT and Wannier bands by $E_F=\mu_0=-2.4083$~eV, the
nearest-band RMS deviation within $|E-E_F|\leq1$~eV is 0.85~meV. This
supports agreement of the interpolated band energies over the displayed
window. The nearest-band RMS metric does not, by itself, bound errors in
Fermi velocities, eigenvectors, or sub-meV spin splittings, nor does it
fix the detailed filling dependence of the critical currents.

\begin{table*}[t]
\caption{Spinor Wannier basis, pairing projector, and exchange
projector.  Each spatial orbital appears as a consecutive two-component
spinor block.  The main calculations use local spin-singlet pairing only in
the active Nb $d_{z^2}$, $d_{x^2-y^2}$, and $d_{xy}$ blocks, while magnetic
proximity exchange is applied only to the top and/or bottom Se $p$ blocks.}
\label{tab:S2_wannier_basis}
\begin{ruledtabular}
\begin{tabular}{lcccc}
Spinor indices & Spatial orbital & Layer & Pairing active? & Exchange projector \\
\hline
0, 1   & Nb $d_{z^2}$       & Nb     & yes & -- \\
2, 3   & Nb $d_{xz}$        & Nb     & no  & -- \\
4, 5   & Nb $d_{yz}$        & Nb     & no  & -- \\
6, 7   & Nb $d_{x^2-y^2}$   & Nb     & yes & -- \\
8, 9   & Nb $d_{xy}$        & Nb     & yes & -- \\
10, 11 & Se$_t$ $p_z$       & top    & no  & $\bD_t$ \\
12, 13 & Se$_t$ $p_x$       & top    & no  & $\bD_t$ \\
14, 15 & Se$_t$ $p_y$       & top    & no  & $\bD_t$ \\
16, 17 & Se$_b$ $p_z$       & bottom & no  & $\bD_b$ \\
18, 19 & Se$_b$ $p_x$       & bottom & no  & $\bD_b$ \\
20, 21 & Se$_b$ $p_y$       & bottom & no  & $\bD_b$ \\
\end{tabular}
\end{ruledtabular}
\end{table*}

\subsection*{S2.2. Layer-selective magnetic proximity}

Magnetic proximity is introduced as a static layer-selective exchange
self-energy acting on the Se-derived Wannier spinor blocks,
\begin{equation}
  H_{\rm ex}
  =
  \sum_{\ell=t,b} P_\ell\,
  \bD_\ell\cdot\boldsymbol{\sigma},
  \label{eq:S2_Hex}
\end{equation}
where $P_t$ and $P_b$ project onto the top and bottom Se $p$ subspaces listed
in Table~\ref{tab:S2_wannier_basis}.  The exchange vector on sublayer $\ell$
is parameterized as
\begin{equation}
  \bD_\ell
  =
  D_\ell
  \left(
  \sin\theta_\ell\cos\phi_\ell,
  \sin\theta_\ell\sin\phi_\ell,
  \cos\theta_\ell
  \right).
  \label{eq:S2_D_parameterization}
\end{equation}
The normal-state Hamiltonian entering the BdG calculation is therefore
\begin{equation}
  H_e(\bk)=H_0(\bk)+H_{\rm ex}.
  \label{eq:S2_He}
\end{equation}
This proximity term should be understood as an effective low-energy exchange
self-energy obtained after integrating out the magnetic environment and
projecting onto the NbSe$_2$ Wannier subspace.  In an explicit magnetic
heterostructure, additional terms such as interface electrostatic potentials,
spin-dependent hybridization, and modified hoppings can appear.  The present
model deliberately keeps only the leading local layer-selective exchange so
that the layer parity and orientation of the magnetic proximity can be varied
in a controlled way.

Unless stated otherwise, the exchange magnitude on each magnetized
sublayer is $1$~meV. For an ideal pure Wannier spin block the term
$\bD\!\cdot\!\boldsymbol{\sigma}$ has eigenvalues $\pm D$, whereas the shift
of an actual low-energy Bloch state is reduced by its Se-layer and orbital
weight. We quantified this on the normal-state Fermi contours at $\mu_0$
($36$ directions around $\Gamma$ and around $K$, $D_b=1$~meV, exact
diagonalization of $H_e$). The bottom-Se weight of the Fermi-surface states
is on average $0.10$--$0.11$ on the $\Gamma$ pocket and $0.14$--$0.16$ on
the $K$ pocket, and reaches $0.20$ on parts of the $K$ pocket. Out-of-plane exchange shifts these states at first order: by
$\pm0.10$~meV on the $\Gamma$ pocket, with the sign of the Ising spin
texture, and on average by $+0.13$ and $-0.16$~meV (at most $0.20$~meV) on
the two spin-split sheets of the $K$ pocket, i.e., a spin splitting of
$0.2$--$0.3$~meV per meV of $D_z$; a shift of $0.10$--$0.20$~meV is the
Zeeman energy $\mu_BB$ of a $1.7$--$3.5$~T field. The largest shift is consistent with the
linear closing of the spectral gap, $0.21$~meV per meV of $D_b$
(Sec.~S4.6). In-plane exchange is quenched by the Ising spin--valley
locking: its shifts are of second order and below $0.04$~meV.
Magnetic-proximity splittings of a few meV are well established in
monolayer TMD/magnetic insulator systems, both experimentally and in
first-principles calculations \cite{ZhaoNatNano17,ZollnerPRB19}. The range
used here, $|D_\ell|\leq1.84$~meV, therefore represents weak proximity, and
the out-of-plane component is bounded by pair breaking (Sec.~S4.2); the
magnitudes are not material-specific predictions for a particular
NbSe$_2$/magnet interface.

\subsection*{S2.3. Pairing projector and calibration of the superconducting interaction}

The superconducting interaction is treated phenomenologically as a local
attractive interaction in selected Nb-derived Wannier blocks.  In the main
calculations the active spatial orbitals are $d_{z^2}, d_{x^2-y^2}, {\rm and}~d_{xy}$,
corresponding to spinor block starts 0, 6, and 8 in
Table~\ref{tab:S2_wannier_basis}; this choice is discussed in Sec.~S6.3.
The pairing matrix is
\begin{equation}
  \hat\Delta
  =
  \Delta\,\Gamma_{\rm pair},
  \qquad
  \Gamma_{\rm pair}
  =
  P_{\rm pair}\otimes i\sigma_y,
  \label{eq:S2_Delta_matrix}
\end{equation}
where $P_{\rm pair}$ projects onto the three active Nb blocks.  The scalar
order parameter $\Delta$ is taken real; its phase is fixed by a global
superconducting gauge choice.

We write Brillouin-zone averages as $N_k^{-1}\sum_{\bk}$, as for an
equal-weight mesh. For the Fermi-surface-adaptive quadrature, this notation
stands for the nested quadrature of Eq.~\eqref{eq:S3_nested} applied to the
same integrand (Sec.~S3.1).
For a trial value of $\Delta$, the anomalous average in the active pairing
channel is denoted by
\begin{multline}
  F_{\rm pair}(\Delta,Q)
  =
  \frac{1}{N_k N_{\rm pair}}
  \sum_{\bk}\sum_{a\in{\rm pair}} \\
  \left\langle
    c_{-\bk+\bQ/2,a\downarrow}
    c_{\bk+\bQ/2,a\uparrow}
    -
    c_{-\bk+\bQ/2,a\uparrow}
    c_{\bk+\bQ/2,a\downarrow}
  \right\rangle/2,
  \label{eq:S2_Fpair_definition}
\end{multline}
with $N_{\rm pair}=3$.  Equivalently, Eq.~\eqref{eq:S2_Fpair_definition} is
the projection of the anomalous density matrix onto
$\Gamma_{\rm pair}$.  The scalar gap equation is
\begin{equation}
  \Delta
  =
  g_{\rm SC}F_{\rm pair}(\Delta,Q),
  \label{eq:S2_gap_equation}
\end{equation}
where $g_{\rm SC}$ is the effective attractive interaction in the selected
pairing channel.

The coupling is calibrated in pristine monolayer NbSe$_2$ at $Q=0$,
$\mu=\mu_0$, and $k_BT=10^{-5}$~eV by imposing $\Delta^\star_0=0.4~{\rm meV}$.
The calibration condition is
\begin{equation}
  g_{\rm SC}
  =
  \frac{\Delta^\star_0}
       {F_{\rm pair}(\Delta^\star_0,Q=0)},
  \label{eq:S2_g_calibration}
\end{equation}
which gives $g_{\rm SC}=0.500~{\rm eV}$ on the uniform
$201\times201+{\rm third9}$ mesh. On the production Fermi-surface-adaptive
quadrature of Sec.~S3.1, the same condition gives
$g_{\rm SC}=0.501$~eV for $L=2$, $3$, and $4$ (the three values differ by
less than $10^{-4}$~eV), and the value for the respective $L$ is used for
all production results. The imposed
$\Delta_0^\star=0.4$~meV is a phenomenological pairing-amplitude scale,
motivated by the superconducting gap $\Delta_{\rm SC}=0.40\pm0.02$~meV
measured by scanning tunneling spectroscopy on monolayer NbSe$_2$ on
bilayer graphene at $0.34$~K \cite{DreherACSNano21}.
It is not a fit of the minimum quasiparticle excitation gap to an experimental
spectral gap. In this multiorbital model, pairing acts only in the selected
Nb subspace, so the scalar amplitude and the spectral gap generally differ.
We define the latter on the sampled Brillouin zone as
\[
 E_{\rm gap}(Q)=\min_{n,\bk:\,E_{n\bk Q}>0} E_{n\bk Q}.
\]
For the pristine reference calculation at $Q=0$, $\mu=\mu_0$, and
$k_BT=10^{-5}$~eV on the uniform $201\times201+{\rm third9}$ mesh, the
recorded values are $\Delta_0^\star=0.400$~meV and
$E_{\rm gap}(0)=0.246$~meV; on the production quadrature,
$E_{\rm gap}(0)=0.237$~meV. This mesh-resolved spectral gap is a separate
observable; identifying it with the imposed pairing amplitude would require
a different calibration. The reference calibration does not impose the
experimental spectral gap; a larger pairing amplitude, which moves the spectral
gap toward it, is tested in Sec.~S6.3.

Figure~\ref{fig:S2_model_calibration}(b) shows the residual
\begin{equation}
  R(\Delta)
  =
  g_{\rm SC}F_{\rm pair}(\Delta,Q=0)-\Delta
  \label{eq:S2_gap_residual}
\end{equation}
on the uniform $201\times201+{\rm third9}$ mesh.  The residual changes
sign through the target value and gives $R(0.4~{\rm meV})=-3.42\times10^{-12}~{\rm eV}$,
corresponding to $|R|/\Delta^\star_0=8.56\times10^{-9}$.  After this
calibration, $g_{\rm SC}$ is held fixed for all exchange-field, carrier,
angular, layer-parity, and temperature calculations; only the gap-magnitude
test of Sec.~S6.3 uses a different calibration target.

The numerical Wannier interpolation and finite-$\bQ$ calculation use
fractional reciprocal-lattice coordinates. We define
\begin{equation}
  \mathbf a_i\cdot\mathbf b_j=2\pi\delta_{ij},
\end{equation}
and distinguish physical wave vectors from dimensionless coordinates:
\begin{equation}
  \bk=\sum_i\kappa_i\mathbf b_i,
  \qquad
  \bQ=\sum_i q_i\mathbf b_i.
\end{equation}
For an integer translation vector $\mathbf n=(n_1,n_2,n_3)$, let
$\bR_{\mathbf n}=\sum_i n_i\mathbf a_i$. The numerical Fourier transform is
\begin{equation}
  H_0(\boldsymbol\kappa)
  =\sum_{\mathbf n}H_0(\bR_{\mathbf n})
    e^{i2\pi\mathbf n\cdot\boldsymbol\kappa}.
\end{equation}
Here $H_0(\boldsymbol\kappa)$ is shorthand for
$H_0(\bk(\boldsymbol\kappa))$.

The lattice vectors in the Wannier Hamiltonian used here are
\begin{align*}
  \mathbf a_1&=3.423\,\hat{\mathbf x}~\text{\AA},\\
  \mathbf a_2&=(-1.711\,\hat{\mathbf x}
               +2.964\,\hat{\mathbf y})~\text{\AA}.
\end{align*}
They give
\begin{align*}
  \mathbf b_1&=(1.836\,\hat{\mathbf x}
                +1.060\,\hat{\mathbf y})~\text{\AA}^{-1},\\
  \mathbf b_2&=2.120\,\hat{\mathbf y}~\text{\AA}^{-1},
\end{align*}
and $A_{\rm uc}=10.144~\text{\AA}^2$.
A scan is specified by the angle $\varphi_Q$ of its direction
$\hat{\mathbf n}$ from $+\hat{\mathbf x}$ and by a scalar fractional parameter
$q$,
\begin{equation}
  \bQ=Q\,\hat{\mathbf n}+Q_\perp\hat{\mathbf n}_\perp,
  \qquad Q=|\mathbf b_2|\,q,
  \label{eq:S2_scanline}
\end{equation}
with $\hat{\mathbf n}=(\cos\varphi_Q,\sin\varphi_Q)$,
$\hat{\mathbf n}_\perp=\hat{\mathbf z}\times\hat{\mathbf n}$, and a fixed
transverse offset $Q_\perp$ that is zero unless stated otherwise. The
fractional components follow from $\bQ=\sum_iq_i\mathbf b_i$. For $Q_\perp=0$,
$\varphi_Q=90^\circ$ gives the scan along $\hat{\mathbf y}=\hat{\mathbf b}_2$,
$\mathbf q=(0,q,0)$ and $Q_y=|\mathbf b_2|q_2$; $\varphi_Q=30^\circ$ is the
scan along $\mathbf b_1$, $\mathbf q=(q,0,0)$; and $\varphi_Q=0^\circ$ is the
scan along the zigzag direction $\hat{\mathbf x}$, for which both fractional
components vary. In all cases a stored value of $q$ is multiplied by
$2.120$ to obtain $Q$ in $\text{\AA}^{-1}$, so that the $Q$ spacing and
the current conversion below do not depend on the direction. All plotted and
quoted momenta in the main text and this Supplemental Material use physical
units.

Define the numerical derivative
$\mathcal J=d\Omega_{\rm SC}/dq$ along the self-consistent branch.
It has units of eV per cell per dimensionless fractional coordinate.
The physical sheet current along $\hat{\mathbf n}$ is
\begin{equation}
  K=\frac{2e}{\hbar A_{\rm uc}}\frac{d\Omega_{\rm SC}}{dQ}
     =\frac{2e}{\hbar A_{\rm uc}|\mathbf b_2|}\mathcal J.
  \label{eq:S2_current_conversion}
\end{equation}
With energies and lengths converted to SI, this gives
\begin{equation}
  K[\mathrm{A/m}]
  =2.264\times10^5\,
    \mathcal J[\mathrm{eV/cell}].
  \label{eq:S2_current_conversion_numeric}
\end{equation}
The elementary charge is $e>0$, and the current polarity follows the
phase-gradient convention used in the BdG Hamiltonian. No monolayer
thickness is introduced. The common positive current factor leaves the zero
crossing in $q$, the diode efficiency, and relative reciprocity errors
unchanged; its physical location is $Q_{\rm eq}=|\mathbf b_2|q_{\rm eq}$.
A nonzero $Q_\perp$ is used only for the tilted textures scanned along
$\hat{\mathbf x}$ (Table~\ref{tab:S7_z}, block v), where $Q_\perp$ is set to the helical
$Q_{\rm eq}$ along $\hat{\mathbf y}$ so that the scan line passes through the
free-energy minimum, where the transverse current vanishes.

All Brillouin-zone averages are built on a $201\times201$ base grid with
the shifted ``third9'' quadrature, which the uniform-mesh reference
calculations use alone.  Here a shift $(s_x,s_y)$ is specified in
units of one base-mesh spacing, and third9 denotes the nine equally weighted
meshes
\begin{equation}
  (s_x,s_y)\in
  \left\{0,\frac13,\frac23\right\}
  \times
  \left\{0,\frac13,\frac23\right\}.
  \label{eq:S2_third9}
\end{equation}
The nine meshes are concatenated before the Brillouin-zone average, giving
$N_k=9\times201^2=363609$ points.  Because the observables are normalized by
the total number of points, this is exactly an equal-weight average over the
nine shifts; modulo reciprocal-lattice periodicity, the union has the same
point spacing as a $603\times603$ uniform quadrature.  The purpose of the
shift average is numerical rather than physical: at small thermal energies
such as the calibration value $k_BT=10^{-5}$~eV, Fermi-surface and BdG sums can show appreciable
mesh-commensurability (shell) effects when a single grid happens to sample or
miss narrow spin-split pockets.  The third9 average strongly suppresses this
sensitivity without modifying the Hamiltonian or introducing an artificial
energy broadening. In the production calculations, the states near the
Fermi level are integrated on top of this base grid with the
Fermi-surface-adaptive quadrature of Sec.~S3.1, which is required to converge
the critical currents.

\subsection*{S2.4. Finite-\texorpdfstring{$\bQ$}{Q} BdG Hamiltonian}

Following the standard Bogoliubov--de Gennes mean-field construction
\cite{BogoliubovJETP58,deGennesBook66}, a superconducting state with
center-of-mass momentum $\bQ$ \cite{FuldeFerrellPR64} pairs electronic states at $\bk+\bQ/2$ and
$-\bk+\bQ/2$.  We use the Nambu spinor
\begin{equation}
  \Psi_{\bk\bQ}
  =
  \begin{pmatrix}
    c_{\bk+\bQ/2} \\
    c^\dagger_{-\bk+\bQ/2}
  \end{pmatrix},
  \label{eq:S2_Nambu_spinor}
\end{equation}
where each entry is a 22-component spinor in the Wannier basis.  The
finite-$\bQ$ BdG Hamiltonian is then, with
$\bk_+=\bk+\bQ/2$ and $\bk_-=-\bk+\bQ/2$,
\begin{equation}
  H_{\rm BdG}(\bk,\bQ)
  =
  \begin{pmatrix}
    H_e(\bk_+)-\mu & \hat\Delta \\
    \hat\Delta^\dagger &
    -H_e^*(\bk_-)+\mu
  \end{pmatrix}.
  \label{eq:S2_BdG}
\end{equation}
For the present 22-state Wannier model, Eq.~\eqref{eq:S2_BdG} is a
$44\times44$ Hermitian matrix at each $\bk$ and $\bQ$.  The quasiparticle
energies and eigenvectors are obtained from
\begin{equation}
  H_{\rm BdG}(\bk,\bQ)
  |\Phi_{n\bk\bQ}\rangle
  =
  E_{n\bk\bQ}|\Phi_{n\bk\bQ}\rangle.
  \label{eq:S2_BdG_eigenproblem}
\end{equation}
In each calculation, $\bQ$ runs along the scan line of
Eq.~\eqref{eq:S2_scanline}. The formulas of Secs.~S2.5 and S2.6 are written
for the scans along $\hat{\mathbf y}$, $\bQ=Q_y\hat{\mathbf y}$; for any
other direction $Q_y$ and $K_y$ are replaced by $Q$ and $K$ along
$\hat{\mathbf n}$.

\subsection*{S2.5. Self-consistency workflow and superconducting stability}

For each value of $Q_y$, the calculation proceeds as follows.  First, the
finite-$Q_y$ BdG Hamiltonian in Eq.~\eqref{eq:S2_BdG} is constructed on the
Brillouin-zone quadrature described in Secs.~S2.3 and S3.1.  Second, the
anomalous average $F_{\rm pair}(\Delta,Q_y)$ is evaluated from the
positive-energy BdG eigenvectors, weighted by $\tanh(\beta E_n/2)$. Particle--hole
symmetry maps an eigenstate with energy $E$ at $\bk$ onto one with $-E$ at
$-\bk$ with the same contribution, and the quadrature is symmetric under
$\bk\rightarrow-\bk$, so this equals half the sum over all $44$ eigenstates,
also when Bogoliubov pockets make some energies of the upper half of the
spectrum negative. It is therefore consistent with the grand potential below,
which sums over all eigenvalues.  Third, the scalar gap equation
\begin{equation}
  G(\Delta,Q_y)
  \equiv
  g_{\rm SC}F_{\rm pair}(\Delta,Q_y)-\Delta
  =0
  \label{eq:S2_self_consistency_residual}
\end{equation}
is solved for the self-consistent value $\Delta^\star(Q_y)$.  The scan
starts at $Q_y=0$ and proceeds outward separately along the positive and
negative branches. At each point the solver is seeded with the converged
value at the neighboring $Q_y$; a previously converged branch interpolated to
that $Q_y$, when one is supplied, serves only as a fallback seed
(Sec.~S3.3). It brackets the local positive root near this seed and then
refines it within the bracket, by bisection in the uniform-mesh tests and by
the Illinois variant of the regula falsi method, which needs fewer
evaluations, in all Fermi-surface-adaptive calculations. This continuation
procedure helps
track a connected branch, but does not by itself guarantee selection of the
lowest-grand-potential root when multiple superconducting solutions exist;
Sec.~S6.5 checks this explicitly.

For clarity, we write the thermodynamic functional in the same normalization
used in the calculations.  Define
\begin{align}
  h_e(\bk,Q_y)
  &=H_e(\bk+\bQ/2)-\mu,\\
  h_h(\bk,Q_y)
  &=H_e(-\bk+\bQ/2)-\mu .
  \label{eq:S2_he_hh_def}
\end{align}
After diagonalizing the full $44\times44$ BdG matrix, the quasiparticle and
Nambu-trace contribution per primitive cell is
\begin{align}
  F_{\rm BdG}(Q_y,\Delta)
  =
  &\frac{1}{N_k}\sum_{\bk}
  \Bigg\{
    \frac14\tr\!\left[h_e(\bk,Q_y)+h_h(\bk,Q_y)\right] \nonumber \\
    &-\frac{1}{2\beta}
    \sum_n
    \ln\!\left[
      2\cosh\!\left(\frac{\beta E_{n\bk Q_y}}{2}\right)
    \right]
  \Bigg\},
  \label{eq:S2_FBdG}
\end{align}
where $\beta=(k_BT)^{-1}$.  The first term is the Nambu trace correction; it
removes the double counting introduced by the particle--hole representation.
The superconducting grand potential per primitive cell is then
\begin{equation}
  \Omega_{\rm SC}(Q_y,\Delta)
  =F_{\rm BdG}(Q_y,\Delta)
  +N_{\rm pair}\frac{\Delta^2}{g_{\rm SC}},
  ~ N_{\rm pair}=3,
  \label{eq:S2_Omega_SC}
\end{equation}
where the factor $N_{\rm pair}$ is required because the same scalar gap is
applied to the three active Nb pairing blocks.  The normal-state reference
$\Omega_{\rm N}(Q_y)$ is evaluated with the identical $Q_y$ and $k$ mesh but with
$\Delta=0$.  We define
\begin{equation}
  \Delta\Omega(Q_y)
  =
  \Omega_{\rm SC}(Q_y,\Delta^\star(Q_y))
  -\Omega_{\rm N}(Q_y).
  \label{eq:S2_DeltaOmega}
\end{equation}

We use two complementary controls to delimit the finite-$Q_y$ branch on which
critical currents are searched.  First, a conservative numerical
amplitude cutoff is imposed,
\begin{equation}
  \Delta^\star(Q_y)>0.2\,\Delta^\star(0).
  \label{eq:S2_gap_stability}
\end{equation}
This threshold is not a thermodynamic phase boundary; it is a
branch-termination safeguard that prevents very small, poorly conditioned residual
roots from being interpreted as a robust superconducting state.  Second, the
comparison with the normal state requires
\begin{equation}
  \Delta\Omega(Q_y)<0,
  \label{eq:S2_thermo_stability}
\end{equation}
so that the superconducting solution is lower in grand potential than the
normal state.  Once $\Delta\Omega>0$, the normal state is thermodynamically
preferred even if the nonlinear gap equation still admits a small nonzero
solution. We retain a point on the selected superconducting branch only
while both Eqs.~\eqref{eq:S2_gap_stability} and
\eqref{eq:S2_thermo_stability} are satisfied. Within the scalar-gap ansatz
the selected root is moreover a local minimum of $\Omega_{\rm SC}(\Delta)$.
Differentiating Eq.~\eqref{eq:S2_Omega_SC} at fixed $\bQ$ gives
$\partial\Omega_{\rm SC}/\partial\Delta=-(2N_{\rm pair}/g_{\rm SC})\,
G(\Delta)$, so a root at which $G$ changes from positive to negative is a
minimum. At sufficiently large $\Delta$, the positive quadratic term
$N_{\rm pair}\Delta^2/g_{\rm SC}$ dominates and $G<0$. The largest simple
positive root therefore has a positive-to-negative crossing and is a local
minimum; a multiple root, as at a fold, requires separate consideration.
At $\bQ=0$ the solver starts from $0.4$~meV and selects the root nearest to
it; Sec.~S6.5 confirms that this is the largest positive root in the
scanned range.
Continuation in $Q$ preserves the character of the root until a fold, where
the branch terminates and the solver may jump to another root; such jumps are
excluded by the extraction rule of Sec.~S2.6. Section~S6.5 verifies this directly at the equilibrium and
critical momenta of representative states. Stability against spatially
varying amplitudes or other pairing channels is outside this ansatz.  A small minimum positive BdG
quasiparticle energy $E_{\rm gap}$ is recorded as a diagnostic but is not, by
itself, used to declare loss of superconductivity, since a gapless
superconducting state can still have $\Delta^\star>0$ and
$\Delta\Omega<0$.

\subsection*{S2.6. Thermodynamic current, critical currents, and diode efficiency}

The supercurrent is obtained by differentiating the self-consistent
superconducting thermodynamic potential,
\begin{equation}
  K_y(Q_y)
  =
  \frac{2e}{\hbar A_{\rm uc}}\frac{d}{dQ_y}
  \Omega_{\rm SC}(Q_y,\Delta^\star(Q_y)).
  \label{eq:S2_thermo_current}
\end{equation}
Because $\Delta^\star(Q_y)$ satisfies the stationary condition
\begin{equation}
  \left.
  \frac{\partial\Omega_{\rm SC}(Q_y,\Delta)}{\partial\Delta}
  \right|_{\Delta=\Delta^\star(Q_y)}
  =0,
  \label{eq:S2_stationary_condition}
\end{equation}
the implicit derivative through $\Delta^\star(Q_y)$ does not contribute to
Eq.~\eqref{eq:S2_thermo_current}; only the explicit $Q_y$ derivative of the
self-consistent free energy is needed.

The equilibrium pair momentum $Q_{\rm eq}$ is obtained from the central zero
crossing of $K_y(Q_y)$ associated with the central free-energy minimum on
the selected branch. The positive and negative
critical currents are defined as
\begin{gather}
  K_c^+=\max_{Q_y\in{\rm selected}}K_y(Q_y), \nonumber
  \\
  K_c^-=\min_{Q_y\in{\rm selected}}K_y(Q_y).
  \label{eq:S2_Jc_definitions}
\end{gather}
Here ``selected'' denotes the branch satisfying the criteria in Sec.~S2.5,
with one further restriction. Near the end of the branch,
$\Delta^\star(Q_y)$ can collapse within a few grid points before the
criteria of Sec.~S2.5 are violated, and a finite-difference derivative taken
across the collapse produces an isolated spike in $K_y$. Starting from
$Q_y=0$ and moving outward, we therefore terminate the search interval at the
first abrupt step, defined as
$|\Delta^\star(Q_{i+1})-\Delta^\star(Q_i)|>0.02\,\Delta^\star(0)$ and more
than three times the preceding step, and also discard the last point before
it. Branches on which $\Delta^\star$ decreases smoothly are followed to their
end, apart from the outermost point, whose derivative is one-sided. The result
is insensitive to these choices: for all calculations reported here,
thresholds of $0.01$, $0.02$, and $0.05\,\Delta^\star(0)$ and factors of $2$, $3$, and $5$ give
the same $\eta$ at the quoted precision, and the reported $K_c^\pm$ are
interior extrema of the remaining interval (Sec.~S3.2).
The extrema estimate intrinsic uniform-state depairing currents within the
chosen pairing ansatz. Their extraction requires that the extrema be
resolved before the numerical branch cutoff; cutoff and grid sensitivity
must be assessed in the convergence analysis. They are not a calculation of
device switching currents, which can depend on vortex entry, contacts, and
spatial inhomogeneity.
The superconducting diode efficiency is
\begin{equation}
  \eta
  =
  \frac{K_c^+-|K_c^-|}
       {K_c^++|K_c^-|}.
  \label{eq:S2_eta}
\end{equation}
For diagnostic purposes we also separate the diode numerator and overall
critical-current scale,
\begin{gather}
  K_{\rm diode}=K_c^+ + K_c^- = K_c^+-|K_c^-|, \nonumber
  \\
  K_{\rm scale}=K_c^+-K_c^- = K_c^+ + |K_c^-|,
  \label{eq:S2_Jdiode_Jscale}
\end{gather}
so that $\eta=K_{\rm diode}/K_{\rm scale}$. Thus $K_{\rm diode}=\Delta K$ and
$K_{\rm scale}=2\bar K_c$ in the notation of the main text. This separation
shows whether a
change of $\eta$ comes from the current asymmetry or from the overall current
scale.


\section*{S3. Numerical convergence and reciprocity controls}

This section quantifies the numerical uncertainty of the finite-$Q$
observables with respect to Brillouin-zone sampling, $Q$-grid spacing, and
the self-consistency criterion, and summarizes the reciprocity floor.
Section~S3.1 shows that the tested uniform Brillouin-zone meshes do not
adequately converge the critical-current asymmetry, introduces the
Fermi-surface-adaptive quadrature used for the production results, and
assesses its accuracy.

\subsection*{S3.1. Brillouin-zone sampling and Fermi-surface-adaptive quadrature}

\emph{Uniform meshes.}
Table~\ref{tab:S3_conv} lists one-sided in-plane states ($\theta=90^\circ$,
$D_b=1$~meV) computed on uniform third9 meshes with base sizes $N=201$ to
$601$, i.e., $603$ to $1803$ points per reciprocal-lattice direction. For each
$N$ the coupling $g_{\rm SC}$ is recalibrated on the same mesh at the
pristine reference point (Sec.~S2.3), so that every mesh reproduces the same
target pairing amplitude.
The equilibrium momentum is converged on all meshes: $Q_{\rm eq}$ varies by
at most $0.05\times10^{-6}~\text{\AA}^{-1}$. It is set by the linear-in-$Q$ part of the free energy
near $Q=0$, where the spectrum is fully gapped and the Brillouin-zone
integrand is smooth. The critical currents are not converged. Between
neighboring meshes, $\eta$ changes by up to a factor of three without a
monotonic trend (by an order of magnitude at $\delta\mu=+5$~meV), and at
$\delta\mu=+5$~meV even $\Delta^\star(0)$ fluctuates between $0.347$ and
$0.408$~meV.

The origin is the structure of the critical states. In every state the
critical momenta lie beyond the momenta at which the quasiparticle gap
closes, so the critical states are gapless and contain small Bogoliubov
Fermi surfaces. For the converged one-sided state at $\mu_0$, for example,
the gap falls below $1~\mu$eV at $Q_y\approx+1.8$ and
$-2.3\times10^{-4}~\text{\AA}^{-1}$, whereas $K_c^\pm$ occur at $+2.54$ and
$-2.88\times10^{-4}~\text{\AA}^{-1}$ ($L=3$). Near these surfaces the integrand varies on the energy scale
$\Delta^\star,k_BT\lesssim0.4$~meV, whereas a uniform cell spans an energy
range $\hbar v_F\,\delta k\approx4$~meV at $N=601$. The Brillouin-zone sum then
depends on how the mesh intersects these surfaces. A direct measure of the
resulting error is the normal-state current
$K_{\rm N}=(2e/\hbar A_{\rm uc})\,d\Omega_{\rm N}/dQ_y$. It vanishes identically in
the continuum limit, because a rigid shift of $\mathbf k$ leaves the normal
state unchanged, but on the uniform meshes it reaches $2$--$20$~A/m
(Table~\ref{tab:S3_conv}). This is comparable to the diode numerator
$|K_c^++K_c^-|\approx2$~A/m. Below we quote its effect on the
efficiency as the spurious fraction $[K_{\rm N}(Q_c^+)+K_{\rm N}(Q_c^-)]/\Delta K$, the
part of the current asymmetry $\Delta K=K_c^+-|K_c^-|$ ($=K_{\rm diode}$ of
Sec.~S2.6) that comes from the
normal-state quadrature error. Brute-force convergence would require several
thousand points per direction.

\emph{Fermi-surface-adaptive quadrature.}
We therefore refine the Brillouin-zone quadrature only where the integrand
varies rapidly, on the scale $k_BT$, and becomes non-analytic as $T\rightarrow0$. Let $Q_l[f]$ denote the midpoint rule for an integrand
$f(\mathbf k)$ on the base mesh subdivided $l$ times into $m\times m$
sub-cells (spacing $\delta k/m^l$). Here the base mesh, $Q_0$, is the
combined third9 quadrature, equivalent to a uniform mesh with $3N$ points per
reciprocal-lattice direction, and $\delta k$ is its spacing. Let $S_l(E)$ be a
$C^\infty$ window
with $S_l=1$ for $|E|\le W_l/4$, $S_l=0$ for $|E|\ge W_l$, and
$W_l=W/m^{l-1}$. The spectral contributions to the Brillouin-zone averages
of Sec.~S2 have the form
$f(\mathbf k)=\sum_n g(E_n(\mathbf k),\psi_n(\mathbf k))$ over BdG
eigenpairs, and we write $(fS_l)(\mathbf k)=\sum_n g\,S_l(|E_n|)$. The
Nambu-trace term of Eq.~\eqref{eq:S2_FBdG}, a smooth function of the
normal-state Hamiltonian, is integrated on $Q_0$ alone. The spectral
integral is evaluated as
\begin{align}
  \langle f\rangle
  &= Q_0[f]+\sum_{l=1}^{L}\Bigl(Q_l[fS_l]-Q_{l-1}[fS_l]\Bigr) \nonumber\\
  &= Q_L[fS_L]+\sum_{l=0}^{L-1}Q_l\bigl[f\,(S_l-S_{l+1})\bigr],
  \label{eq:S3_nested}
\end{align}
with $S_0\equiv1$. The second form shows the logic. The rapidly varying part of
the integrand near $E=0$ is integrated on the finest mesh, with spacing
$\delta k/m^L$. Each coarser mesh only integrates
$f(S_l-S_{l+1})$, which is smooth on that mesh provided the taper of
$S_{l+1}$ spans several of its cells; for such integrands the periodic
midpoint rule converges faster than any power of the spacing. Each
correction in Eq.~\eqref{eq:S3_nested} is evaluated only on the cells of
level $l-1$ that $S_l$ can reach. These are selected from the normal-state
bands at $\pm\mathbf k$ on a $3\times3$ stencil per cell, with the band
intervals padded by one quarter of their spread and by an additional
$1$~meV. On each refined cell, the fine-mesh geometric weights sum to the
weight of the coarse point they replace. Each correction therefore carries
zero net geometric weight before multiplication by the energy-dependent
windows, so the total geometric weight, which sets the normalization $N_k$
of the average, is unchanged. Inside a refined cell, $\mathcal H(\mathbf k)$ is
obtained from a Taylor expansion about the cell center, constructed once
and independent of $\mathbf Q$. The expansion order is chosen such that
$\max|\mathcal H_{\rm Taylor}-\mathcal H|<10^{-9}$~eV over the cell
enlarged by $|q_i|/2\le5\times10^{-4}$ in each fractional component
(Sec.~S2.3; fifth order at $N=201$, error $7\times10^{-11}$~eV).

Two limits make the scheme testable. For $W\to\infty$ and odd $m$,
Eq.~\eqref{eq:S3_nested} reduces exactly to the uniform mesh with spacing
$\delta k/m^L$; we verified this to a relative $10^{-14}$ for $L=1$ and $2$. For
finite $W$, provided that the selected cells contain the full support of the
windows and the Taylor-interpolation error is negligible, which the padded
cell selection and the Taylor tolerance above are designed to ensure, the
additional error relative to the uniformly refined mesh comes from
integrating the smooth remainder on the coarser meshes. We measured it
directly at $N=201$, $L=1$, and $m=3$, where the refined points coincide with
the uniform third9 mesh with $N=603$ ($1809$ points per direction),
at fixed $(Q_y,\Delta)$ pairs bracketing the critical region. The
error decreases rapidly with the ratio of $W$ to the energy spread
$\epsilon_c$ of a Fermi-surface cell ($\epsilon_c=11.8$~meV at $N=201$,
90th percentile). The relative error of the pairing kernel
[Eq.~\eqref{eq:S2_Fpair_definition}] is $3\times10^{-7}$,
$2\times10^{-8}$, and $3\times10^{-9}$ for $W/\epsilon_c=4$, $6$, and $8$.
The corresponding error of the nonreciprocal free-energy difference
$\Omega(Q_y)-\Omega(-Q_y)$ is $1.1\times10^{-2}$, $1.8\times10^{-3}$, and
$3\times10^{-4}$ at the
near-critical pair $(Q_y,\Delta)=(\pm3.2\times10^{-4}~\text{\AA}^{-1},
0.25~{\rm meV})$. For comparison, the uniform $N=201$ mesh differs from
$N=603$ by a relative $0.17$ in the same quantity. We use $m=3$ and
$W=72$~meV$\,\approx6\epsilon_c$.

\begin{figure*}[t]
  \centering
  \includegraphics[width=\textwidth]{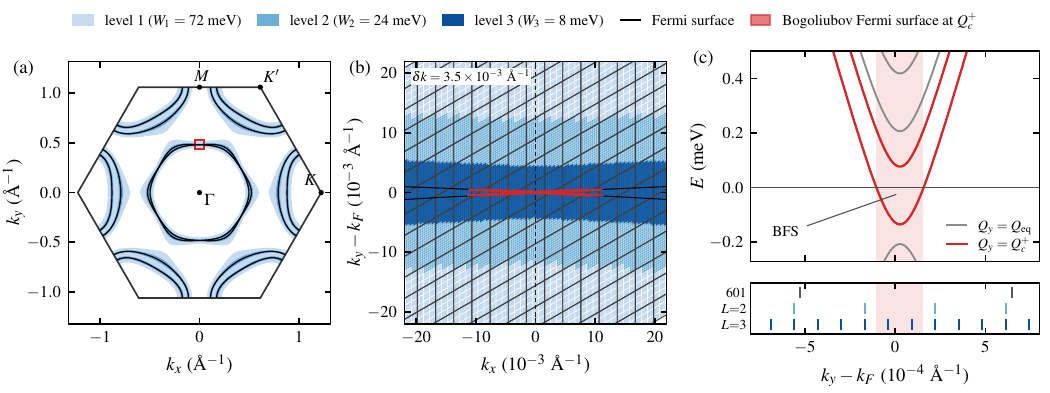}
  \caption{Fermi-surface-adaptive quadrature for the one-sided in-plane state
  ($D_b=1$~meV, $\mu_0$; $N=201$, $m=3$, $L=3$).
  (a) Regions refined at levels $1$, $2$, and $3$ (windows $W_l=72$, $24$,
  and $8$~meV), approximated by the regions in which a normal-state band lies
  within $W_l$ of $E_F$, and the Fermi surface. The square marks the region
  shown in (b).
  (b) Cells near the crossing of the two spin-split $\Gamma$ sheets on the
  $\Gamma$--$M$ line ($k_F=0.481~\text{\AA}^{-1}$), selected with the
  production criterion. Dark lines: base cells (spacing $\delta k$); white
  lines: level-1 and level-2 cells; level-3 cells ($\delta k/27$) are not
  outlined. Red: Bogoliubov Fermi surface at $Q_y=Q_c^+$. Dashed line: the cut
  shown in (c).
  (c) BdG quasiparticle energies along this cut at $Q_y=Q_{\rm eq}$ and
  $Q_y=Q_c^+=2.50\times10^{-4}~\text{\AA}^{-1}$
  ($\Delta^\star=0.398$ and $0.367$~meV; $L=2$ branch). The shaded interval
  is the Bogoliubov Fermi surface (BFS). Bottom: $k$ points on the cut for the
  uniform $N=601$ mesh ($1803$ points per reciprocal-lattice direction) and for
  the quadrature with $L=2$ and $3$ ($5427$ and $16281$).}
  \label{fig:S3_quadrature}
\end{figure*}

Figure~\ref{fig:S3_quadrature} shows the resulting quadrature for the
one-sided in-plane state. The refined cells form nested shells around the
normal-state Fermi surface [Fig.~\ref{fig:S3_quadrature}(a)]. At $N=201$ and
$L=3$, the first level refines $33\%$ of the base cells, and the three levels
refine $1.2\times10^5$, $4.3\times10^5$, and $1.5\times10^6$ cells, so that
near the Fermi surface each base cell is divided into up to $27\times27$
cells [Fig.~\ref{fig:S3_quadrature}(b)]. This resolution is set by the
Bogoliubov Fermi surfaces of the critical states. At $Q_c^+$, a pocket forms
where the two spin-split $\Gamma$ sheets cross on the $\Gamma$--$M$ line. It is
$2.6\times10^{-4}~\text{\AA}^{-1}$ wide along $\Gamma$--$M$ and
$2.3\times10^{-2}~\text{\AA}^{-1}$ long [Fig.~\ref{fig:S3_quadrature}(c)], and
it contains no point of the base mesh, $2$ points of the uniform $N=601$ mesh,
$30$ points at $L=2$, and $282$ points at $L=3$.

\begin{figure*}[t]
  \centering
  \includegraphics[width=0.85\textwidth]{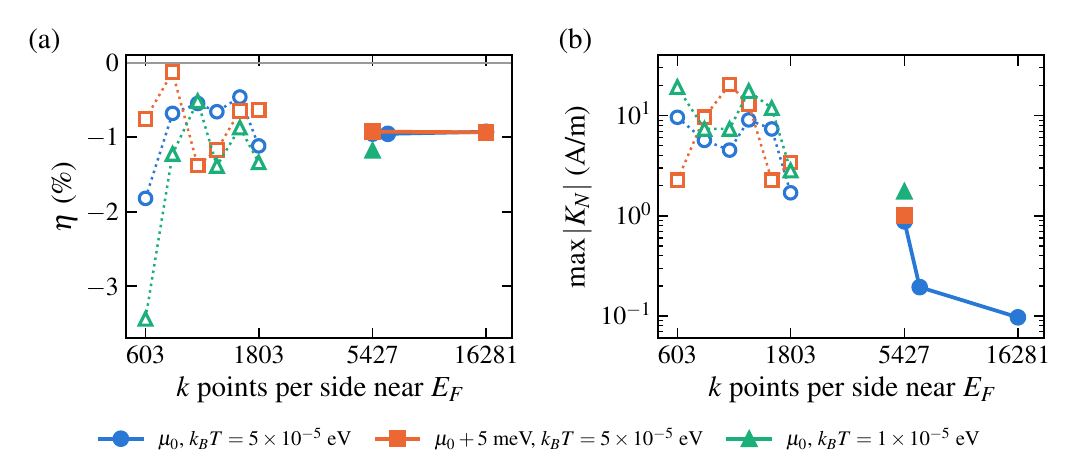}
  \caption{Brillouin-zone convergence of the diode efficiency.
  (a) $\eta$ and (b) maximum spurious normal-state current $|K_{\rm N}|$ versus
  the number of $k$ points per reciprocal-lattice direction near $E_F$,
  for one-sided in-plane exchange. Open symbols: uniform third9 meshes,
  $N=201$--$601$ ($3N$ points per direction). Filled symbols:
  Fermi-surface-adaptive quadrature [Eq.~\eqref{eq:S3_nested}] with
  $m=3$ on base meshes $N=201$ ($L=2$: $5427$; $L=3$: $16281$) and $N=233$
  ($L=2$: $6291$). The $L=2$ quadrature resolves the Fermi surface three
  times more finely than the largest uniform mesh ($5427$ versus $1803$
  points per direction) with about $1.5$ times as many $k$ points in
  total. Values are listed in
  Table~\ref{tab:S3_conv}.}
  \label{fig:S3_conv}
\end{figure*}

\begin{table*}[t]
\caption{Brillouin-zone convergence for one-sided in-plane exchange
($\theta=90^\circ$, $D_b=1$~meV). $N$ is the base mesh (third9);
$N_{\rm F}$ is the number of $k$ points per reciprocal-lattice direction in
the energy window around $E_F$ ($3N$ for uniform meshes and $3N\cdot3^L$
for the Fermi-surface-adaptive quadrature). $g_{\rm SC}$ is recalibrated
for every quadrature. $\Delta^\star(0)$ is in meV, $Q_{\rm eq}$ in
$10^{-6}~\text{\AA}^{-1}$, $K_c^\pm$ and $K_{\rm N}$ in A/m. $K_c^\pm$ are
obtained from $d\Omega_{\rm SC}/dQ_y$ with the rule of Sec.~S2.6; FS-refined
runs use $dq=2\times10^{-6}$ ($L=2$) and $4\times10^{-6}$ ($L=3$).}
\label{tab:S3_conv}
\begin{ruledtabular}
\begin{tabular}{llcccccccc}
State & Quadrature & $N$ & $N_{\rm F}$ & $\Delta^\star(0)$ & $Q_{\rm eq}$ & $K_c^+$ & $K_c^-$ &
$\eta$ (\%) & $\max|K_{\rm N}|$\\
\hline
$\mu_0$, $k_BT=5\times10^{-5}$~eV
 & uniform & 201 & 603 & 0.3985 & $-17.83$ & 96.96 & $-100.56$ & $-1.82$ & 9.6\\
 & uniform & 261 & 783 & 0.3985 & $-17.87$ & 96.47 & $-97.79$ & $-0.68$ & 5.7\\
 & uniform & 333 & 999 & 0.3985 & $-17.86$ & 99.14 & $-100.23$ & $-0.55$ & 4.5\\
 & uniform & 401 & 1203 & 0.3984 & $-17.86$ & 94.33 & $-95.58$ & $-0.66$ & 9.1\\
 & uniform & 501 & 1503 & 0.3985 & $-17.88$ & 95.38 & $-96.27$ & $-0.46$ & 7.4\\
 & uniform & 601 & 1803 & 0.3984 & $-17.85$ & 95.75 & $-97.91$ & $-1.12$ & 1.7\\
 & FS, $L=2$ & 201 & 5427 & 0.3985 & $-17.86$ & 95.64 & $-97.48$ & $-0.95$ & 0.88\\
 & FS, $L=2$ & 233 & 6291 & 0.3985 & $-17.85$ & 95.88 & $-97.73$ & $-0.96$ & 0.19\\
 & FS, $L=3$ & 201 & 16281 & 0.3985 & $-17.86$ & 95.95 & $-97.75$ & $-0.93$ & 0.10\\
\hline
$\mu_0+5$~meV, $k_BT=5\times10^{-5}$~eV
 & uniform & 201 & 603 & 0.3542 & $-17.82$ & 80.68 & $-81.91$ & $-0.76$ & 2.3\\
 & uniform & 261 & 783 & 0.3999 & $-17.86$ & 100.99 & $-101.24$ & $-0.13$ & 9.6\\
 & uniform & 333 & 999 & 0.4079 & $-17.81$ & 93.66 & $-96.28$ & $-1.38$ & 20.4\\
 & uniform & 401 & 1203 & 0.3470 & $-17.82$ & 82.34 & $-84.29$ & $-1.17$ & 13.0\\
 & uniform & 501 & 1503 & 0.3569 & $-17.84$ & 80.30 & $-81.35$ & $-0.65$ & 2.3\\
 & uniform & 601 & 1803 & 0.3689 & $-17.84$ & 85.89 & $-86.98$ & $-0.63$ & 3.4\\
 & FS, $L=2$ & 201 & 5427 & 0.3671 & $-17.83$ & 85.18 & $-86.77$ & $-0.93$ & 1.0\\
 & FS, $L=3$ & 201 & 16281 & 0.3677 & $-17.83$ & 84.93$^{a}$ & $-86.53^{a}$ & $-0.93^{a}$ & --\\
\hline
$\mu_0$, $k_BT=10^{-5}$~eV
 & uniform & 201 & 603 & 0.3997 & $-17.87$ & 118.55 & $-126.99$ & $-3.44$ & 19.2\\
 & uniform & 261 & 783 & 0.3999 & $-17.87$ & 116.62 & $-119.51$ & $-1.23$ & 7.4\\
 & uniform & 333 & 999 & 0.3998 & $-17.87$ & 124.66 & $-125.95$ & $-0.52$ & 7.4\\
 & uniform & 401 & 1203 & 0.3999 & $-17.87$ & 114.92 & $-118.15$ & $-1.39$ & 17.5\\
 & uniform & 501 & 1503 & 0.3998 & $-17.87$ & 115.96 & $-118.00$ & $-0.87$ & 11.9\\
 & uniform & 601 & 1803 & 0.3998 & $-17.87$ & 119.20 & $-122.44$ & $-1.34$ & 2.8\\
 & FS, $L=2$ & 201 & 5427 & 0.3998 & $-17.87$ & 117.24 & $-120.04$ & $-1.18$ & 1.8\\
\end{tabular}
\end{ruledtabular}
\raggedright
$^{a}$\,From $d(\Omega_{\rm SC}-\Omega_{\rm N})/dQ_y$. For the other converged
runs this definition and $d\Omega_{\rm SC}/dQ_y$ agree to within $0.07$
percentage points in $\eta$; at $L=3$ ($\mu_0$) they agree to $0.002$.
\end{table*}

\emph{Convergence and production protocol.}
With the adaptive quadrature the diode efficiency converges
(Fig.~\ref{fig:S3_conv} and Table~\ref{tab:S3_conv}). At $\mu_0$ and
$k_BT=5\times10^{-5}$~eV, $L=2$ on base meshes $N=201$ and $233$ gives
$\eta=-0.95\%$ and $-0.96\%$, and $L=3$ gives $-0.93\%$. The critical
currents agree to within about $0.3$~A/m, $\Delta^\star(0)$ to within
$10^{-4}$~meV, and the recalibrated couplings to within $10^{-4}$~eV
($g_{\rm SC}=0.500676$, $0.500716$, and $0.500721$~eV). The spurious current $K_{\rm N}$ drops to $0.9$, $0.2$, and
$0.1$~A/m. At $\mu_0+5$~meV, $L=2$ and $L=3$ give $\eta=-0.93\%$ to within
$0.01$ percentage points, and the large fluctuations of $\Delta^\star(0)$
on the uniform meshes disappear. The $L=3$ values,
$\eta(\mu_0)=-0.93\%$ and $\eta(\mu_0+5~{\rm meV})=-0.93\%$ at
$k_BT=5\times10^{-5}$~eV, differ from every $L=2$ result by at most $0.09$
percentage points, and by at most $0.03$ when the current is taken from
$d\Omega_{\rm SC}/dQ_y$ as in the rest of this work.
At $k_BT=5\times10^{-5}$~eV, the uniform-mesh results scatter around these
values by up to $\pm0.9$ percentage points; at $k_BT=10^{-5}$~eV the scatter
reaches $2.3$ percentage points. In particular, the apparent near-cancellation
of $\eta$ at $\delta\mu=+5$~meV on uniform meshes is a sampling artifact:
the converged efficiencies at $\mu_0$ and $\mu_0+5$~meV are equal within
their uncertainty.

Unless stated otherwise, diode efficiencies for in-plane and tilted exchange
are therefore computed on the $N=201$ third9 base mesh with $m=3$, $L=2$, and
$W=72$~meV,
with $g_{\rm SC}$ calibrated on the same quadrature, at
$k_BT=5\times10^{-5}$~eV and $dq=4\times10^{-6}$ (or $2\times10^{-6}$ where
noted). Doubling the $Q$ spacing from $2\times10^{-6}$ changes
$\eta$ by about $0.02$ percentage points (Sec.~S3.2). For the states tested above, these
comparisons suggest numerical uncertainties of approximately $0.1$ percentage
points in $\eta$ and $0.3$~A/m in $K_c^\pm$. For purely out-of-plane exchange, the
critical-current asymmetry requires finer Brillouin-zone resolution: $L=2$
is not sufficient there, and those states are computed with $L=3$ unless
marked, and checked against $L=4$ at
$D_z=1$~meV (Sec.~S4.6). One gap-equation evaluation takes $14$~s on $128$ cores
($4.9$~million points in total), compared with $5$~s for the uniform
$N=601$ mesh and $50$~s for $L=3$. The production temperature
corresponds to $T\approx0.58$~K, about $0.3\,T_c$ for the model
$T_c\approx1.9$~K estimated from $\Delta^\star(T)$ (Sec.~S6.2). At
$k_BT=10^{-5}$~eV the in-plane state was computed only with $L=2$, which
gives $\eta=-1.18\%$; an $L=3$ comparison was not made at this temperature,
but the convergence diagnostics are as good as for the production states
(Sec.~S6.2).

\subsection*{S3.2. \texorpdfstring{$Q$}{Q}-grid resolution and critical-current extraction}

Table~\ref{tab:S3_dq} tests the $Q$ spacing on the production quadrature,
with the coarser grids taken from the same runs. The stored
$\Omega_{\rm SC}(Q_y)$, written with 13 significant digits, resolves its
variation between neighboring $Q_y$ only to about $10^{-11}$~eV per cell,
which is insufficient for this test. The stored current, itself a central
difference of the full-precision $\Omega_{\rm SC}$, is therefore integrated
numerically to reconstruct $\Omega_{\rm SC}(Q_y)$ up to a constant, and the
reconstruction is differentiated by central differences using every second
or fourth point. The reconstructed free energy is not identical to the
directly evaluated one. This procedure tests the sensitivity to coarser
differentiation and extremum sampling, while retaining the native-grid
derivative error and introducing an additional reconstruction error.
Going from $dq=2\times10^{-6}$ to the production value $4\times10^{-6}$
($\delta Q_y=4.2$ and $8.5\times10^{-6}~\text{\AA}^{-1}$) changes $K_c^\pm$
by at most $0.13$~A/m and $\eta$ by about $0.02$ percentage points, and
$Q_{\rm eq}$ by at most $0.01\times10^{-6}~\text{\AA}^{-1}$. Doubling the
spacing again, to $dq=8\times10^{-6}$, shifts $\eta$ by about $0.08$
percentage points (AP). The observed sensitivity to coarsening is smaller
than the estimated quadrature uncertainty of approximately $0.1$ percentage
points (Sec.~S3.1). The same
subsampling applied to the out-of-plane states at $L=3$ of
Table~\ref{tab:S7_z} (not tabulated) changes $\eta$ by at most $0.06$
percentage points when the spacing is doubled ($D_b=1$~meV, $-1.447\%\rightarrow-1.389\%$), including the cusp at
$k_BT=10^{-5}$~eV ($-3.372\%\rightarrow-3.406\%$). These subsampling tests
indicate a $Q$-grid sensitivity of about $0.06$ percentage points or less for
the tested out-of-plane states. The comparison provides an empirical resolution estimate rather
than a strict bound on the continuum-grid error: two spacings do not
establish the asymptotic regime, in particular where the sampled extremum
moves relative to a cusp.

\begin{table}[t]
\caption{$Q$-grid convergence on the production quadrature
($N=201$, third9, $m=3$, $L=2$, $k_BT=5\times10^{-5}$~eV unless noted).
$dq$ is the fractional spacing, and the first row of each block is the
native grid of the run. Units as in Table~\ref{tab:S3_conv}.}
\label{tab:S3_dq}
\begin{ruledtabular}
\begin{tabular}{lccccc}
State & $dq$ & $Q_{\rm eq}$ & $K_c^+$ & $K_c^-$ & $\eta$ (\%)\\
\hline
One-sided, $\mu_0$ & $2\times10^{-6}$ & $-17.86$ & 95.64 & $-97.48$ & $-0.952$\\
 & $4\times10^{-6}$ & $-17.86$ & 95.51 & $-97.39$ & $-0.974$\\
 & $8\times10^{-6}$ & $-17.86$ & 95.25 & $-97.13$ & $-0.976$\\
\hline
One-sided, & $2\times10^{-6}$ & $-17.83$ & 85.18 & $-86.77$ & $-0.925$\\
$\delta\mu=+5$~meV & $4\times10^{-6}$ & $-17.84$ & 85.08 & $-86.65$ & $-0.918$\\
 & $8\times10^{-6}$ & $-17.84$ & 84.70 & $-86.22$ & $-0.891$\\
\hline
One-sided, & $2\times10^{-6}$ & $-17.87$ & 117.24 & $-120.04$ & $-1.176$\\
$k_BT=10^{-5}$~eV & $4\times10^{-6}$ & $-17.87$ & 117.20 & $-119.99$ & $-1.178$\\
 & $8\times10^{-6}$ & $-17.87$ & 117.00 & $-119.87$ & $-1.211$\\
\hline
AP, $\mu_0$ & $4\times10^{-6}$ & $-35.71$ & 94.18 & $-97.82$ & $-1.895$\\
 & $8\times10^{-6}$ & $-35.71$ & 93.65 & $-97.43$ & $-1.977$\\
 & $1.6\times10^{-5}$ & $-35.71$ & 92.63 & $-96.37$ & $-1.980$\\
\end{tabular}
\end{ruledtabular}
\end{table}

The extraction rule of Sec.~S2.6 matters only where $\Delta^\star(Q_y)$
collapses abruptly at the end of a branch. There, a finite-difference
derivative across the collapse produces an isolated spike in $K_y$, and a
naive extremum search can return it as a critical current. On uniform meshes
this happened, for example, for the AP state at $dq=5\times10^{-7}$, where it
gave $K_c^-=-136.2$~A/m instead of the interior extremum. On the production
quadrature, all reported $K_c^\pm$ are interior extrema. Varying the collapse
threshold between $0.01$ and $0.05\,\Delta^\star(0)$ and the abruptness
factor between $2$ and $5$ leaves every $\eta$ in this work unchanged at the
quoted precision.

\subsection*{S3.3. Self-consistency criterion and seeding}

The gap equation, Eq.~\eqref{eq:S2_self_consistency_residual}, is solved to
an absolute residual $|G|<\epsilon$ with $\epsilon=10^{-8}$~eV in production.
At each $Q_y$, the solver starts from the converged value at the neighboring
$Q_y$. When an earlier branch is available, its value interpolated to that
$Q_y$ is used as a fallback seed. The tolerance test of
Table~\ref{tab:S3_tol} was performed on the uniform $N=201$ third9 mesh at
$k_BT=10^{-5}$~eV. The residual criterion has the same definition on the
production quadrature, although the sensitivity of the root,
$\delta\Delta^\star\simeq\delta G/\partial_\Delta G$, also depends on the
local slope of the gap equation and grows near a fold. The present test
therefore establishes tolerance insensitivity for the uniform-mesh benchmark
calculations. Tightening $\epsilon$
to $10^{-9}$ and $10^{-10}$~eV changes $\Delta^\star(0)$ by at most
$10^{-4}$~meV, $Q_{\rm eq}$ by at most $0.01\times10^{-6}~\text{\AA}^{-1}$,
$K_c^\pm$ by less than
$0.01$~A/m, and $\eta$ by less than $0.01$ percentage points. For these
benchmark calculations, the residual tolerance is not a limiting source of
numerical error.

\begin{table}[t]
\caption{Self-consistency criterion (uniform $N=201$ third9 mesh,
$dq=10^{-6}$, $k_BT=10^{-5}$~eV). $\epsilon$ is the absolute residual
tolerance of the gap equation; units as in Table~\ref{tab:S3_conv}. The
one-sided rows were computed with an earlier version of the one-sided
exchange Hamiltonian with a shorter hopping range, and therefore
differ from the uniform $N=201$ row of Table~\ref{tab:S3_conv}; all other
results in this work use the full Hamiltonian. Only the dependence on
$\epsilon$ is tested here.}
\label{tab:S3_tol}
\begin{ruledtabular}
\begin{tabular}{lcccccc}
State & $\epsilon$ (eV) & $\Delta^\star(0)$ & $Q_{\rm eq}$ & $K_c^+$ & $K_c^-$ & $\eta$ (\%)\\
\hline
One-sided, $\mu_0$ & $10^{-8}$ & 0.3951 & $-17.88$ & 117.36 & $-125.41$ & $-3.32$\\
 & $10^{-9}$ & 0.3951 & $-17.88$ & 117.36 & $-125.41$ & $-3.32$\\
 & $10^{-10}$ & 0.3951 & $-17.88$ & 117.36 & $-125.41$ & $-3.32$\\
\hline
One-sided, & $10^{-8}$ & 0.3521 & $-17.85$ & 108.24 & $-108.02$ & $+0.10$\\
$\delta\mu=+5$~meV & $10^{-9}$ & 0.3520 & $-17.85$ & 108.24 & $-108.02$ & $+0.10$\\
 & $10^{-10}$ & 0.3520 & $-17.86$ & 108.24 & $-108.02$ & $+0.10$\\
\hline
AP, $\mu_0$ & $10^{-8}$ & 0.4000 & $-35.73$ & 116.85 & $-130.27$ & $-5.43$\\
 & $10^{-9}$ & 0.4000 & $-35.73$ & 116.85 & $-130.27$ & $-5.43$\\
\end{tabular}
\end{ruledtabular}
\end{table}

\subsection*{S3.4. Numerical reciprocity floor}

The reciprocity floor is established in Sec.~S1.3 with the no-exchange and P
states: $\max|\Omega_{\rm SC}^{\rm odd}|\approx8\times10^{-14}$~eV/cell,
$\max|K^{\rm even}|/\max|K|\approx3\times10^{-5}$, and $|\eta|<10^{-3}\%$.
On the production quadrature, the out-of-plane one-sided state scanned along
$\hat{\mathbf y}$, which is reciprocal by $M_x\mathcal T$ (Sec.~S4.1), gives
$|Q_{\rm eq}|<0.01\times10^{-6}~\text{\AA}^{-1}$ and $|\eta|=0.003\%$
(Table~\ref{tab:S4_states}). Symmetry checks
compare states computed on the same quadrature. These checks bound
numerical symmetry breaking, but they do not bound symmetry-preserving
quadrature errors in the critical currents or the diode efficiency
(Sec.~S3.1).


\section*{S4. Helical response and layer parity: complete results}

\begin{figure*}[t]
  \centering
  \includegraphics[width=\textwidth]{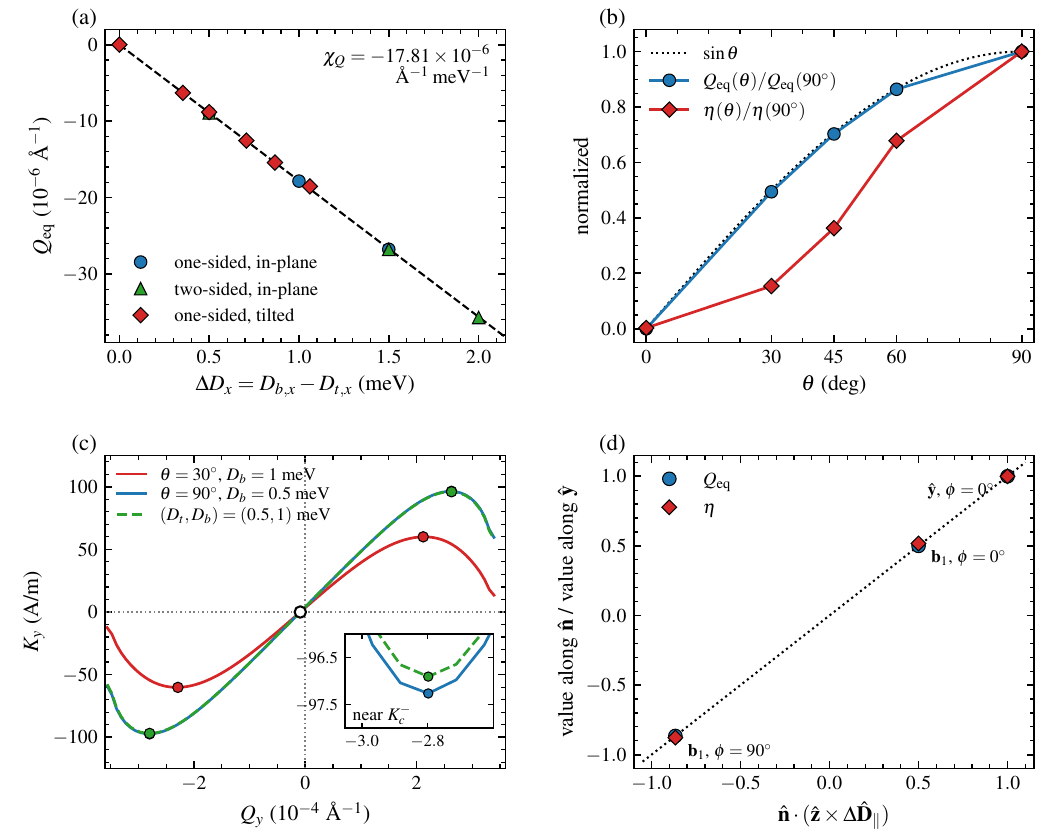}
  \caption{Helical momentum and diode efficiency at $\mu=\mu_0$
  (production quadrature, $k_BT=5\times10^{-5}$~eV).
  (a) $Q_{\rm eq}$ versus $\Delta D_x$ [Eq.~\eqref{eq:S4_dDx}] for the
  following states scanned along $\hat{\mathbf y}$ (Table~\ref{tab:S4_states}): one-sided
  in-plane ($D_b=0.5$, $1$, $1.5$~meV), two-sided in-plane
  ($(D_t,D_b)=(0.5,1)$, $(-0.5,1)$, and AP), and one-sided tilted
  ($\theta=0^\circ$--$60^\circ$ at $D_b=1$~meV, and $\theta=45^\circ$ at
  $0.5$ and $1.5$~meV). The dashed line is Eq.~\eqref{eq:S4_collapse}.
  (b) $Q_{\rm eq}(\theta)/Q_{\rm eq}(90^\circ)$ and
  $\eta(\theta)/\eta(90^\circ)$ for the one-sided polar-angle scan at
  $D_b=1$~meV, $\phi=0^\circ$. The dotted curve is $\sin\theta$.
  (c) Sheet current $K_y(Q_y)$ for three states with
  $Q_{\rm eq}\simeq-8.9\times10^{-6}~\text{\AA}^{-1}$. The two in-plane
  states, which differ only in $D_+$, are indistinguishable on this scale;
  the inset enlarges the region near $K_c^-$. Filled circles mark $K_c^\pm$,
  and the open circle marks the common zero crossing.
  (d) Vector form of the helical response. $Q_{\rm eq}$ and $\eta$ along a
  scan direction $\hat{\mathbf n}$, normalized to the values for
  $\hat{\mathbf n}=\hat{\mathbf y}$ and $\phi=0^\circ$, versus the projection
  $\hat{\mathbf n}\cdot(\hat{\mathbf z}\times\Delta\hat{\mathbf D}_\parallel)$.
  The dotted line is the projection law. Momenta are in
  $10^{-6}~\text{\AA}^{-1}$ in (a) and $10^{-4}~\text{\AA}^{-1}$ in (c).}
  \label{fig:S4_helical}
\end{figure*}

This section collects the $\mu=\mu_0$ calculations behind the helical
magnetoelectric response and the layer-parity control of the main text. All
states use the production quadrature of Sec.~S3.1 ($N=201$, third9, $m=3$,
$L=2$, $W=72$~meV; $L=3$ for the out-of-plane states of Sec.~S4.6 unless
marked, and in addition for the tilted textures with $D_z>1$~meV),
$k_BT=5\times10^{-5}$~eV, $dq=4\times10^{-6}$ (or
$2\times10^{-6}$ where noted), and $g_{\rm SC}$ calibrated on the same
quadrature (Sec.~S2.3). The observables are listed in
Table~\ref{tab:S4_states} and summarized in Fig.~\ref{fig:S4_helical}. To
place one-sided, two-sided, and tilted textures on a common footing, we use
the $x$ component of the top--bottom exchange contrast,
\begin{equation}
  \Delta D_x\equiv D_{b,x}-D_{t,x},
  \qquad
  D_{\ell,x}=D_\ell\sin\theta_\ell\cos\phi_\ell ,
  \label{eq:S4_dDx}
\end{equation}
and the out-of-plane component $D_z=D_b\cos\theta$ of the one-sided
textures. For the collinear in-plane textures, $\Delta D_x$ reduces to
$\Delta D=D_b-D_t=2D_-$ [Eq.~\eqref{eq:S1_exchange_parity}].

All states scanned along $\hat{\mathbf y}$ fall on a single line
[Fig.~\ref{fig:S4_helical}(a)],
\begin{equation}
  Q_{\rm eq}\simeq\chi_Q\,\Delta D_x,
  \qquad
  \chi_Q=-17.81\times10^{-6}~\text{\AA}^{-1}\,{\rm meV}^{-1},
  \label{eq:S4_collapse}
\end{equation}
where $\chi_Q$ is a least-squares fit through the origin over the states
of Fig.~\ref{fig:S4_helical}(a); including the four further tilted states
of Sec.~S4.1 changes it by $0.09\times10^{-6}~\text{\AA}^{-1}\,{\rm meV}^{-1}$.
The deviation is below $0.09\times10^{-6}~\text{\AA}^{-1}$ for all in-plane
textures and reaches $0.34$ and $0.53\times10^{-6}~\text{\AA}^{-1}$ for the
tilted states with $\theta=45^\circ$ and $D_b=1.5$ and $1.75$~meV. The collapse holds,
to this accuracy, irrespective of the layer-even exchange $D_+$ and of the
out-of-plane component. The diode efficiency behaves differently. For in-plane textures
it is also approximately a function of $\Delta D_x$ alone over the tested
configurations, and is linear in it
[Fig.~1(c) of the main text]. An out-of-plane component, which leaves
$Q_{\rm eq}$ nearly unchanged, strongly suppresses it and, for
$D_z\gtrsim1.1$~meV and at low temperature (Sec.~S6.2), reverses its sign. The following subsections
document the individual scans.

\begin{table*}[t]
\caption{Finite-$Q$ observables at $\mu=\mu_0$ (production
quadrature, $k_BT=5\times10^{-5}$~eV). Exchange magnitudes $D_t$ and $D_b$
are in meV. For the two-sided textures ($\theta=90^\circ$, $\phi=0^\circ$),
$D_t$ and $D_b$ are signed $x$ components, so that a negative value denotes
$\mathbf D_\ell\parallel-\hat{\mathbf x}$. The column ``$\mathbf Q\parallel$''
gives the scan direction; $\mathbf b_1$ lies at $30^\circ$ to
$\hat{\mathbf x}$, and for these rows $Q_{\rm eq}$ is the component along the
scan. $\Delta D_x$ [Eq.~\eqref{eq:S4_dDx}] is in meV, $Q_{\rm eq}$ in
$10^{-6}~\text{\AA}^{-1}$, $K_c^\pm$ in A/m, and $\Delta^\star(0)$ in meV.
Entries marked ``sym.'' are fixed by symmetry and were not recomputed on the
production quadrature (Secs.~S1 and S4.1). Rows marked ($L=3$) use the
three-level quadrature, all others $L=2$. $^\dagger$\,$dq=2\times10^{-6}$;
all other rows $dq=4\times10^{-6}$.}
\label{tab:S4_states}
\begin{ruledtabular}
\begin{tabular}{lccccccccccc}
State & $D_t$ & $D_b$ & $\theta$ & $\phi$ & $\mathbf Q\parallel$ &
$\Delta D_x$ & $Q_{\rm eq}$ & $K_c^+$ & $K_c^-$ & $\eta$ (\%) &
$\Delta^\star(0)$\\
\hline
\multicolumn{12}{l}{\emph{(i) One-sided, polar angle}}\\
\quad Bottom only & 0 & 1 & $0^\circ$ & $0^\circ$ & $\hat{\mathbf y}$ & 0 & 0.00 & 49.35 & $-$49.35 & $-$0.003 & 0.385 \\
\quad Bottom only & 0 & 1 & $30^\circ$ & $0^\circ$ & $\hat{\mathbf y}$ & 0.50 & $-$8.82 & 60.00 & $-$60.17 & $-$0.147 & 0.390 \\
\quad Bottom only & 0 & 1 & $45^\circ$ & $0^\circ$ & $\hat{\mathbf y}$ & 0.71 & $-$12.55 & 71.29 & $-$71.78 & $-$0.346 & 0.394 \\
\quad Bottom only & 0 & 1 & $60^\circ$ & $0^\circ$ & $\hat{\mathbf y}$ & 0.87 & $-$15.43 & 83.14 & $-$84.22 & $-$0.646 & 0.397 \\
\quad Bottom only$^\dagger$ & 0 & 1 & $90^\circ$ & $0^\circ$ & $\hat{\mathbf y}$ & 1.00 & $-$17.86 & 95.64 & $-$97.48 & $-$0.952 & 0.398 \\
\quad\quad ($L=3$) & 0 & 1 & $90^\circ$ & $0^\circ$ & $\hat{\mathbf y}$ & 1.00 & $-$17.86 & 95.95 & $-$97.75 & $-$0.927 & 0.398 \\
\hline
\multicolumn{12}{l}{\emph{(ii) One-sided, exchange strength}}\\
\quad Bottom only & 0 & 0.5 & $90^\circ$ & $0^\circ$ & $\hat{\mathbf y}$ & 0.50 & $-$8.93 & 96.33 & $-$97.27 & $-$0.483 & 0.399 \\
\quad Bottom only & 0 & 1.5 & $90^\circ$ & $0^\circ$ & $\hat{\mathbf y}$ & 1.50 & $-$26.78 & 94.60 & $-$97.37 & $-$1.442 & 0.398 \\
\quad Bottom only & 0 & 0.5 & $45^\circ$ & $0^\circ$ & $\hat{\mathbf y}$ & 0.35 & $-$6.31 & 90.02 & $-$90.60 & $-$0.319 & 0.398 \\
\quad Bottom only & 0 & 1.5 & $45^\circ$ & $0^\circ$ & $\hat{\mathbf y}$ & 1.06 & $-$18.53 & 43.83 & $-$43.88 & $-$0.061 & 0.381 \\
\quad\quad ($L=3$) & 0 & 1.5 & $45^\circ$ & $0^\circ$ & $\hat{\mathbf y}$ & 1.06 & $-$18.55 & 44.01 & $-$44.04 & $-$0.034 & 0.381 \\
\quad Bottom only & 0 & 1.75 & $45^\circ$ & $0^\circ$ & $\hat{\mathbf y}$ & 1.24 & $-$21.47 & 28.30 & $-$28.13 & $+$0.309 & 0.366 \\
\quad\quad ($L=3$) & 0 & 1.75 & $45^\circ$ & $0^\circ$ & $\hat{\mathbf y}$ & 1.24 & $-$21.51 & 28.44 & $-$28.24 & $+$0.354 & 0.366 \\
\quad Bottom only & 0 & 1.5 & $30^\circ$ & $0^\circ$ & $\hat{\mathbf y}$ & 0.75 & $-$13.05 & 23.05 & $-$22.92 & $+$0.289 & 0.359 \\
\quad\quad ($L=3$) & 0 & 1.5 & $30^\circ$ & $0^\circ$ & $\hat{\mathbf y}$ & 0.75 & $-$13.07 & 23.17 & $-$23.02 & $+$0.330 & 0.359 \\
\quad Bottom only ($L=3$) & 0 & 1.84 & $45^\circ$ & $0^\circ$ & $\hat{\mathbf y}$ & 1.30 & $-$22.67 & 22.74 & $-$22.46 & $+$0.615 & 0.358 \\
\quad Bottom only & 0 & 1.414 & $60^\circ$ & $0^\circ$ & $\hat{\mathbf y}$ & 1.22 & $-$21.73 & 70.76 & $-$71.62 & $-$0.605 & 0.394 \\
\hline
\multicolumn{12}{l}{\emph{(iii) Two-sided, in-plane ($\mathbf D_\ell\parallel\pm\hat{\mathbf x}$)}}\\
\quad P & 1 & 1 & $90^\circ$ & $0^\circ$ & $\hat{\mathbf y}$ & 0 & sym. & -- & -- & 0 (sym.) & -- \\
\quad Two-sided & 0.5 & 1 & $90^\circ$ & $0^\circ$ & $\hat{\mathbf y}$ & 0.50 & $-$8.93 & 95.97 & $-$96.91 & $-$0.488 & 0.398 \\
\quad Two-sided & $-$0.5 & 1 & $90^\circ$ & $0^\circ$ & $\hat{\mathbf y}$ & 1.50 & $-$26.78 & 94.94 & $-$97.72 & $-$1.441 & 0.399 \\
\quad AP & $-$1 & 1 & $90^\circ$ & $0^\circ$ & $\hat{\mathbf y}$ & 2.00 & $-$35.71 & 94.18 & $-$97.82 & $-$1.895 & 0.398 \\
\hline
\multicolumn{12}{l}{\emph{(iv) Direction of $\mathbf Q$}}\\
\quad Bottom only & 0 & 1 & $90^\circ$ & $90^\circ$ & $\hat{\mathbf y}$ & 0 & sym. & -- & -- & 0 (sym.) & -- \\
\quad Bottom only & 0 & 1 & $90^\circ$ & $0^\circ$ & $\mathbf b_1$ & 1.00 & $-$8.93 & 95.81 & $-$96.75 & $-$0.493 & 0.398 \\
\quad Bottom only & 0 & 1 & $90^\circ$ & $90^\circ$ & $\mathbf b_1$ & 0 & $+$15.45 & 97.20 & $-$95.59 & $+$0.835 & 0.398 \\
\end{tabular}
\end{ruledtabular}
\end{table*}

\subsection*{S4.1. Polar angle and the out-of-plane limit}

For one-sided bottom-layer exchange with $D_b=1$~meV and $\phi=0^\circ$, the
ratio $Q_{\rm eq}/\sin\theta$ is $-17.65$, $-17.75$, $-17.82$, and
$-17.86\times10^{-6}~\text{\AA}^{-1}$ for $\theta=30^\circ$, $45^\circ$,
$60^\circ$, and $90^\circ$ (Table~\ref{tab:S4_states}, block i). It is
constant to within $0.21\times10^{-6}~\text{\AA}^{-1}$, and the small
reduction at small $\theta$ is consistent
with the higher-order mixed terms discussed below. The helical displacement is thus
controlled by the in-plane component $D_b\sin\theta$
[Fig.~\ref{fig:S4_helical}(b)].

The out-of-plane limit is reciprocal along $\hat{\mathbf y}$ by symmetry.
Besides $\sigma_h$, the $1H$ structure with
$\mathbf a_1\parallel\hat{\mathbf x}$ has the vertical mirror
\begin{equation}
  M_x:\quad(x,y,z)\rightarrow(-x,y,z),
\end{equation}
which maps each Se sublayer onto itself. An axial vector transforms as
$(D_x,D_y,D_z)\rightarrow(D_x,-D_y,-D_z)$ under $M_x$. Combining this with
time reversal gives
\begin{align}
  M_x\mathcal T:\quad
  &(D_{\ell,x},D_{\ell,y},D_{\ell,z})\rightarrow
  (-D_{\ell,x},D_{\ell,y},D_{\ell,z}),\nonumber\\
  &(Q_x,Q_y)\rightarrow(Q_x,-Q_y) .
  \label{eq:S4_MxT}
\end{align}
Any texture with $D_{t,x}=D_{b,x}=0$ is therefore invariant under
$M_x\mathcal T$ and satisfies $\Omega_{\rm SC}(Q_y)=\Omega_{\rm SC}(-Q_y)$ along
$\hat{\mathbf y}$, for arbitrary $y$ and $z$ exchange components on the two
layers. Only the $x$ components can generate nonreciprocity along
$\hat{\mathbf y}$. Together with the $\sigma_h\mathcal T$ rule of Sec.~S1.4,
the only term linear in both $Q_y$ and the exchange field is $Q_y\Delta D_x$,
and Eq.~\eqref{eq:S4_collapse} is the corresponding leading response. Mixed
terms such as $Q_y\Delta D_xD_z^2$ are allowed at higher order and can
account for deviations from the leading linear response. The computed ratio
$Q_{\rm eq}/\Delta D_x$ decreases in magnitude by up to
$0.43\times10^{-6}~\text{\AA}^{-1}\,{\rm meV}^{-1}$ in the tilted states,
growing with $D_z$: $0.06$ at $D_z=0.71$~meV, $0.17$ at $0.87$~meV, and
$0.32$--$0.43$ for $D_z\geq1.06$~meV, in these units.
On the production quadrature, the out-of-plane state gives
$|Q_{\rm eq}|<0.01\times10^{-6}~\text{\AA}^{-1}$ and $\eta=-0.003\%$ along
$\hat{\mathbf y}$, which sets the numerical floor for this symmetry class
(Sec.~S3.4). Equation~\eqref{eq:S4_MxT} leaves the $\hat{\mathbf x}$
direction unconstrained; the out-of-plane state is nonreciprocal along
$\hat{\mathbf x}$ (Sec.~S4.6).

The efficiency does not follow $\sin\theta$ [Fig.~\ref{fig:S4_helical}(b)].
For $\theta=60^\circ$, $45^\circ$, and $30^\circ$, $\eta$ is $0.78$, $0.51$,
and $0.31$ times the in-plane value at the same $\Delta D_x$,
$\eta_\parallel(\Delta D_x)=-0.955\%\times\Delta D_x/{\rm meV}$ (Sec.~S4.2).
At $\theta=45^\circ$, the ratio is $0.94$ for $D_b=0.5$~meV and drops to
$0.03$ for $D_b=1.5$~meV. The latter state, computed with $L=3$ because its
efficiency is small, has $Q_{\rm eq}=-18.55\times10^{-6}~\text{\AA}^{-1}$,
nearly the same as the one-sided in-plane state with $D_b=1$~meV
($-17.86\times10^{-6}~\text{\AA}^{-1}$), but an efficiency $27$ times
smaller ($-0.034\%$ against $-0.927\%$, both at $L=3$). With $L=2$ the
values are $-0.061\%$ and $-0.952\%$, a factor of $16$. The suppression
thus exceeds an order of magnitude at both refinement levels, but its
precise factor depends on the convergence of this small residual efficiency,
which changes from $-0.061\%$ to $-0.034\%$ between $L=2$ and $L=3$ and
was not computed at $L=4$. For this state the spurious fraction is $0.014$, and the efficiency from $d(\Omega_{\rm SC}-\Omega_{\rm N})/dQ$
is $-0.035\%$. The suppression is controlled by
$D_z$ [Fig.~1(d) of the main text]. The out-of-plane exchange is not
protected by Ising spin--valley locking and is pair breaking. It reduces the
mean critical current $\bar K_c=(K_c^++|K_c^-|)/2$ to $0.94$, $0.87$, $0.74$,
$0.62$, and $0.46$ of the in-plane value for $D_z=0.35$, $0.5$,
$0.71$, $0.87$, and $1.06$~meV. The current asymmetry $K_c^+-|K_c^-|$ is
reduced even more strongly, to $0.89$, $0.68$, $0.38$, $0.19$, and
$0.015$ ($0.027$ with $L=2$) of its in-plane value at the same
$\Delta D_x$; for the last state $\Delta K=-0.030$~A/m ($-0.053$~A/m with
$L=2$), against $-1.96$~A/m from the in-plane slope of Sec.~S4.2. The
suppression follows the closing of the spectral gap at the equilibrium
momentum: $E_{\rm gap}(Q_{\rm eq})=0.163$, $0.131$, $0.088$, $0.055$, and
$0.007$~meV for the same states, against $0.209$~meV for the in-plane state
with $D_b=1$~meV and $0.2365$~meV for the AP$_z$ texture, whose exchange
has no measurable effect on the condensate (Sec.~S4.6). The values at $D_z=0.5$, $0.71$, and
$0.87$~meV are computed directly at $\mathbf Q=Q_{\rm eq}\hat{\mathbf y}$
(Table~\ref{tab:S7_z}, block v); the others, here and below, are obtained by
linear interpolation of $E_{\rm gap}(Q)$ on the $Q$ grid, which
underestimates the cusp maximum at $Q_{\rm eq}$ by up to a few $\mu$eV.

At larger $D_z$ the efficiency changes sign. The tilted states with
$\theta=45^\circ$, $D_b=1.75$~meV ($D_z=1.24$~meV) and $\theta=30^\circ$,
$D_b=1.5$~meV ($D_z=1.30$~meV), computed with $L=3$, give $\eta=+0.354\%$
and $+0.330\%$ ($+0.309\%$ and $+0.289\%$ with $L=2$), i.e., $-0.30$ and
$-0.46$ times $\eta_\parallel(\Delta D_x)$; $d(\Omega_{\rm SC}-\Omega_{\rm N})/dQ$
gives $+0.360\%$ and $+0.337\%$. Their current asymmetry,
$K_c^+-|K_c^-|=+0.20$ and $+0.15$~A/m, is $-0.087$ and $-0.110$ ($-0.076$
and $-0.096$ with $L=2$) times its in-plane value at the same $\Delta D_x$.
Linear interpolation of this ratio between the two states with $\theta=45^\circ$ and
$D_b=1.5$ and $1.75$~meV puts the zero at $D_z\approx1.09$~meV ($1.11$~meV
with $L=2$). Both states are gapless along the whole branch, $\bar K_c$
drops to $0.29$ and $0.24$ of the in-plane value, and the selected solution
has the lowest grand potential in the scanned amplitude range at the tested
momenta (Sec.~S6.5); the temperature
dependence of the reversal is given in Sec.~S6.2. Their helical
momenta, $Q_{\rm eq}=-21.51$ and $-13.07\times10^{-6}~\text{\AA}^{-1}$,
nevertheless remain within $0.53\times10^{-6}~\text{\AA}^{-1}$ of
Eq.~\eqref{eq:S4_collapse}.

The suppression and the reversal are governed primarily by $D_z$. Two further states
repeat the $D_z$ of earlier ones with a different in-plane component
(Table~\ref{tab:S4_states}, block ii). The state with $\theta=60^\circ$ and
$D_b=1.414$~meV ($D_z=0.71$~meV, $\Delta D_x=1.22$~meV, $L=2$) gives
$\eta=-0.605\%$. Its ratios to the in-plane values, $\eta/\eta_\parallel=0.517$
and $\bar K_c/\bar K_{c\parallel}=0.738$, agree with $0.512$ and $0.742$ for
$\theta=45^\circ$, $D_b=1$~meV ($D_z=0.71$~meV, $\Delta D_x=0.71$~meV), and
so does the interpolated $E_{\rm gap}(Q_{\rm eq})$, $0.086$ against
$0.087$~meV. The state with $\theta=45^\circ$ and $D_b=1.84$~meV
($D_z=1.30$~meV, $\Delta D_x=1.30$~meV, $L=3$) gives $\eta=+0.615\%$, with
ratios $-0.495$ and $0.234$ against $-0.461$ and $0.240$ for
$\theta=30^\circ$, $D_b=1.5$~meV ($\Delta D_x=0.75$~meV); its current
asymmetry is $-0.115$ times the in-plane value, against $-0.110$, and it is
gapless along the whole branch. With $\Delta D_x$ changed by a factor of
$1.7$ at fixed $D_z$, the normalized ratios thus differ by at most $0.04$,
supporting the approximate collapse versus $D_z$ shown in Fig.~1(d) of the
main text. The helical
momenta, $-21.73$ and $-22.67\times10^{-6}~\text{\AA}^{-1}$, lie $0.08$ and
$0.50\times10^{-6}~\text{\AA}^{-1}$ from Eq.~\eqref{eq:S4_collapse}.

\emph{Pocket-resolved current asymmetry.} The following decomposition
identifies the pocket contributions to a mirrored-momentum diagnostic of the
current asymmetry for the in-plane state with $D_b=1$~meV and the tilted
states with $\theta=45^\circ$, $D_b=1.5$ and $1.75$~meV, and
$\theta=30^\circ$, $D_b=1.5$~meV, all at $L=3$. Because $\partial\Omega_{\rm SC}/\partial\Delta=0$
at self-consistency, the sheet current can be evaluated at fixed
$\Delta^\star(Q)$ and split term by term. The current decomposed here is
taken from $d(\Omega_{\rm SC}-\Omega_{\rm N})/dQ$, which removes the
normal-state quadrature error $K_{\rm N}$ (Sec.~S3.1); for these states it
changes $K_c^+-|K_c^-|$ from that of Eq.~\eqref{eq:S2_current_conversion} by
at most $0.01$~A/m. For each signed eigenvalue $E_n$
in the upper half of the BdG spectrum, the free-energy term
$-\beta^{-1}\ln[2\cosh(\beta E_n/2)]$, with $\beta=1/k_BT$, splits as
\begin{equation}
  -\frac{E_n}{2}-\beta^{-1}\ln\left(1+e^{-\beta E_n}\right).
  \label{eq:S4_split}
\end{equation}
The second term gives the quasiparticle back-flow $K^{\rm qp}$, carried by
thermally excited quasiparticles and by the occupied states of Bogoliubov
pockets ($E_n<0$). The remainder, including the pairing term and the
normal-state reference, is the condensate part $K^{\rm c}=K-K^{\rm qp}$.
The current density is also resolved on the base $\bk$ mesh, with the
refined cells folded onto their base cell, and summed over the regions
nearest to $\Gamma$ and to $K$ or $K'$. Only the sum over $K$ and $K'$ is
meaningful, because the Nambu spinor at $\bk$ contains the electron at
$\bk+\bQ/2$ and the hole at $-\bk+\bQ/2$.

The critical momenta lie at slightly different distances from $Q_{\rm eq}$:
$+272.2$ and $-270.4\times10^{-6}~\text{\AA}^{-1}$ in-plane, and $+196.6$
and $-193.4\times10^{-6}~\text{\AA}^{-1}$ for $\theta=45^\circ$, $D_b=1.5$~meV. The total
current is flat there, but its two parts are not; each changes by about
$0.46$~A/m per $10^{-6}~\text{\AA}^{-1}$, with opposite signs. The parts
are therefore compared at the mirrored momenta $Q_{\rm eq}\pm fs_c$, where
$s_c$ is the mean critical distance and $\Delta^\star$ is interpolated from
the scan:
\begin{equation}
  A_X(f)=X(Q_{\rm eq}+fs_c)+X(Q_{\rm eq}-fs_c)-2X(Q_{\rm eq}).
  \label{eq:S4_mirror}
\end{equation}
Subtracting $2X(Q_{\rm eq})$ removes the equilibrium counterflow between the
pockets. At $Q_{\rm eq}$, where $K=0$, the $\Gamma$ pocket carries
$+1.02$~A/m in-plane ($+1.10$~A/m for $\theta=45^\circ$, $D_b=1.5$~meV), and the $K$ and $K'$ pockets
carry the opposite current. For the total current, $X=K$, we write
$A_{\rm tot}\equiv A_K$; at $f=1$ it equals $K_c^+-|K_c^-|$ up to the offset
of the critical momenta: $-1.79$~A/m against $\Delta K=-1.79$~A/m in-plane,
$-0.008$~A/m against $\Delta K=-0.031$~A/m for
$\theta=45^\circ$, $D_b=1.5$~meV, and $+0.205$ and $+0.152$~A/m against
$+0.204$ and $+0.155$~A/m in the two sign-reversed states. All differences
are small compared with the pocket contributions below.

\begin{figure*}[t]
  \centering
  \includegraphics[width=\textwidth]{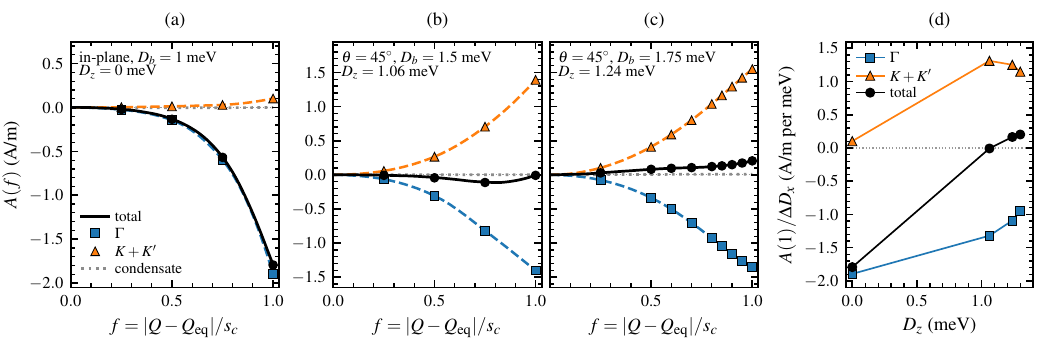}
  \caption{Pocket-resolved helical current asymmetry ($L=3$,
  $k_BT=5\times10^{-5}$~eV, scans along $\hat{\mathbf y}$). (a)--(c)
  Mirrored asymmetry $A_X(f)$ [Eq.~\eqref{eq:S4_mirror}] for (a) the in-plane
  state with $D_b=1$~meV and the tilted states with $\theta=45^\circ$ and (b)
  $D_b=1.5$~meV and (c) $D_b=1.75$~meV: total current (solid line, from the
  scan; circles, from the decomposition), quasiparticle back-flow in the
  $\Gamma$ (squares) and $K+K'$ (triangles) regions, and condensate part
  (dotted). Dashed lines are monotone interpolations in $f^2$, a guide to
  the eye. (d) The pocket contributions and the total at $f=1$ per meV of
  $\Delta D_x$ versus $D_z$ for the four states of
  Table~\ref{tab:S4_pockets}.}
  \label{fig:S4_pockets}
\end{figure*}

Table~\ref{tab:S4_pockets} lists the decomposition at $f=1$, and
Fig.~\ref{fig:S4_pockets} shows its dependence on $f$. In all four states,
the condensate contribution to the mirrored asymmetry is numerically small:
$|A_{\rm c}|<4\times10^{-3}$~A/m at every sampled $f$ in the first two states
of Table~\ref{tab:S4_pockets} and $\leq0.01$~A/m in the other two, so that
$K^{\rm c}(Q)-K^{\rm c}(Q_{\rm eq})$ is nearly odd about $Q_{\rm eq}$. In
the first two states it has the same slope, $0.46$~A/m per
$10^{-6}~\text{\AA}^{-1}$. The asymmetry is therefore carried by
quasiparticle back-flow, and it builds up
close to the critical momenta: in-plane, $A_{\rm tot}=-0.02$, $-0.13$, $-0.57$, and
$-1.79$~A/m for $f=0.25$, $0.5$, $0.75$, and $1$. The pocket asymmetries
quoted below include both condensate and quasiparticle contributions; their
back-flow parts agree with them to within $0.01$~A/m. In the in-plane state
the asymmetry comes from the $\Gamma$ pocket ($-1.90$~A/m), with a small contribution of
opposite sign from the $K$ and $K'$ pockets ($+0.10$~A/m). Effective models
restricted to the $K$ and $K'$ pockets~\cite{BankierPRB25,LevchenkoKhodasArXiv26}
therefore do not capture the dominant contribution in this setting. In the tilted
state with $\theta=45^\circ$ and $D_b=1.5$~meV, the back-flow of the
$\Gamma$ pocket changes little, and the
asymmetry of this pocket decreases in magnitude to $-1.40$~A/m. The back-flow in the $K$ and $K'$
pockets, in contrast, doubles: the out-of-plane exchange depairs these
pockets. The total asymmetry of these pockets remains opposite in sign
to that of the $\Gamma$ pocket and grows to $+1.39$~A/m, which cancels the $\Gamma$
contribution to within $0.01$~A/m. The same back-flow lowers $\bar K_c$:
with an unchanged condensate slope, the larger back-flow makes the current
peak at a smaller $|Q-Q_{\rm eq}|$. The suppression of $\eta$ in this state
is therefore a compensation between the two pocket families, not a removal
of the helical asymmetry.

At larger $D_z$ the balance tips [Fig.~\ref{fig:S4_pockets}(d)]. The
back-flow in the $K$ and $K'$ pockets no longer grows: it is $30$--$33$~A/m
on each side in all three tilted states. Per meV of $\Delta D_x$, the
asymmetry of these pockets decreases only slightly, from $+1.31$~A/m at $D_z=1.06$~meV to
$+1.25$ and $+1.15$~A/m at $D_z=1.24$ and $1.30$~meV. The asymmetry of the
$\Gamma$ pocket per meV of $\Delta D_x$ keeps decreasing in magnitude, from $-1.90$~A/m
in-plane to $-1.32$, $-1.09$, and $-0.95$~A/m. The $K$ and $K'$ pockets
therefore dominate, and $A_{\rm tot}=+0.21$ and $+0.15$~A/m for $D_z=1.24$ and
$1.30$~meV, the sign of the reversed efficiency.

\begin{table*}[t]
\caption{Pocket-resolved helical current asymmetry at the mirrored
momenta $Q_{\rm eq}\pm s_c$ [$f=1$ in Eq.~\eqref{eq:S4_mirror}; $L=3$,
$k_BT=5\times10^{-5}$~eV, scans along $\hat{\mathbf y}$]. $s_c$ in
$10^{-6}~\text{\AA}^{-1}$, $E_{\rm gap}(Q_{\rm eq})$ in meV, and currents in
A/m. $A_\Gamma$ and $A_{K+K'}$ are the total contributions of the two pocket
families, so that $A_{\rm tot}=A_{\rm c}+A_{\rm qp}=A_\Gamma+A_{K+K'}$ up to
rounding; their back-flow parts differ from them by at most $0.01$~A/m. The
last two columns give the back-flow of each pocket family at
$Q_{\rm eq}+s_c$ and $Q_{\rm eq}-s_c$.}
\label{tab:S4_pockets}
\begin{ruledtabular}
\begin{tabular}{lcccccccc}
State & $s_c$ & $E_{\rm gap}(Q_{\rm eq})$ & $A_{\rm tot}$ & $A_{\rm c}$ & $A_\Gamma$ & $A_{K+K'}$ & $K^{\rm qp}_\Gamma$ & $K^{\rm qp}_{K+K'}$\\
\hline
In-plane, $D_b=1$~meV & 271.3 & 0.209 & $-1.79$ & 0.00 & $-1.90$ & $+0.10$ & $-13.5$, $+11.6$ & $-15.2$, $+15.3$\\
$\theta=45^\circ$, $D_b=1.5$~meV & 195.0 & 0.007 & $-0.008$ & 0.00 & $-1.40$ & $+1.39$ & $-14.1$, $+12.6$ & $-31.7$, $+32.8$\\
$\theta=45^\circ$, $D_b=1.75$~meV & 156.9 & 0 & $+0.205$ & 0.01 & $-1.34$ & $+1.55$ & $-12.6$, $+10.9$ & $-31.4$, $+32.7$\\
$\theta=30^\circ$, $D_b=1.5$~meV & 139.9 & 0 & $+0.152$ & 0.01 & $-0.71$ & $+0.86$ & $-11.1$, $+10.1$ & $-30.2$, $+31.0$\\
\end{tabular}
\end{ruledtabular}
\end{table*}

\subsection*{S4.2. Exchange strength and the layer-even exchange}

Table~\ref{tab:S4_states}, blocks ii and iii, contains two pairs of in-plane
states with the same $\Delta D_x$ but different layer-even exchange $D_+$:
one-sided $D_b=0.5$~meV against $(D_t,D_b)=(0.5,1)$~meV
($\Delta D_x=0.5$~meV), and one-sided $D_b=1.5$~meV against
$(D_t,D_b)=(-0.5,1)$~meV ($\Delta D_x=1.5$~meV). Within each pair,
$Q_{\rm eq}$ agrees to $0.01\times10^{-6}~\text{\AA}^{-1}$, $K_c^\pm$ to
$0.4$~A/m, and $\eta$ to $0.005$ percentage points
($-0.483\%$ against $-0.488\%$, and $-1.442\%$ against $-1.441\%$). The
current curves of the first pair cannot be distinguished on the scale of
Fig.~\ref{fig:S4_helical}(c). For the tested in-plane configurations,
$Q_{\rm eq}$ and $\eta$ therefore depend primarily on $\Delta D_x=2D_-$;
changing the layer-even exchange $D_+$ between $0.25$ and $0.75$~meV
produces only the small differences quoted above.

Both observables are also linear in $\Delta D_x$. Over the six in-plane
states, $0.5\leq\Delta D_x\leq2$~meV, $Q_{\rm eq}/\Delta D_x$ lies between
$-17.86$ and $-17.85\times10^{-6}~\text{\AA}^{-1}\,{\rm meV}^{-1}$, and
$\eta/\Delta D_x$ between $-0.95$ and $-0.98\%$ per meV; a fit through the
origin gives $\eta_\parallel=-0.955\%\times\Delta D_x/{\rm meV}$. The current
asymmetry $K_c^+-|K_c^-|=-1.85\pm0.03$~A/m per meV of $\Delta D_x$ is equally
linear, while $\bar K_c=96.4\pm0.4$~A/m does not depend on the exchange.
In-plane exchange is weakly pair breaking because of Ising spin--valley
locking, so that the critical-current scale is that of the no-exchange
superconductor; the AP$_z$ texture, whose exchange has no measurable effect
on the condensate (Sec.~S4.6), gives $\bar K_c=97.1$~A/m along
$\hat{\mathbf x}$. The spectral gap at $Q_{\rm eq}$ falls only from
$0.2365$~meV (AP$_z$) to $0.209$ and $0.156$~meV for one-sided in-plane
exchange $D_b=1$ and $1.5$~meV.

Out-of-plane exchange, in contrast, is strongly pair breaking. At
$\theta=0^\circ$, $D_b=1$~meV reduces $\bar K_c$ to $49.0$~A/m ($L=3$) and
reduces the spectral gap to $27~\mu$eV at $Q=0$. At $D_b=1.5$~meV, the state
is gapless at all $Q$, and its condensation energy is $0.04$ times that at
$D_b=1$~meV. The normal state is then consistent with a local minimum of
$\Omega_{\rm SC}(\Delta)$ ($G<0$ at the smallest sampled amplitudes), but
the superconducting solution lies below it
at every $Q$ of the branch and has the lowest grand potential in the
scanned amplitude range at the tested momenta (Sec.~S6.5). The one-sided scans are therefore restricted to
$D_z\leq1.5$~meV and, except for the tilted states with $D_b=1.75$ and
$1.84$~meV of Sec.~S4.1, to $D_b\leq1.5$~meV, i.e., $\Delta D_x\leq1.5$~meV; the two-sided
textures extend the in-plane contrast to $2$~meV.

\subsection*{S4.3. Parallel, antiparallel, and single-layer textures}

The P state is reciprocal by $\sigma_h\mathcal T$ (Sec.~S1). On the uniform
mesh, the independently computed P state satisfied this to within the
no-exchange floor (Sec.~S1.3). It was not recomputed on the production
quadrature. The AP state has the largest contrast, $\Delta D_x=2$~meV, and
gives $Q_{\rm eq}=-35.71\times10^{-6}~\text{\AA}^{-1}$ and
$\eta=-1.895\%$, twice the one-sided values ($-17.86$ and $-0.952\%$).
At a common spacing $dq=4\times10^{-6}$, the ratio
$\eta_{\rm AP}/\eta_{\rm one\text{-}sided}$ is $2.06$, $1.95$, and $2.01$ at
$\delta\mu=-5$, $0$, and $+5$~meV (Sec.~S5).

The horizontal mirror maps a top-layer exchange $D_{t,x}$ onto a
bottom-layer exchange $-D_{t,x}$ [Eq.~\eqref{eq:S1_axial_mirror}]. Top-only
exchange along $+\hat{\mathbf x}$ is therefore symmetry-equivalent to
bottom-only exchange along $-\hat{\mathbf x}$, and both have
$\Delta D_x=-1$~meV. Time reversal maps $(\mathbf D_t,\mathbf D_b,Q_y)$ to
$(-\mathbf D_t,-\mathbf D_b,-Q_y)$ and therefore requires
\begin{equation}
  K_y(Q_y;\mathbf D_t,\mathbf D_b)
  =
  -K_y(-Q_y;-\mathbf D_t,-\mathbf D_b).
  \label{eq:S4_reversal}
\end{equation}
The magnetization-reversed one-sided and AP configurations therefore have
$Q_{\rm eq}=+17.86$ and $+35.71\times10^{-6}~\text{\AA}^{-1}$ and
$\eta=+0.952\%$ and $+1.895\%$, respectively, with the critical currents transformed as $K_c^+\rightarrow-K_c^-$ and
$K_c^-\rightarrow-K_c^+$. On the uniform mesh, the corresponding
independent calculations reproduced Eq.~\eqref{eq:S4_reversal} to within
$0.01$~A/m between the critical momenta.

\subsection*{S4.4. Vector form of the helical response}

The layer-preserving subgroup of $D_{3h}$ contains $C_3$ and the vertical
mirrors. Under this subgroup, the only invariant that couples the polar
in-plane vector $\mathbf Q_\parallel$ to the in-plane axial exchange
contrast $\Delta\mathbf D_\parallel=\mathbf D_{b,\parallel}-\mathbf
D_{t,\parallel}$ at linear order is
$\mathbf Q_\parallel\cdot(\hat{\mathbf z}\times\Delta\mathbf D_\parallel)$.
The scalar $\mathbf Q_\parallel\cdot\Delta\mathbf D_\parallel$ is odd under
$M_x$. The $C_3$ symmetry also makes the quadratic superfluid stiffness
isotropic at zeroth order in the exchange. The equilibrium momentum is
therefore
\begin{equation}
  \mathbf Q_{\rm eq}\simeq\chi_Q\,\hat{\mathbf z}\times\Delta\mathbf D_\parallel ,
  \label{eq:S4_vector}
\end{equation}
which reduces to Eq.~\eqref{eq:S4_collapse} for
$\Delta\mathbf D_\parallel\parallel\hat{\mathbf x}$. The helical momentum is
perpendicular to the exchange contrast and rotates rigidly with it.

Table~\ref{tab:S4_states}, block iv, tests Eq.~\eqref{eq:S4_vector} with
scans along $\mathbf b_1$. For an isotropic quadratic term, the minimum along
a line through the origin with direction $\hat{\mathbf n}$ lies at
$\mathbf Q_{\rm eq}\cdot\hat{\mathbf n}$. Equation~\eqref{eq:S4_vector} then
predicts $-17.86\sin30^\circ=-8.93$ and $+17.86\cos30^\circ=+15.47$ (in
$10^{-6}~\text{\AA}^{-1}$) for $\phi=0^\circ$ and $90^\circ$. The computed
values are $-8.93$ and $+15.45$. The diode efficiency follows the same
projection law closely: $\eta=-0.493\%$ and $+0.835\%$,
against $-0.476\%$ and $+0.825\%$ from projecting the $\hat{\mathbf y}$ value
[Fig.~\ref{fig:S4_helical}(d)]. To a good approximation, the in-plane
efficiency along $\hat{\mathbf n}$ is thus proportional to
$\hat{\mathbf n}\cdot(\hat{\mathbf z}\times\Delta\mathbf D_\parallel)$. The ratios of computed to projected values ($1.036$ and $1.012$) are
partly a $Q$-grid effect: the $\mathbf b_1$ scans use $dq=4\times10^{-6}$,
and projecting the $\hat{\mathbf y}$ value at the same spacing
($-0.974\%$, Table~\ref{tab:S3_dq}) gives $-0.487\%$ and $+0.844\%$, and
ratios of $1.012$ and $0.990$. An anisotropy of the
depairing regions, allowed by $C_3$ at higher order in $Q$, is therefore at
most at this level.

\subsection*{S4.5. Summary of the helical channel}

For in-plane exchange, the helical momentum and the diode efficiency are
both linear responses to the single parameter
$\Delta\mathbf D_\parallel=2\mathbf D_{-,\parallel}$. Both are governed by
$\hat{\mathbf z}\times\Delta\mathbf D_\parallel$ (the momentum as a vector,
the efficiency through its projection on the scan direction), and both are
insensitive to the layer-even exchange within the numerical accuracy of the
tested configurations. The out-of-plane exchange separates them. It leaves
$\mathbf Q_{\rm eq}$ on the in-plane line to within
$0.53\times10^{-6}~\text{\AA}^{-1}$ but, through pair breaking, suppresses
$\eta$ by more than an order of magnitude (a factor of $27$ at $L=3$ and
$16$ at $L=2$ for $\theta=45^\circ$, $D_b=1.5$~meV) and, at low
temperature (Sec.~S6.2), reverses it for $D_z\gtrsim1.1$~meV: the depaired $K$ and $K'$ pockets carry a back-flow
asymmetry that compensates or, at larger $D_z$, exceeds the one carried by
the $\Gamma$ pocket (Sec.~S4.1). A finite helical momentum is therefore
not a quantitative proxy for the rectification. The symmetry relations of
the model ($\sigma_h$, $\mathcal T$, $\sigma_h\mathcal T$, and $M_x\mathcal T$)
are reproduced at the level of the numerical floor.

\subsection*{S4.6. Out-of-plane exchange: nonreciprocity along \texorpdfstring{$\hat{\mathbf x}$}{x}}

For purely out-of-plane exchange, $D_{\ell,x}=D_{\ell,y}=0$, the magnetic
point group of a one-sided texture is $3m'$. It contains $C_3$ and the three
vertical mirrors combined with time reversal, $\sigma_v\mathcal T$, but not
$\sigma_h$ or $C_2'$. The operation $M_x\mathcal T$ maps
$(Q_x,Q_y)\rightarrow(Q_x,-Q_y)$ [Eq.~\eqref{eq:S4_MxT}]. A linear term
$\mathbf Q\cdot\mathbf v$ in $\Omega_{\rm SC}$ is therefore allowed only for
$\mathbf v\parallel\hat{\mathbf x}$, and $C_3$ then forces $\mathbf v=0$.
The point $\mathbf Q=0$ is therefore stationary for any $D_z$. Symmetry alone
does not exclude finite-$\mathbf Q$ minima that break $C_3$. $C_3$ does,
however, make the quadratic term isotropic, so a positive curvature along
one direction makes $\mathbf Q=0$ a local minimum in the whole $\mathbf Q$
plane. The computed free energy is lowest at $\mathbf Q=0$ over the whole
superconducting branch in every purely out-of-plane scan
(Table~\ref{tab:S7_z}, blocks i--iv, and the out-of-plane rows of
Table~\ref{tab:S6_gap}). These form a collection of scans rather than a full
parameter grid: along $\varphi_Q=0^\circ$ for one- and two-sided textures
with $D_b=0.25$--$1.5$~meV (for $D_b=1$~meV also at $k_BT=10^{-5}$ and
$10^{-4}$~eV, and for $D_b=1$ and $1.5$~meV also with the larger gap of
Sec.~S6.3), and along
$\varphi_Q=10^\circ$, $20^\circ$, and $90^\circ$ for $D_b=1$~meV at
$k_BT=5\times10^{-5}$~eV. Since $\Omega_{\rm SC}(\mathbf Q)$ is invariant
under $C_3$ and under $M_x\mathcal T$, which maps $\varphi_Q$ to
$-\varphi_Q$, the half-lines $\varphi_Q$ and $\varphi_Q+180^\circ$ of these
four scans cover the inequivalent wedge $0^\circ\leq\varphi_Q\leq60^\circ$ in
$10^\circ$ steps for $D_b=1$~meV. We therefore take
\begin{equation}
  \mathbf Q_{\rm eq}=0 .
  \label{eq:S4_trig_Qeq}
\end{equation}
The lowest odd invariant is cubic,
\begin{equation}
  \Omega_{\rm SC}^{\rm odd}(\mathbf Q)
  \propto D_z\,(Q_x^3-3Q_xQ_y^2)
  = D_z\,|\mathbf Q|^3\cos3\varphi_Q ,
  \label{eq:S4_trig_cubic}
\end{equation}
where $\varphi_Q$ is measured from $\hat{\mathbf x}$, the normal of the
mirror $M_x$. All odd invariants up to order $|\mathbf Q|^7$ share the
$\cos3\varphi_Q$ dependence. The diode efficiency along $\hat{\mathbf n}$ is
then expected to vary as $\cos3\varphi_Q$. It vanishes exactly along the
mirror lines ($\varphi_Q=30^\circ,90^\circ,\ldots$), which includes the
$\hat{\mathbf y}$ scans of Sec.~S4.1, and $\eta(60^\circ)=-\eta(0^\circ)$ and
$\eta(45^\circ)=-\eta(15^\circ)$ by symmetry. The horizontal mirror keeps
$D_z$ and exchanges the layers. A two-sided out-of-plane texture with $D_{t,z}=D_{b,z}$ is therefore
$\sigma_h$-symmetric, whereas one with $D_{t,z}=-D_{b,z}$ preserves
$\sigma_h\mathcal T$ and is reciprocal. This channel is controlled by the layer-even
exchange $D_{+,z}$, opposite to the helical channel, which is controlled by
$D_{-,x}$.

\emph{Converged result.} For $D_b=1$~meV, $\theta=0^\circ$, scanned along
$\hat{\mathbf x}$ ($\varphi_Q=0^\circ$), the three-level refinement ($L=3$)
gives $Q_{\rm eq}=0.05\times10^{-6}~\text{\AA}^{-1}$, which is zero within
the $Q$ grid, together with $K_c^+=48.26$~A/m, $K_c^-=-49.68$~A/m, and
$\eta=-1.447\%$. The definition through $d(\Omega_{\rm SC}-\Omega_{\rm N})/dQ$
gives $-1.437\%$, and the spurious fraction is $0.01$. A four-level refinement ($L=4$, $g_{\rm SC}$
recalibrated on the same quadrature) gives $K_c^+=48.26$~A/m,
$K_c^-=-49.68$~A/m, and $\eta=-1.452\%$ ($-1.445\%$ from
$d(\Omega_{\rm SC}-\Omega_{\rm N})/dQ$); at common $Q$ points the currents of
the two refinements agree to a relative $10^{-4}$. The $L=3$ and $L=4$ values therefore agree
within $0.01$ percentage points at fixed $Q$ spacing. For this symmetry
class, the $L=2$ quadrature is not sufficient. It gives $\eta=-1.668\%$ from
$d\Omega_{\rm SC}/dQ$ and $-1.107\%$ from $d(\Omega_{\rm SC}-\Omega_{\rm N})/dQ$,
and its spurious fraction is $0.23$--$0.84$ for all
out-of-plane scans along $\hat{\mathbf x}$ with $D_z\leq1$~meV (not
tabulated; it shows up as the difference between $\eta$ and
$\eta_{\Delta\Omega}$ in Table~\ref{tab:S7_z}), compared with at most $0.07$
for the in-plane states.
For $D_b=1.5$~meV, whose critical momenta are three times smaller
(Table~\ref{tab:S6_stab}), the spurious fraction is only $0.01$. The normal-state quadrature error has the same $3m'$
symmetry as the physical effect, so the angular dependence alone cannot
separate the two. The $L=2$ angular ratios are therefore only a symmetry
check; the quantitative test of the angular dependence uses $L=3$ (below).

\emph{Why the efficiency exceeds the in-plane value.} The out-of-plane state
has a larger efficiency than the one-sided in-plane state of the same
strength ($-1.447\%$ against $-0.927\%$, both at $L=3$), although its helical
momentum vanishes. Two facts explain this. First, $Q_{\rm eq}$ is set by the
linear coefficient of $\Omega_{\rm SC}(Q)$, Eq.~(4) of the main text, whereas
$\eta$ is set by the odd part of $\Omega_{\rm SC}$ about its minimum, which
first enters at cubic order. $C_3$ forbids the former but allows the latter,
Eq.~\eqref{eq:S4_trig_cubic}. Second, writing $\eta=\Delta K/(2\bar K_c)$, the
two states have comparable current asymmetries, $\Delta K=-1.42$ and
$-1.80$~A/m, but very different current scales, $\bar K_c=48.97$ and
$96.85$~A/m. The ratio of efficiencies,
$1.56=(1.42/1.80)\times(96.85/48.97)$, comes from the current scale. In-plane
exchange is ineffective against Ising-locked pairs, whereas out-of-plane
exchange acts as a Zeeman field along the Ising axis. It reduces the spectral
gap at $\mathbf Q=0$ to $27~\mu$eV (from $0.24$~meV) and halves the
critical current.

Microscopically, the out-of-plane exchange adds to the Ising spin splitting
in one valley and subtracts from it in the other. The Fermi pockets at $K$
and $K'$ are trigonally warped with opposite orientations. In the
time-reversal-symmetric state their odd contributions to $\Omega_{\rm SC}$
cancel. The exchange-induced valley imbalance leaves a net term with the
$\cos3\varphi_Q$ form of Eq.~\eqref{eq:S4_trig_cubic}, but no net
displacement of the condensate, since each pocket is $C_3$ symmetric about
its valley. This is the mechanism of magnetochiral anisotropy in gated
MoS$_2$ and of the out-of-plane-field diode effect in few-layer NbSe$_2$
\cite{WakatsukiSciAdv17,HoshinoPRB18,BauriedlNatCommun22}, and a cubic term
of this form has been invoked for the field-free diode effect of
NbSe$_2$/CrBr$_3$ heterostructures \cite{SunArXiv26}. The same
exchange also acts on the helical (in-plane) diode effect of tilted textures
(Sec.~S4.1). Along $\hat{\mathbf y}$ it contributes no asymmetry on its own,
by $M_x\mathcal T$, but it depairs the $K$ and $K'$ pockets. Their
back-flow then carries an asymmetry of opposite sign to that of the
$\Gamma$ pocket and compensates or, for $D_z\gtrsim1.1$~meV and at low
temperature, exceeds it.

\emph{Layer parity.} The horizontal mirror extends the selection rule of
Sec.~S1.4 to arbitrary collinear textures. With
$D_{\pm,i}=(D_{b,i}\pm D_{t,i})/2$ for each Cartesian component $i$,
Eq.~\eqref{eq:S1_axial_mirror} and time reversal give
\begin{multline}
  \sigma_h\mathcal T:\quad
  (D_{+,\parallel},D_{-,\parallel},D_{+,z},D_{-,z})\\
  \rightarrow(D_{+,\parallel},-D_{-,\parallel},-D_{+,z},D_{-,z}),\quad
  \mathbf Q\rightarrow-\mathbf Q .
  \label{eq:S4_parity}
\end{multline}
$\Omega_{\rm SC}(\mathbf Q)$ can therefore be odd in $\mathbf Q$ only through
the layer-odd in-plane exchange $D_{-,\parallel}$, which drives the helical
channel (Secs.~S4.1--S4.5), and the layer-even out-of-plane exchange
$D_{+,z}$, which drives the trigonal channel. The components
$D_{+,\parallel}$ and $D_{-,z}$ cannot rectify on their own.

Table~\ref{tab:S7_z}, block ii, tests this for the out-of-plane channel. The
parallel textures $D_t=D_b=0.25$ and $0.5$~meV have the same $D_{+,z}$ as the
one-sided textures $D_b=0.5$ and $1$~meV, but no $D_{-,z}$. Symmetry alone
does not make the two identical: $\sigma_h$ allows $D_{-,z}$ to enter at
even order, so their agreement measures the size of these terms. They
reproduce the one-sided states: over the whole branch, the differences in
$\Delta^\star(Q_x)$ and $K_x(Q_x)$ are at most $2\times10^{-5}$ and
$3.2\times10^{-5}$ of the maximum of $\Delta^\star$ and of $|K_x|$ along the
branch, and $\eta$ agrees to $0.001$ percentage points. The antiparallel texture
$D_t=-D_b=-1$~meV has $D_{+,z}=0$ and $D_{-,z}=1$~meV. It is reciprocal,
$K_c^+=|K_c^-|=97.07$~A/m and $|\eta|<10^{-3}\%$, as
Eq.~\eqref{eq:S4_parity} requires. Its $\Delta^\star(0)=0.399$~meV and
$E_{\rm gap}(0)=0.2365$~meV agree with the no-exchange values calibrated at
$k_BT=10^{-5}$~eV ($0.400$ and $0.237$~meV) to within $1~\mu$eV, the
reduction expected from the higher temperature; the weakest in-plane state,
$D_b=0.5$~meV, has the same $\Delta^\star(0)$. A layer-odd out-of-plane
exchange of $1$~meV thus has no measurable effect on the condensate. The
normal-state free-energy differences between the parallel and one-sided
states, $2.7\times10^{-8}$ and $1.09\times10^{-7}$~eV per cell for
$D_{-,z}=0.25$ and $0.5$~meV, scale approximately as $D_{-,z}^2$. This
follows from $\sigma_h$. With spin--orbit coupling,
$\sigma_h$ (which leaves the in-plane $\bk$ unchanged) acts as
$z\rightarrow-z$ combined with $i s_z$, so the Bloch states at every $\bk$
are $\sigma_h$ eigenstates with eigenvalues $\pm i$. The exchange
$D_{-,z}s_z(n_b-n_t)$ is $\sigma_h$ odd and therefore has no diagonal matrix
elements in this basis, so it shifts nondegenerate energies only at second
order. Its off-diagonal elements connect opposite mirror eigenvalues. At the
Fermi level this requires the small admixture of $\sigma_h$-odd orbitals to
the Nb-$d$-dominated bands (Sec.~S6.3); otherwise it reaches only
bands far from $E_F$. A first-order effect is possible only where bands of
opposite mirror eigenvalue are degenerate, i.e., at the crossings of the
spin-split bands on the nodal lines of the Ising splitting ($\Gamma$--M),
and it is small for the same reason. The computed normal-state free-energy differences support this
second-order interpretation. The superconducting observables change only
slightly over the investigated range: $\Delta^\star$ differs by at most
$6\times10^{-9}$~eV along the branch, comparable to the self-consistency
tolerance (Sec.~S3.3), so its scaling is not resolved. Because the
parallel and one-sided states differ only at this level and are computed on
the same quadrature, the converged one-sided value at $L=3$ applies to the
parallel texture with the same $D_{+,z}$.

\emph{Pair breaking by the four components.} The components also differ in
their effect on the condensate. Relative to the AP$_z$ value,
$E_{\rm gap}(0)=0.2365$~meV, the layer-even out-of-plane exchange reduces
the spectral gap linearly, $E_{\rm gap}(0)\simeq(0.238-0.421|D_{+,z}|/{\rm
meV})$~meV for $|D_{+,z}|\leq0.5$~meV, and the state is gapless at
$D_{+,z}=0.75$~meV (Table~\ref{tab:S7_z}). It acts as a Zeeman field along
the Ising axis, and $\bar K_c$ falls from $97$ to $93.9$, $83.6$, $49.0$,
and $5.6$~A/m for $D_{+,z}=0.125$, $0.25$, $0.5$, and $0.75$~meV. Ising
spin--valley locking protects the condensate against in-plane exchange,
which reduces the gap much more weakly. For the in-plane states the relevant gap is that at the
equilibrium momentum, since $\mathbf Q=0$ is Doppler shifted away from the
helical minimum. Relative to the AP$_z$ value, $E_{\rm gap}(Q_{\rm eq})$ is
reduced by less than $0.01$, by $0.03$, and by $0.08$~meV for the one-sided
textures with $D_{+,\parallel}=D_{-,\parallel}=0.25$, $0.5$, and
$0.75$~meV, faster than linearly in the exchange. The reduction is
controlled by $D_{+,\parallel}$: the two-sided textures with
$(D_{+,\parallel},D_{-,\parallel})=(0.75,0.25)$ and $(0.25,0.75)$~meV give
$0.08$ and $0.01$~meV, and the AP texture, $(0,1)$~meV, gives $0.01$~meV.
These reductions are accurate to the few-$\mu$eV interpolation error stated
in Sec.~S4.1.
A cross term $D_{+,\parallel}D_{-,\parallel}$ is forbidden by $\sigma_h$, and
$\bar K_c=96.4\pm0.4$~A/m is unchanged throughout. The layer-odd
out-of-plane exchange has no measurable effect.
Only $D_{+,z}$ is pair breaking at linear order, which is why it suppresses
the helical channel of tilted textures (Sec.~S4.1) while the in-plane
components leave the trigonal channel intact (below).

\emph{Tilted textures along $\hat{\mathbf x}$.} For the one-sided tilted
textures, the scan along $\hat{\mathbf x}$ is shifted to pass through
$\mathbf Q_{\rm eq}=Q_{\rm eq}\hat{\mathbf y}$ (Table~\ref{tab:S7_z},
block v). $M_x$ keeps $D_x$ and $Q_y$ and reverses $D_z$ and $Q_x$, so that
$\eta_x$ is odd in $D_z$; $M_x\mathcal T$ reverses $D_x$ and $Q_y$ and keeps
$D_z$ and $Q_x$, so that $\eta_x$ is even in $D_x$. An in-plane component
alone gives no rectification along $\hat{\mathbf x}$. At $L=3$, the
$\theta=45^\circ$ texture ($D_x=D_z=0.71$~meV) gives $\eta=-0.735\%$
($-0.738\%$ from $d(\Omega_{\rm SC}-\Omega_{\rm N})/dQ$). The purely
out-of-plane $L=3$ values at $D_z=0.5$ and $1$~meV, interpolated as a power
of $D_z$ (exponent $1.98$), give $-0.729\%$ at $D_z=0.71$~meV, a ratio of
$1.01$. At $L=2$, where the tilted and purely out-of-plane states carry
similar quadrature errors, the ratios are $1.01$, $1.00$, and $0.92$ for
$\theta=30^\circ$, $45^\circ$, and $60^\circ$. The mean critical currents along
$\hat{\mathbf x}$ likewise follow $D_z$. An in-plane exchange of up to
$0.87$~meV thus leaves the trigonal channel essentially unchanged, while an
out-of-plane component strongly suppresses the helical one. The mutual
influence is asymmetric because only $D_{+,z}$ is pair breaking at linear order.

\emph{Angular dependence.} At $L=3$, $\eta(10^\circ)=-1.280\%$ and
$\eta(20^\circ)=-0.771\%$ ($-1.270\%$ and $-0.766\%$ from
$d(\Omega_{\rm SC}-\Omega_{\rm N})/dQ$), so that
$\eta(\varphi_Q)/\eta(0^\circ)=0.884$ and $0.532$, against
$\cos3\varphi_Q=0.866$ and $0.500$. Both excesses follow from a small
$\cos9\varphi_Q$ harmonic, the next term allowed by $C_3$, $M_x\mathcal T$,
and the oddness in $\mathbf Q$. The form
$\eta(\varphi_Q)\propto\cos3\varphi_Q+a_9\cos9\varphi_Q$ with $a_9=-0.021$,
fixed by the ratio at $20^\circ$, predicts $0.885$ at $10^\circ$, within
$0.001$ of the computed value. Along
the mirror line ($\varphi_Q=90^\circ$) the efficiency vanishes to
$-0.003\%$ (Fig.~2(c) of the main text).

\emph{Exchange strength.} At $L=3$, $D_b=0.5$~meV gives $\eta=-0.367\%$
($-0.366\%$ from $d(\Omega_{\rm SC}-\Omega_{\rm N})/dQ$). With $-1.447\%$ at
$D_b=1$~meV and $-3.354\%$ at $1.5$~meV, $\eta/D_b^2=-1.47$, $-1.45$, and
$-1.49\%$/meV$^2$, with mean $-1.47\%$/meV$^2$. Over this range the
efficiency grows as the square of the exchange; it remains odd in $D_{+,z}$
by time reversal. The $D_b=1.5$~meV value
is taken at $L=2$, where the spurious fraction is $0.01$. Its critical
momenta are three times smaller than at $1$~meV. The leading
symmetry-allowed odd error in the normal-state free energy is cubic in $Q$,
so its contribution to the current is quadratic and is reduced at these
smaller momenta. The quadratic growth combines a current
asymmetry that peaks near $1$~meV ($\Delta K=-0.61$, $-1.42$, and
$-0.38$~A/m) with a
collapsing current scale ($\bar K_c=83.6$, $49.0$, and $5.6$~A/m) as $D_{+,z}$
closes the spectral gap. The efficiency of the trigonal channel exceeds that
of the helical channel, $-0.955\%\times\Delta D_x/{\rm meV}$, for exchange
above $0.65$~meV.
At smaller exchange the growth is steeper. $D_b=0.25$~meV gives
$\eta=-0.058\%$ ($-0.055\%$ from $d(\Omega_{\rm SC}-\Omega_{\rm N})/dQ$, $L=3$),
below the $-0.092\%$ of the quadratic law; between $0.25$ and $0.5$~meV,
$\eta\propto D_b^{2.7}$. No term linear in the exchange is resolved. The
current asymmetry also changes sign along the branch: at $D_b=0.25$~meV,
$K_x(Q)+K_x(-Q)$ is positive, up to $+0.45$~A/m, while the spectrum is gapped,
and turns negative once the gap closes at $|Q_x|\approx2.4\times10^{-4}~\text{\AA}^{-1}$,
before the critical momenta. The diode polarity is therefore set in the gapless regime, by the negative
mirrored asymmetry near the critical momenta.
As a function of the critical-current suppression, which a spin valve
measures directly through its P$_z$ and AP$_z$ states (Table~\ref{tab:S7_z},
block ii), the efficiency is $-0.06$, $-0.37$, $-1.45$, and $-3.4\%$ at
$\bar K_c/\bar K_c({\rm AP}_z)=0.97$, $0.86$, $0.50$, and $0.06$
($D_b=0.25$, $0.5$, $1$, and $1.5$~meV).

\emph{Temperature.} At $L=3$, $\eta=-3.372\%$, $-1.447\%$, and $-0.323\%$ for
$k_BT=10^{-5}$, $5\times10^{-5}$, and $10^{-4}$~eV ($0.12$, $0.58$, and
$1.16$~K), a tenfold change, against $2.4$ for the helical channel
(Sec.~S6.2). The current asymmetry changes by a factor of $35$
($\Delta K=-4.38$, $-1.42$, and $-0.125$~A/m) and the current scale by $3.4$
($\bar K_c=65.0$, $49.0$, and $19.4$~A/m). The spectral gap at $\mathbf Q=0$ is
$36$ and $27~\mu$eV at the two lower temperatures, and the state is gapless
at $10^{-4}$~eV. At $k_BT=10^{-5}$~eV, $K_x(Q_x)$ develops a cusp on the
$+Q_x$ side at $Q_x\approx1.9\times10^{-4}~\text{\AA}^{-1}$, where $\Delta^\star$ starts to fall
linearly, while the $-Q_x$ side peaks smoothly at a larger $|K|$. The
asymmetry thus originates in the nearly gapless, exchange-split
quasiparticles, whose structure on the scale of tens of $\mu$eV is washed out
by thermal smearing.

\begin{table*}[t]
\caption{Out-of-plane, two-sided out-of-plane, and tilted textures
(production quadrature, $\mathbf Q\parallel\hat{\mathbf x}$ unless noted,
$k_BT=5\times10^{-5}$~eV unless noted). $D_t$, $D_b$, and
$D_{+,z}=(D_{b,z}+D_{t,z})/2$ are $z$ components in meV. $\eta_{\Delta\Omega}$
is the efficiency from $d(\Omega_{\rm SC}-\Omega_{\rm N})/dQ$; a large
difference from $\eta$ signals an unconverged quadrature (Sec.~S4.6).
$\Delta^\star(0)$ and $E_{\rm gap}(0)$ are in meV; for block v,
$E_{\rm gap}$ is taken at the origin of the scan line,
$\mathbf Q=Q_{\rm eq}\hat{\mathbf y}$. Other units as in
Table~\ref{tab:S4_states}.}
\label{tab:S7_z}
\begin{ruledtabular}
\begin{tabular}{lccccccccccc}
State & $D_t$ & $D_b$ & $D_{+,z}$ & $L$ & $Q_{\rm eq}$ & $K_c^+$ & $K_c^-$ &
$\eta$ (\%) & $\eta_{\Delta\Omega}$ (\%) & $\Delta^\star(0)$ & $E_{\rm gap}(0)$\\
\hline
\multicolumn{12}{l}{\emph{(i) One-sided, $\theta=0^\circ$}}\\
\quad Bottom only & 0 & 0.25 & 0.125 & 3 & $-$0.01 & 93.82 & $-$93.93 & $-$0.058 & $-$0.055 & 0.398 & 0.186\\
\quad Bottom only & 0 & 0.5 & 0.25 & 2 & $-$0.02 & 83.01 & $-$83.73 & $-$0.435 & $-$0.098 & 0.397 & 0.135\\
 & 0 & 0.5 & 0.25 & 3 & $-$0.01 & 83.34 & $-$83.95 & $-$0.367 & $-$0.366 & 0.397 & 0.135\\
\quad Bottom only & 0 & 1 & 0.5 & 2 & 0.06 & 47.97 & $-$49.60 & $-$1.668 & $-$1.107 & 0.385 & 0.027\\
 & 0 & 1 & 0.5 & 3 & 0.05 & 48.26 & $-$49.68 & $-$1.447 & $-$1.437 & 0.385 & 0.027\\
 & 0 & 1 & 0.5 & 4 & 0.05 & 48.26 & $-$49.68 & $-$1.452 & $-$1.445 & 0.385 & 0.027\\
\quad Bottom only & 0 & 1.5 & 0.75 & 2 & 0.34 & 5.40 & $-$5.78 & $-$3.354 & $-$3.425 & 0.312 & 0.000\\
\hline
\multicolumn{12}{l}{\emph{(ii) Two-sided, $\theta=0^\circ$}}\\
\quad P$_z$ & 0.25 & 0.25 & 0.25 & 2 & $-$0.01 & 83.01 & $-$83.73 & $-$0.435 & $-$0.097 & 0.397 & 0.135\\
\quad P$_z$ & 0.5 & 0.5 & 0.5 & 2 & 0.06 & 47.97 & $-$49.60 & $-$1.667 & $-$1.107 & 0.385 & 0.027\\
\quad AP$_z$ & $-$1 & 1 & 0 & 2 & 0.00 & 97.07 & $-$97.07 & 0.000 & 0.000 & 0.399 & 0.236\\
\hline
\multicolumn{12}{l}{\emph{(iii) Scan direction, one-sided $D_b=1$~meV}}\\
\quad $\varphi_Q=10^\circ$ & 0 & 1 & 0.5 & 2 & 0.05 & 48.17 & $-$49.61 & $-$1.471 & $-$0.994 & 0.385 & 0.027\\
 & 0 & 1 & 0.5 & 3 & 0.05 & 48.44 & $-$49.70 & $-$1.280 & $-$1.270 & 0.385 & 0.027\\
\quad $\varphi_Q=20^\circ$ & 0 & 1 & 0.5 & 2 & 0.05 & 48.67 & $-$49.53 & $-$0.883 & $-$0.614 & 0.385 & 0.027\\
 & 0 & 1 & 0.5 & 3 & 0.05 & 48.92 & $-$49.67 & $-$0.771 & $-$0.766 & 0.385 & 0.027\\
\quad $\varphi_Q=90^\circ$ ($\hat{\mathbf y}$) & 0 & 1 & 0.5 & 2 & 0.00 & 49.35 & $-$49.35 & $-$0.003 & $-$0.004 & 0.385 & 0.027\\
\hline
\multicolumn{12}{l}{\emph{(iv) Temperature, one-sided $D_b=1$~meV}}\\
\quad $k_BT=10^{-5}$~eV & 0 & 1 & 0.5 & 2 & $-$0.05 & 62.76 & $-$67.52 & $-$3.652 & $-$2.794 & 0.400 & 0.035\\
 & 0 & 1 & 0.5 & 3 & $-$0.05 & 62.79 & $-$67.18 & $-$3.372 & $-$3.323 & 0.400 & 0.036\\
\quad $k_BT=10^{-4}$~eV & 0 & 1 & 0.5 & 2 & 0.06 & 19.12 & $-$19.32 & $-$0.529 & $-$0.227 & 0.304 & 0.000\\
 & 0 & 1 & 0.5 & 3 & 0.06 & 19.30 & $-$19.43 & $-$0.323 & $-$0.309 & 0.305 & 0.000\\
\hline
\multicolumn{12}{l}{\emph{(v) One-sided tilted, $|\mathbf D_b|=1$~meV ($z$ components listed), scan line through $Q_{\rm eq}\hat{\mathbf y}$}}\\
\quad $\theta=30^\circ$ & 0 & 0.87 & 0.43 & 2 & 0.01 & 58.63 & $-$60.15 & $-$1.277 & $-$0.776 & 0.390 & 0.055\\
\quad $\theta=45^\circ$ & 0 & 0.71 & 0.35 & 2 & $-$0.02 & 70.16 & $-$71.37 & $-$0.853 & $-$0.416 & 0.394 & 0.088\\
 & 0 & 0.71 & 0.35 & 3 & $-$0.02 & 70.45 & $-$71.50 & $-$0.735 & $-$0.738 & 0.394 & 0.088\\
\quad $\theta=60^\circ$ & 0 & 0.50 & 0.25 & 2 & $-$0.02 & 82.77 & $-$83.43 & $-$0.402 & $-$0.065 & 0.397 & 0.131\\
\end{tabular}
\end{ruledtabular}
\end{table*}


\section*{S5. Carrier dependence}

This section documents the chemical-potential dependence of the in-plane
states. All calculations use the production quadrature and
$k_BT=5\times10^{-5}$~eV. For each texture, the exchange field and
$g_{\rm SC}$ are held fixed as $\mu$ is varied.
Here $\delta\mu=\mu-\mu_0$, and the Brillouin-zone refinement is
rebuilt for each $\mu$ without recalibrating the pairing interaction.

\begin{figure*}[t]
  \centering
  \includegraphics[width=0.92\textwidth]{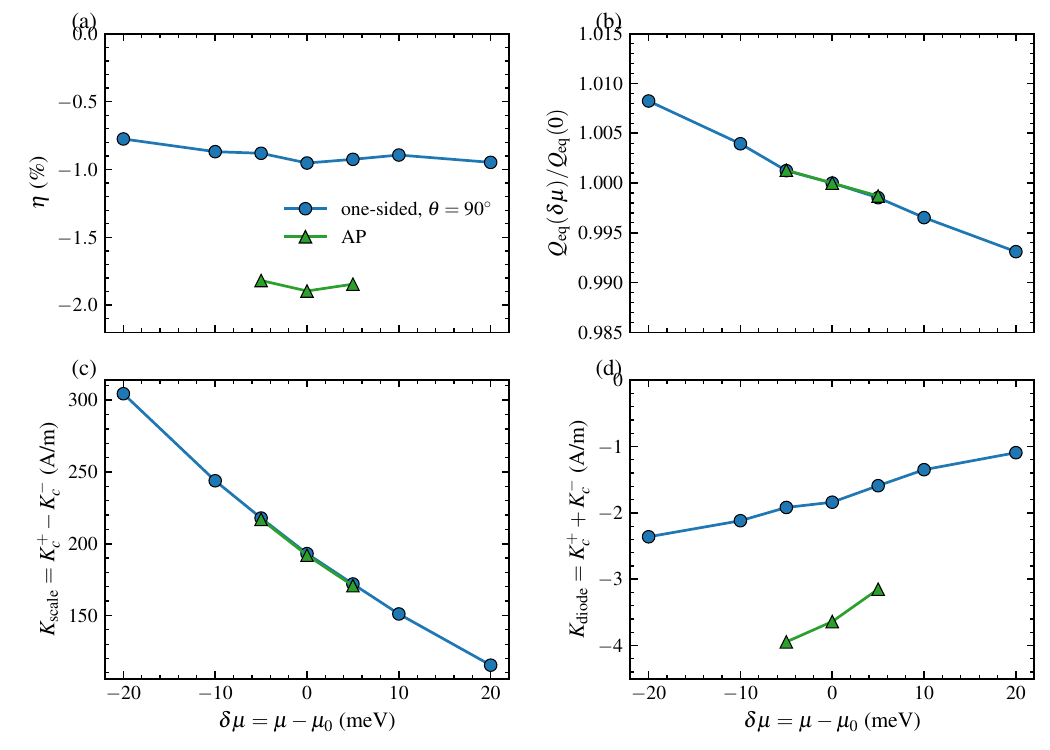}
  \caption{Carrier dependence of the finite-$Q$ observables
  (production quadrature, $k_BT=5\times10^{-5}$~eV) for the one-sided
  in-plane state ($D_b=1$~meV, $\theta=90^\circ$, $\phi=0^\circ$) and the
  in-plane AP texture ($D_{b,x}=1$~meV, $D_{t,x}=-1$~meV).
  (a) Diode efficiency $\eta$.
  (b) Equilibrium momentum normalized to its value at $\delta\mu=0$.
  (c) Critical-current scale $K_{\rm scale}=K_c^+-K_c^-=2\bar K_c$.
  (d) Diode numerator $K_{\rm diode}=K_c^++K_c^-$.}
  \label{fig:S5_carrier}
\end{figure*}

\subsection*{S5.1. Chemical-potential dependence}

Table~\ref{tab:S5_mu} and Fig.~\ref{fig:S5_carrier} show the one-sided
in-plane state over $-20\leq\delta\mu\leq20$~meV and the AP texture at
$\delta\mu=-5$, $0$, and $+5$~meV. The observables respond differently
to filling. $Q_{\rm eq}$ varies weakly and approximately linearly,
from $-18.00$ to $-17.73\times10^{-6}~\text{\AA}^{-1}$, with a similar
relative dependence for the AP texture
[Fig.~\ref{fig:S5_carrier}(b)]. The current scale,
$K_{\rm scale}=2\bar K_c$, decreases by a factor of $2.6$, from
$305$ to $115$~A/m, while $\Delta^\star(0)$ decreases from $0.557$
to $0.290$~meV. At fixed $g_{\rm SC}$, this gap variation reflects
the filling dependence of the electronic states participating in pairing.
The diode numerator $K_{\rm diode}$ decreases in magnitude from
$2.36$ to $1.09$~A/m, approximately tracking the current scale.
Consequently, the efficiency remains between $-0.78\%$ and
$-0.95\%$ over the scan, with no sign change
[Fig.~\ref{fig:S5_carrier}(a)].

For the AP texture, $\eta=-1.82\%$, $-1.90\%$, and $-1.85\%$ at
$\delta\mu=-5$, $0$, and $+5$~meV, respectively. At the common
spacing $dq=4\times10^{-6}$, these are $2.06$, $1.95$, and $2.01$
times the one-sided efficiencies at the same fillings
(Table~\ref{tab:S3_dq} for $\delta\mu=0$ and $+5$~meV).
The approximate doubling expected from the leading dependence on
$\Delta D_x$ at $\mu_0$ (Sec.~S4.2) therefore persists at all three
fillings tested for the AP texture.

\begin{table*}[t]
\caption{Carrier scans (production quadrature,
$k_BT=5\times10^{-5}$~eV). $\delta\mu$ and $\Delta^\star(0)$ are in meV,
$Q_{\rm eq}$ in $10^{-6}~\text{\AA}^{-1}$, and currents in A/m.
$K_{\rm diode}=K_c^++K_c^-$ and $K_{\rm scale}=K_c^+-K_c^-$
[Eq.~\eqref{eq:S2_Jdiode_Jscale}]. $^\dagger$\,$dq=2\times10^{-6}$; all other
rows $dq=4\times10^{-6}$. Derived quantities are evaluated from unrounded
currents and rounded independently for display.}
\label{tab:S5_mu}
\begin{ruledtabular}
\begin{tabular}{rccccccc}
$\delta\mu$ & $\Delta^\star(0)$ & $Q_{\rm eq}$ & $K_c^+$ & $K_c^-$ &
$K_{\rm diode}$ & $K_{\rm scale}$ & $\eta$ (\%)\\
\hline
\multicolumn{8}{l}{\emph{One-sided, $\theta=90^\circ$, $D_b=1$~meV}}\\
$-20$ & 0.557 & $-18.00$ & 151.09 & $-153.45$ & $-2.36$ & 304.54 & $-0.775$\\
$-10$ & 0.470 & $-17.93$ & 120.87 & $-122.99$ & $-2.12$ & 243.86 & $-0.869$\\
$-5$ & 0.432 & $-17.88$ & 107.98 & $-109.90$ & $-1.92$ & 217.88 & $-0.881$\\
$0^\dagger$ & 0.398 & $-17.86$ & 95.64 & $-97.48$ & $-1.84$ & 193.12 & $-0.952$\\
$+5^\dagger$ & 0.367 & $-17.83$ & 85.18 & $-86.77$ & $-1.59$ & 171.95 & $-0.925$\\
$+10$ & 0.339 & $-17.80$ & 74.85 & $-76.20$ & $-1.35$ & 151.05 & $-0.893$\\
$+20$ & 0.290 & $-17.73$ & 57.13 & $-58.23$ & $-1.09$ & 115.36 & $-0.948$\\
\hline
\multicolumn{8}{l}{\emph{AP}}\\
$-5$ & 0.432 & $-35.76$ & 106.52 & $-110.46$ & $-3.94$ & 216.98 & $-1.818$\\
$0$ & 0.398 & $-35.71$ & 94.18 & $-97.82$ & $-3.64$ & 192.00 & $-1.895$\\
$+5$ & 0.367 & $-35.66$ & 83.80 & $-86.95$ & $-3.15$ & 170.75 & $-1.845$\\
\end{tabular}
\end{ruledtabular}
\end{table*}

\subsection*{S5.2. Filling dependence on uniform meshes}

On uniform meshes, $\eta$ appears to depend strongly on filling. At
$k_BT=5\times10^{-5}$~eV (Table~\ref{tab:S3_conv}), the one-sided state gives
$-0.68\%$ at $\mu_0$ but $-0.13\%$ at $\delta\mu=+5$~meV on the $N=261$
mesh, whereas the $N=333$ mesh gives the opposite trend, $-0.55\%$ and
$-1.38\%$. The convergence tests in Sec.~S3.1 identify these large,
mesh-dependent variations as Brillouin-zone sampling artifacts.
On the Fermi-surface-adaptive quadrature, the efficiencies at
$\mu_0$ and $\mu_0+5$~meV differ by only $0.03$ percentage points.
Across the full $40$~meV interval, $K_{\rm diode}$ and
$K_{\rm scale}$ vary smoothly and approximately in proportion,
so that $\eta$ remains between $-0.78\%$ and $-0.95\%$.
Within this model and over the sampled fillings, carrier filling
therefore changes the current scale by a factor of $2.6$ but the
rectification efficiency by less than $0.2$ percentage points.


\section*{S6. Superconducting stability, temperature dependence, and model robustness}

\subsection*{S6.1. Representative finite-\texorpdfstring{$Q$}{Q} superconducting branch}

\begin{figure*}[t]
  \centering
  \includegraphics[width=\textwidth]{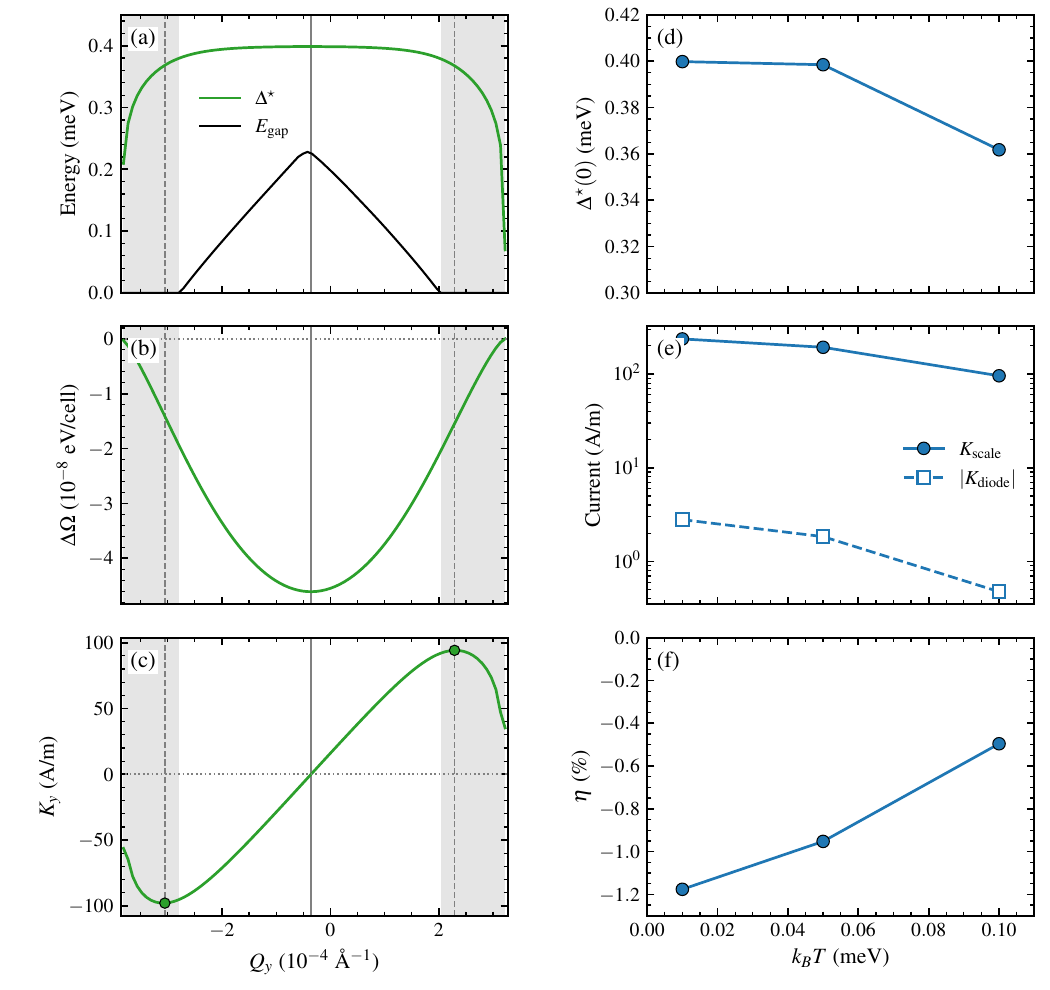}
  \caption{Superconducting branch and temperature dependence
  (production quadrature).
  (a)--(c) AP state at $\mu_0$, $k_BT=5\times10^{-5}$~eV:
  (a) $\Delta^\star(Q_y)$ and the quasiparticle spectral gap $E_{\rm gap}(Q_y)$,
  (b) condensation energy $\Delta\Omega(Q_y)$, and (c) sheet current
  $K_y(Q_y)$ with $K_c^\pm$ marked. The solid vertical line is $Q_{\rm eq}$,
  dashed lines are $Q_c^\pm$, and shaded regions are gapless
  ($E_{\rm gap}<5~\mu$eV).
  (d)--(f) Temperature dependence of the one-sided in-plane state at $\mu_0$:
  (d) $\Delta^\star(0)$, (e) $K_{\rm scale}$ and $|K_{\rm diode}|$ on a
  logarithmic scale, and (f) $\eta$.}
  \label{fig:S6}
\end{figure*}

Figure~\ref{fig:S6}(a)--(c) shows the full self-consistent branch of the AP
state at $\mu_0$ on the production quadrature. The pairing amplitude is
nearly constant, $\Delta^\star\simeq0.398$~meV, over the central region. It
decreases once the quasiparticle gap closes and drops steeply near the
branch ends. The quasiparticle spectral gap $E_{\rm gap}$ is approximately
$0.23$~meV near $Q_{\rm eq}$ and decreases approximately linearly with
$|Q_y-Q_{\rm eq}|$, consistent with the Doppler shift. The spectrum becomes
numerically gapless near $Q_y=+2.04$ and
$-2.80\times10^{-4}~\text{\AA}^{-1}$, using the criterion
$E_{\rm gap}<5~\mu$eV indicated by the shaded regions. At both points the
gap falls from $7$--$8~\mu$eV to below $0.04~\mu$eV between neighboring
$Q$ points, so the identified grid points are unchanged for thresholds
between $0.04$ and $7~\mu$eV. The condensation
energy $\Delta\Omega$ is negative over the whole branch and has its minimum
at $Q_{\rm eq}$. The current is linear around $Q_{\rm eq}$ and reaches its
extrema inside the gapless regions, at $Q_c^+=+2.29$ and
$Q_c^-=-3.05\times10^{-4}~\text{\AA}^{-1}$. The critical currents of this
branch therefore correspond to gapless depairing, where $\Delta^\star$ is
still close to its zero-current value. This is why their Brillouin-zone
convergence is more demanding than that of $Q_{\rm eq}$ (Sec.~S3.1). The
asymmetry of the two gap-closing momenta about $Q_{\rm eq}$ already signals
the direction-dependent depairing that underlies the diode effect.

\subsection*{S6.2. Temperature dependence}

Figure~\ref{fig:S6}(d)--(f) shows the one-sided in-plane state at $\mu_0$ for
$k_BT=10^{-5}$, $5\times10^{-5}$, and $10^{-4}$~eV, with $g_{\rm SC}$ fixed.
$\Delta^\star(0)$ is $0.400$, $0.398$, and $0.362$~meV, while
$Q_{\rm eq}$ ($-17.87$, $-17.86$, and $-17.81\times10^{-6}~\text{\AA}^{-1}$)
changes by only $0.06\times10^{-6}~\text{\AA}^{-1}$. A fit of the pairing
amplitude $\Delta^\star(0,T)$ to the temperature dependence of the
weak-coupling BCS gap, with the zero-temperature amplitude and $T_c$ as free
parameters, estimates $T_c\approx1.9$~K, close to the onset temperature
reported for single-layer NbSe$_2$~\cite{UgedaNatPhys15}. The three
temperatures are therefore approximately $0.06$, $0.3$, and $0.6\,T_c$. Over the same range, $K_{\rm scale}$ falls by a factor
of $2.5$, from $237.3$ to $95.8$~A/m, and $|K_{\rm diode}|$ by a factor of
$5.8$, from $2.79$ to $0.48$~A/m. The efficiency therefore decreases in magnitude from
$-1.18\%$ to $-0.95\%$ to $-0.50\%$. Temperature thus changes $\eta$ by a factor of
about $2.4$ while the helical momentum changes by a relative $3\times10^{-3}$, another instance of the separation between
the two observables. The values at $k_BT=10^{-5}$ and $10^{-4}$~eV were
obtained with $L=2$. There, $\eta$ and the efficiency from
$d(\Omega_{\rm SC}-\Omega_{\rm N})/dQ$ agree to within $0.006$ percentage points ($-1.176\%$
against $-1.171\%$, and $-0.496\%$ against $-0.502\%$), and the spurious
fraction is at most $0.02$. These values thus pass
the same diagnostics as the production states of Sec.~S3.1, although no
$L=3$ comparison was made at these temperatures. The out-of-plane channel is
far more sensitive to temperature (Sec.~S4.6).

\emph{Sign-reversed helical state.} The tilted state with $\theta=45^\circ$
and $D_b=1.75$~meV (Sec.~S4.1) was recomputed at the same three
temperatures ($L=3$, same $g_{\rm SC}$). It gives $\eta=+0.300\%$,
$+0.354\%$, and $-0.012\%$ at $0.12$, $0.58$, and $1.16$~K ($+0.318\%$,
$+0.360\%$, and $-0.013\%$ from $d(\Omega_{\rm SC}-\Omega_{\rm N})/dQ$), i.e.,
$-0.21$, $-0.30$, and $+0.02$ times the in-plane efficiency at the same
$\Delta D_x$ and temperature. The current scale falls from
$\bar K_c=43.9$ to $28.3$ to $6.9$~A/m ($0.37$, $0.29$, and $0.14$ of the
in-plane values), the current asymmetry from $K_c^+-|K_c^-|=+0.26$ to
$+0.20$ to $-0.002$~A/m, and $\Delta^\star(0)$ from $0.394$ to $0.366$ to
$0.240$~meV; the state is gapless along the whole branch at all three
temperatures. The reversed sign thus persists on cooling, whereas at
$1.16$~K the computed efficiency is small and has the in-plane sign. At
$0.58$~K, the pocket decomposition attributes the reversal to back-flow from
the depaired $K$ and $K'$ pockets (Sec.~S4.1). Its suppression on warming is
consistent with thermal smearing of low-energy quasiparticle structure on the
scale of tens of $\mu$eV, as in the trigonal channel (Sec.~S4.6). The
reversal is therefore confined to the lower temperatures studied. Its
helical momentum, $Q_{\rm eq}=-20.66$, $-21.51$, and
$-22.63\times10^{-6}~\text{\AA}^{-1}$, changes by $2.0\times10^{-6}~\text{\AA}^{-1}$
over this range, more than in the in-plane state but still less than a tenth of
its magnitude.

\subsection*{S6.3. Gap magnitude and choice of pairing subspace}

\emph{Gap magnitude.} To test how the results depend on the size of the
superconducting gap, $g_{\rm SC}$ was recalibrated on the production
quadrature so that the no-exchange pairing amplitude is $\Delta^\star=0.6$~meV at
$k_BT=10^{-5}$~eV, instead of $0.4$~meV ($g_{\rm SC}=0.52605$ against
$0.50068$~eV at $L=2$, and $0.52606$ against $0.50072$~eV at $L=3$). Three
states were recomputed at $k_BT=5\times10^{-5}$~eV
(Table~\ref{tab:S6_gap}). The helical momentum is nearly independent of the pairing amplitude:
$Q_{\rm eq}=-17.87$ against $-17.86\times10^{-6}~\text{\AA}^{-1}$ for the
one-sided in-plane state. Its current scale grows by a factor of $1.71$ and
its current asymmetry by $1.38$, so that $\eta$ decreases in magnitude from $-0.974\%$ to
$-0.785\%$ (both with $dq=4\times10^{-6}$; Table~\ref{tab:S3_dq}). The out-of-plane state with $D_b=1$~meV behaves differently.
Its current asymmetry and current scale both grow by a factor of about
$2.6$, and $\eta$ changes only from $-1.447\%$ to $-1.508\%$. With the larger
gap the trigonal channel is therefore $1.92$ times stronger than the helical
one, against $1.49$ at $\Delta^\star=0.4$~meV (in-plane values at $L=2$ and
$dq=4\times10^{-6}$ in both cases; $1.52$ with the $dq=2\times10^{-6}$
value of Table~\ref{tab:S6_gap}). The trigonal efficiency is not a function of $D_z/\Delta^\star$
alone. The out-of-plane state with $D_b=1.5$~meV at $\Delta^\star=0.6$~meV,
which has the same ratio as $D_b=1$~meV at $0.4$~meV, gives $\eta=-2.428\%$.
Among these cases, the largest efficiency occurs close to the depairing
limit:
$D_b=1.5$~meV at $\Delta^\star=0.4$~meV, which is gapless at $\mathbf Q=0$
with $\bar K_c=5.6$~A/m and is the lowest-grand-potential scalar-gap
solution found at the tested momenta (Sec.~S6.5), gives $-3.354\%$, whereas the same exchange with the
larger gap gives $-2.428\%$ with $\bar K_c=85.2$~A/m. The qualitative
conclusions of the main text---the weak sensitivity of the helical momentum
to the pairing amplitude and the larger trigonal efficiency in these
comparisons---thus hold for both calibration amplitudes.

\begin{table*}[t]
\caption{Dependence on the gap magnitude (production quadrature,
$k_BT=5\times10^{-5}$~eV). $\Delta^\star_{\rm N}$ is the calibrated
no-exchange pairing amplitude at $k_BT=10^{-5}$~eV; the in-plane state is scanned along
$\hat{\mathbf y}$ and the out-of-plane states along $\hat{\mathbf x}$.
$\Delta^\star_{\rm N}$, $\Delta^\star(0)$, and $E_{\rm gap}(0)$ in meV; other
units as in Table~\ref{tab:S4_states}. For the in-plane state $\mathbf Q=0$
is not the equilibrium, so $E_{\rm gap}(0)$ there includes the Doppler shift
of the helical state. $^\dagger$\,$dq=2\times10^{-6}$; all other rows
$dq=4\times10^{-6}$.}
\label{tab:S6_gap}
\begin{ruledtabular}
\begin{tabular}{lcccccccccc}
State & $\Delta^\star_{\rm N}$ & $L$ & $\Delta^\star(0)$ & $E_{\rm gap}(0)$ &
$Q_{\rm eq}$ & $K_c^+$ & $K_c^-$ & $\bar K_c$ & $\eta$ (\%) &
$\eta_{\Delta\Omega}$ (\%)\\
\hline
In-plane, $D_b=1$~meV$^\dagger$ & 0.4 & 2 & 0.398 & 0.189 & $-17.86$ & 95.64 & $-97.48$ & 96.56 & $-0.952$ & $-0.980$\\
 & 0.6 & 2 & 0.600 & 0.337 & $-17.87$ & 163.46 & $-166.05$ & 164.75 & $-0.785$ & $-0.801$\\
\hline
Out-of-plane, $D_b=1$~meV & 0.4 & 3 & 0.385 & 0.027 & 0.05 & 48.26 & $-49.68$ & 48.97 & $-1.447$ & $-1.437$\\
 & 0.6 & 3 & 0.599 & 0.153 & $-0.04$ & 123.14 & $-126.92$ & 125.03 & $-1.508$ & $-1.506$\\
\hline
Out-of-plane, $D_b=1.5$~meV & 0.4 & 2 & 0.312 & 0.000 & 0.34 & 5.40 & $-5.78$ & 5.59 & $-3.354$ & $-3.425$\\
 & 0.6 & 3 & 0.593 & 0.049 & 0.05 & 83.09 & $-87.23$ & 85.16 & $-2.428$ & $-2.418$\\
\end{tabular}
\end{ruledtabular}
\end{table*}

\emph{Pairing subspace.} The main calculations restrict local singlet
pairing to Nb $d_{z^2}$, $d_{x^2-y^2}$, and $d_{xy}$. This is the standard
choice in models of Ising superconductivity in monolayer NbSe$_2$, where
the same three orbitals and an intra-orbital attraction are used
\cite{HeCommPhys18}, and it is consistent with the conventional
phonon-mediated pairing obtained from first principles
\cite{LianNanoLett18}. In our Wannier model these $\sigma_h$-even orbitals
carry $0.73$ of the spectral weight at $E_F$ ($d_{z^2}$ $0.30$,
$d_{x^2-y^2}$ and $d_{xy}$ $0.22$ each), the Se $p$ orbitals $0.27$, and the
$\sigma_h$-odd $d_{xz}$ and $d_{yz}$ only $0.001$ (orbital-projected density of
states at $\mu_0$ on a $150\times150$ mesh, Gaussian broadening of $5$ to
$20$~meV; the fractions change by less than $0.003$). The Se weight belongs to
the same hybridized bands, so the anomalous amplitude on the Se orbitals is
generated self-consistently even without a pairing interaction there. Additional
on-site, orbital-diagonal singlet pairing can be chosen to transform in the
fully symmetric representation, with equal amplitudes within each orbital
doublet and symmetry-related amplitudes on the two Se sublayers. Such an
extension would preserve the symmetry constraints of Secs.~S1 and S4,
provided the superconducting solution retains these symmetries. It could,
however, modify the band-resolved gaps and the magnitudes of $Q_{\rm eq}$,
$K_c^\pm$, and $\eta$. These effects have not been tested; calibrating an
overall pairing amplitude does not fix the full gap distribution. The interaction is singlet only. The triplet
correlations that the exchange and the Ising spin--orbit coupling induce
are contained in the BdG Green's function, but an attractive triplet
channel, which can reshape the in-plane diode
response~\cite{LevchenkoKhodasArXiv26}, is not included.

\begin{figure*}[t]
  \centering
  \includegraphics[width=\textwidth]{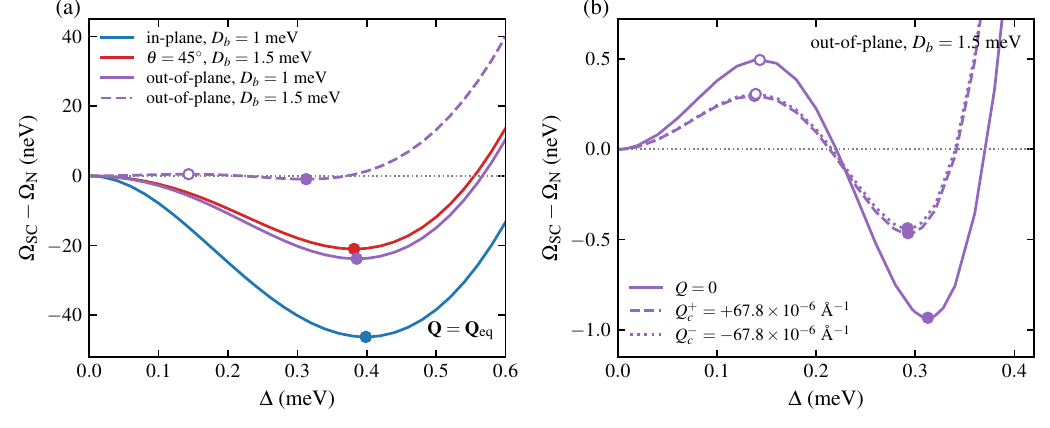}
  \caption{Stability of the selected solutions within the scalar-gap
  ansatz (production quadrature, $k_BT=5\times10^{-5}$~eV). Grand potential
  relative to the normal state, $\Omega_{\rm SC}(\Delta)-\Omega_{\rm N}$, per
  primitive cell at fixed $\mathbf Q$. (a) Four states of
  Table~\ref{tab:S6_stab} (in-plane; tilted, $\theta=45^\circ$,
  $D_b=1.5$~meV; out-of-plane, $D_b=1$ and $1.5$~meV) at the equilibrium
  momentum ($\mathbf Q=0$ for the out-of-plane states). (b) Out-of-plane state with $D_b=1.5$~meV at
  $\mathbf Q=0$ and at the critical momenta. Filled circles are roots of $G$
  that are minima of $\Omega_{\rm SC}(\Delta)$ (the production
  $\Delta^\star$); open circles are roots that are maxima.}
  \label{fig:S7}
\end{figure*}

\subsection*{S6.4. Scope of the proximity model}

The layer-selective on-site exchange of Eq.~\eqref{eq:S2_Hex} is an
effective proximity self-energy acting on the Se $p$ orbitals. The model
does not include an explicit magnetic substrate, interface reconstruction,
charge transfer, or interface-induced changes of the hopping amplitudes. In
a real heterostructure, charge transfer would shift $\mu$, and interface
hybridization would add layer-odd hopping terms that break $\sigma_h$ even
without magnetism. Both effects are allowed by symmetry, and the second
would contribute to $Q_{\rm eq}$ only together with time-reversal breaking.

The induced exchange of a real interface is also not confined to the
interfacial Se layer. It generally extends to the Nb $d$ orbitals, which
carry most of the Fermi-surface weight. Because Nb lies in the $\sigma_h$
plane, $\sigma_h$ maps the Nb site onto itself, so that an exchange field
on Nb transforms as the layer-even components: an in-plane Nb field behaves as
$D_{+,\parallel}$ and an out-of-plane Nb field as $D_{+,z}$
[Eq.~\eqref{eq:S4_parity}]. The selection rules are therefore unchanged. For
in-plane magnetization, an Nb field cannot rectify on its own; it adds pair
breaking, which Ising spin--valley locking weakens, and the helical channel
remains controlled by the layer-odd Se contrast. For out-of-plane
magnetization, it contributes directly to the trigonal channel and to its
pair breaking. The relative size of the Nb and Se components, and with it
the magnitude of $\eta$, depends on the interface and has to be determined
for a specific heterostructure.

The selection rules of Secs.~S1 and S4 hold for any extension of the model
that preserves the relevant magnetic symmetries: $\sigma_h\mathcal T$ for the
reciprocity of parallel in-plane and antiparallel out-of-plane textures, and $M_x\mathcal T$ and $C_3$ for the
trigonal channel. Interface terms that break them, such as an asymmetric
interface potential or Rashba coupling on one side, can add responses; the
exact reciprocity of these textures assumes mirror-equivalent interfaces.
For a one-sided interface, a layer-odd potential $V_-$ allows, for example,
a term $\mathbf Q\cdot(\hat{\mathbf z}\times\mathbf D_{+,\parallel})V_-$, so
that the layer-even in-plane exchange can then contribute in proportion to
$V_-$; we have not computed its size.
The magnitudes of $Q_{\rm eq}$ and $\eta$ are model-specific. The results should therefore be
read as microscopic design principles within a controlled effective model,
not as parameter-free predictions for a particular heterostructure. The
uniform-amplitude finite-$Q$ ansatz also excludes vortex entry,
inhomogeneous depairing, and contact effects. The computed $K_c^\pm$ are
therefore intrinsic depairing currents of the uniform model; measured
switching currents also depend on these effects and on the device geometry. Field-free rectification with a zero-momentum
condensate, the signature of the trigonal channel, has been reported in
NbSe$_2$/CrBr$_3$ heterostructures with out-of-plane magnetization, with
efficiencies of several tens of percent \cite{SunArXiv26}. A quantitative
comparison with the intrinsic values computed here would require the
interface-specific exchange and the switching mechanism of the device.

\emph{Broadening and layer number.} The calculations are for clean
monolayers. In the states compared here, the helical sign reversal and the
larger trigonal efficiency are present at $k_BT\leq5\times10^{-5}$~eV but
absent at $10^{-4}$~eV (Secs.~S4.6 and S6.2). This temperature dependence
indicates sensitivity to low-energy quasiparticle structure. Disorder or
phenomenological spectral broadening may also modify these features, but
neither has been included, so no quantitative broadening threshold is
established. The symmetry selection rules continue to apply to extensions
that preserve the relevant magnetic symmetries. They are derived for the
monolayer; analogous rules follow by symmetry, without further calculation,
for ideal $2H$ stacks with symmetry-equivalent interfaces and
correspondingly symmetric exchange profiles. In a $2H$ bilayer, inversion
relates the two layers, leaves axial vectors unchanged, and reverses
$\mathbf Q$. A texture with parallel
exchange on the two outer Se layers is then inversion symmetric and
reciprocal for any orientation, whereas an antiparallel texture may rectify.
For out-of-plane exchange this is the reverse of the monolayer rule
(Sec.~S4.6). Ideal $2H$ stacks with an odd number of layers retain
$\sigma_h$ through the middle layer and follow the monolayer rules, so that
out-of-plane spin valves should show an even--odd dependence on the layer
number. By the same argument, a uniform out-of-plane field rectifies in
even-layer $2H$ stacks only through inversion breaking by the substrate or
the surfaces. The diode effect observed in a bilayer constriction, with
efficiencies above $60\%$~\cite{BauriedlNatCommun22}, therefore points to such
inversion breaking or to extrinsic contributions. The absence of a
correlation between the efficiency and the current direction reported for
etched constrictions~\cite{BauriedlNatCommun22} is consistent with this. The
angular rule of the trigonal channel is best tested in field-free devices
without etched constrictions.

\subsection*{S6.5. Stability of the selected solutions within the scalar-gap ansatz}

Section~S2.5 shows that the largest simple positive root of the gap equation is a
local minimum of $\Omega_{\rm SC}(\Delta)$ and that continuation in $Q$
preserves the character of the selected root up to a fold. We verified
directly that the selected root is the largest positive one and examined the other
solutions. For six
states that cover the two channels, the most strongly depaired cases, and
the sign-reversed helical states (one-sided in-plane $D_b=1$~meV with
$L=2$; tilted $\theta=45^\circ$, $D_b=1.5$ and $1.75$~meV, and
$\theta=30^\circ$, $D_b=1.5$~meV, with $L=3$; one-sided out-of-plane
$D_b=1$~meV with $L=3$; one-sided out-of-plane $D_b=1.5$~meV with $L=2$),
the residual $G(\Delta)$
and $\Omega_{\rm SC}(\Delta)$ were tabulated at the equilibrium momentum and
at both critical momenta, on a grid from $0.002$ to $0.6$~meV refined around
the production $\Delta^\star$. For the in-plane and tilted states the grid
point nearest to $Q_{\rm eq}$ was used, and for the out-of-plane states
$\mathbf Q=0$; the critical momenta are those of the production branches.

\begin{table}[!t]
\caption{Stability scan (production quadrature,
$k_BT=5\times10^{-5}$~eV). For each state the rows are the equilibrium
momentum (nearest grid point, or $\mathbf Q=0$) and $Q_c^+$, $Q_c^-$. $Q$
in $10^{-6}~\text{\AA}^{-1}$; $\Delta^\star$, the production gap, and
$E_{\rm gap}$ at $\Delta^\star$ in meV; $\Delta\Omega=\Omega_{\rm
SC}(\Delta^\star)-\Omega_{\rm N}$ in neV per primitive cell. In all rows
$\Delta^\star$ is the largest positive root of $G$ in the scanned range
($0.002$--$0.6$~meV) and a minimum of $\Omega_{\rm SC}(\Delta)$. $\Delta_2$
is the only other positive root (a maximum), and
$\Delta\Omega_2$ is its grand potential relative to the normal state.}
\label{tab:S6_stab}
\begin{ruledtabular}
\begin{tabular}{lrcccc}
State & $Q$ & $\Delta^\star$ & $E_{\rm gap}$ & $\Delta\Omega$ &
$\Delta_2$ ($\Delta\Omega_2$)\\
\hline
In-plane, & $-17.0$ & 0.3985 & 0.209 & $-46.24$ & --\\
$D_b=1$~meV, & $250.1$ & 0.3672 & 0 & $-14.79$ & --\\
$L=2$ & $-288.3$ & 0.3677 & 0 & $-14.11$ & --\\
\hline
Tilted $45^\circ$, & $-17.0$ & 0.3817 & 0.006 & $-20.99$ & --\\
$D_b=1.5$~meV, & $178.1$ & 0.3279 & 0 & $-9.67$ & --\\
$L=3$ & $-212.0$ & 0.3297 & 0 & $-9.94$ & --\\
\hline
Tilted $45^\circ$, & $-25.4$ & 0.3667 & 0 & $-12.86$ & --\\
$D_b=1.75$~meV, & $135.7$ & 0.3207 & 0 & $-7.07$ & --\\
$L=3$ & $-178.1$ & 0.3201 & 0 & $-7.13$ & --\\
\hline
Tilted $30^\circ$, & $-17.0$ & 0.3595 & 0 & $-10.10$ & --\\
$D_b=1.5$~meV, & $127.2$ & 0.3186 & 0 & $-5.92$ & --\\
$L=3$ & $-152.6$ & 0.3183 & 0 & $-5.97$ & --\\
\hline
Out-of-plane, & 0 & 0.3854 & 0.027 & $-23.78$ & --\\
$D_b=1$~meV, & $203.5$ & 0.3322 & 0 & $-10.88$ & --\\
$L=3$ & $-203.5$ & 0.3347 & 0 & $-10.76$ & --\\
\hline
Out-of-plane, & 0 & 0.3125 & 0 & $-0.93$ & 0.143 ($+0.49$)\\
$D_b=1.5$~meV, & $67.8$ & 0.2925 & 0 & $-0.46$ & 0.138 ($+0.30$)\\
$L=2$ & $-67.8$ & 0.2920 & 0 & $-0.44$ & 0.139 ($+0.31$)\\
\end{tabular}
\end{ruledtabular}
\end{table}

Figure~\ref{fig:S7} and Table~\ref{tab:S6_stab} summarize the result. The
finite-difference derivative $\partial\Omega_{\rm SC}/\partial\Delta$
between neighboring grid points equals $-(2N_{\rm pair}/g_{\rm SC})G$,
prefactor included, to within $1.3\times10^{-3}$ of its largest magnitude, and
integrating $-(2N_{\rm pair}/g_{\rm SC})G$ over $\Delta$ reproduces
$\Omega_{\rm SC}-\Omega_{\rm N}$ to within $4\times10^{-3}$ of its range. This
confirms the identity of Sec.~S2.5, and with it the normalization of
Eq.~\eqref{eq:S2_Omega_SC}, on the production quadrature. In all eighteen cases
the production $\Delta^\star$ has $|G|<0.7\times10^{-8}$~eV, within the
tolerance of Sec.~S3.3, and lies within $0.18~\mu$eV of a root of $G$. Within the
scanned amplitude range, this root is the largest positive root, a local
minimum of $\Omega_{\rm SC}(\Delta)$, and the solution of lowest grand
potential, the normal state included. Because $\Omega_{\rm SC}$ is
stationary at the root, its error at fixed $Q$ is quadratic in the small
offset of $\Delta^\star$. The effect on the finite-difference current also
depends on the variation of this energy error between neighboring $Q$
points.

For all states except the out-of-plane state with $D_b=1.5$~meV,
including the two sign-reversed tilted states, which are gapless already at
$Q_{\rm eq}$, this is the only positive root found. $G>0$ at every scanned
$\Delta$ with $0<\Delta<\Delta^\star$, consistent with a local maximum at
the normal state; $\Omega_{\rm SC}$ decreases over the sampled interval up
to $\Delta^\star$
[Fig.~\ref{fig:S7}(a) and Table~\ref{tab:S6_stab}]. The out-of-plane state with $D_b=1.5$~meV is
qualitatively different [Fig.~\ref{fig:S7}(b)]. There $G<0$ at the
smallest positive amplitudes sampled, consistent with a local minimum at the
normal state, and a second root near $0.14$~meV is a maximum of
$\Omega_{\rm SC}$ that separates the normal state from the selected
solution. This is the structure expected for a Zeeman field
along the Ising axis near the Pauli limit, where the transition becomes first
order at low temperature. At the tested momenta the selected solution
nevertheless has the lowest grand potential within the scanned range. It lies $0.93$~neV per primitive cell below the normal state
at $\mathbf Q=0$ and $0.44$--$0.46$~neV below it at $Q_c^\pm$, while the
maximum between them lies $0.3$--$0.5$~neV above the normal state. The
largest trigonal efficiency of Sec.~S6.3, $-3.354\%$, therefore belongs to the lowest-grand-potential
scalar-gap solution found at the tested momenta, although its condensation
energy is only $0.04$ times that at $D_b=1$~meV.

The selected root is thus a local minimum and the
lowest-grand-potential scalar-gap solution found at the sampled
equilibrium and critical momenta. Between these momenta the
production branch lies below the normal state at every sampled
$Q$ (Sec.~S2.5), and the extraction rule of Sec.~S2.6 excludes
abrupt steps of $\Delta^\star(Q)$. These checks support a smooth
superconducting branch, but do not exclude a competing solution
that appears only between the tested momenta. The reported
$K_c^\pm$ are the depairing extrema of this selected branch.
Stability against spatially varying amplitudes or other pairing
channels is not tested here (Sec.~S2.5).



\end{document}